\documentclass[twoside]{article}
\usepackage{geometry}
\usepackage{ifthen}
\usepackage{hyperref}
\usepackage{algorithm}
\usepackage{algpseudocode}

\usepackage{caption}
\usepackage{subcaption}

\usepackage{calc}
\usepackage{tikz}
\usepackage{amsmath, amsfonts, bm, amssymb, tabularx}
\usepackage{latexsym, mathtools}
\usepackage{Shorthands}

\usepackage[round]{natbib}

\usepackage{cleveref}
\usepackage{thmtools} 
\usepackage{thm-restate}

\title{Performance-Robustness Tradeoffs in Adversarially Robust Control and Estimation}

\author{Bruce D.\ Lee, Thomas T.C.K.\ Zhang, Hamed Hassani, and Nikolai Matni}

\date{Department of Electrical and Systems Engineering, University of Pennsylvania}

\newcounter{tac}
\begin{document}

\maketitle
 \begin{abstract}
   While $\mathcal{H}_\infty$ methods can introduce robustness against worst-case perturbations, their nominal performance under conventional stochastic disturbances is often drastically reduced. Though this fundamental tradeoff between nominal performance and robustness is known to exist, it is not well-characterized in quantitative terms. Toward addressing this issue, we borrow the increasingly ubiquitous notion of adversarial training from machine learning to construct a class of controllers which are optimized for disturbances consisting of mixed stochastic and worst-case components. We find that this problem admits a linear time invariant optimal controller that has a form closely related to suboptimal $\mathcal{H}_\infty$ solutions. We then provide a quantitative performance-robustness tradeoff analysis in two analytically tractable cases: state feedback control, and state estimation. In these special cases, we demonstrate that the severity of the tradeoff depends in an interpretable manner upon system-theoretic properties such as the spectrum of the  controllability gramian, the spectrum of the observability gramian, and the stability of the system. This provides practitioners with general guidance for determining how much robustness to incorporate based on a priori system knowledge. We empirically validate our results by comparing the performance of our controller against standard baselines, and plotting tradeoff curves.
\end{abstract}
\section{Introduction}

Modern control systems, from mobile robotics to power plants, require controllers that are simultaneously fast, performant, and robust. Many control schemes attempt to achieve these desiderata by combining them into a single objective function and optimizing it, leading to a natural tradeoff. A controller optimized for speed and efficiency may perform poorly in the face of unmodeled phenomena. For instance, Linear-Quadratic Gaussian (LQG) controllers (a special case of $\calH_2$ controllers) explicitly prioritize nominal performance by penalizing the expectation of a quadratic function of the state and input. However, such controllers can be arbitrarily fragile to small perturbations of the dynamics \citep{doyle1978guaranteed}. Replacing the LQG objective with one that considers the response of the system to worst-case dynamic uncertainty and external disturbances results in robust control methods, such as $\calH_\infty$ and $\mathcal{L}_1$ methods \cite{zhou1996robust,dahleh1994control}. Such controllers are provably robust, however they tend to be overly conservative in practice.

Toward balancing the performance of nominal and robust controllers, various approaches have been introduced, most notably mixed $\calH_2/\calH_\infty$ methods. However, the resulting controllers are often complicated to express and compute \citep{carsten1995mixed}, and lack a quantitative characterization of the tradeoff between performance and robustness as a function of system properties. 
Toward addressing these issues, we take inspiration from the notion of adversarial robustness from machine learning \citep{biggio2013evasion,goodfellow2014explaining,madry2017towards, carlini2017towards} and formulate a controller synthesis problem that balances performance and robustness. The goal of adversarial robustness in machine learning is to minimize the expected error under the presence of worst-case norm-bounded perturbations to the data, where the perturbations can depend on the underlying stochasticity of the problem, such as the data distribution and additive noise. We consider an analogous adversarially robust output feedback control problem where we aim to minimize an expected quadratic cost subject to linear dynamics driven by process noise composed of two components: a zero-mean stochastic noise term and a norm-bounded adversarial term. We show that the solution to this problem may be expressed in terms of the solutions to a set of coupled Discrete Algebraic Riccati equations (DAREs). For several special cases, this allows for novel quantitative performance-robustness tradeoff bounds to be computed in which system parameters manifest in an interpretable way. 

\subsection{Contributions}
Toward analyzing the adversarially robust control problem we propose, we first show that when viewed through the lens of dynamic games \citep{basar1991hinf}, adversarially robust LQ control relates to a control problem introduced in \cite{doyle1989optimal}. We show that the optimal solution to the state feedback version of this problem is given by a central static suboptimal $\calH_\infty$ controller, with suboptimality level $\gamma$ depending on both the stochastic noise statistics and the budget given to the adversary. Furthermore, both the worst-case adversary and the corresponding optimal controller can be computed from the solution of a DARE. We then leverage the state feedback solution to provide sufficient conditions for a solution to the output feedback control problem, along with an algorithm to synthesize this controller. Unlike the state feedback setting, the output feedback adversarially robust controller is distinct from a central suboptimal $\calH_\infty$ controller, and involves finding the solution to a set of coupled Riccati equations. 

We leverage the solution to the adversarially robust control problem to quantify the performance-robustness tradeoffs  analytically in two simplified settings: state feedback LQ control, and state estimation. In these settings, we show an interpretable dependence on underlying system-theoretic properties such as controllability, observability, and stability. For the state feedback control setting, we show that the cost gap incurred by the adversarially robust controller in the nominal setting, relative to that achieved by the nominal controller, is upper bounded by $O(\sigma_w^2 \gamma^{-4} \nu^{-1})$. In this expression, $\sigma_w^2$ is the covariance of the additive noise distribution, $\gamma$ is the suboptimality level of the suboptimal $\calH_\infty$ controller, and $\nu$ is the smallest singular value of the controllability gramian. On the other hand, the cost gap is lower bounded by $\Omega\paren{\sigma_w^2 \gamma^{-4} \eta^2 }$, where $\eta$ 
is the largest singular value of the controllability gramian for closed-loop system under the nominal LQ controller with disturbances as the input. These results quantitatively show that systems with uniformly good controllability have small performance-robustness tradeoffs, while those that have a highly controllable mode in the nominal closed-loop system (when viewing disturbances as inputs) lead to large performance-robustness tradeoffs. Similar bounds are shown for the state estimation setting, with controllability gramians replaced by observability gramians.

We conclude with numerical experiments. These simulations indicate that the proposed algorithm for synthesizing the controller is effective, and that the controller performs well relative to standard baselines in stabilizing an inverted pendulum. We additionally illustrate that the analytical trends described above are present in numerical experiments. We demonstrate that in the setting of output feedback control, the impact of poor controllability and poor observability compound to increase the severity of the performance-robustness tradeoff.

\subsection{Related Work}

The mixed stochastic/worst-case problem that we consider is not the only way to strike a balance between the performance of stochastic and robust control methods. Most closely related is \cite{doyle1989optimal}, which considers a similar problem from a deterministic perspective, in which disturbances are composed of both a bounded power component, and a bounded power spectrum component. 
A set description of disturbances that also interpolates between $\calH_2$ and $\calH_\infty$ approaches is proposed in \cite{paganini1993set}.
The class of all stabilizing controllers subject to a $\calH_\infty$ norm constraint is characterized in \cite{glover1988state}, while minimizing an $\calH_2$ objective subject to a $\calH_\infty$ constraint is addressed in \cite{bernstein1988LQG, rotea1991h2}. While conceptually appealing, these methods often lack a simple closed-form stationary solution. A solution to risk-sensitive LQG control which interpolates between $\calH_2$ and $\calH_\infty$ solutions is offered in in \cite{whittle1981risk}. Other recent work also attempts to reduce the conservatism of robust control through risk aware approaches \citep{tsiamis2021linear, chapman2022toward} or regret-minimization \citep{goel2020regret, goel2021competitive, hazan2020nonstochastic}.  None of the aforementioned methods characterize the performance-robustness tradeoffs of the resulting controllers.

Analogous recent work in the machine learning community has analyzed performance-robustness tradeoffs in adversarially robust learning, including precise characterizations of the generalization errors of standard versus adversarially trained models under various theoretical assumptions \citep{zhang2019theoretically,javanmard2020precise, hassani2022curse}, and ``no free lunch'' theorems for obtaining adversarially robust models \citep{tsipras2018robustness, dohmatob2019generalized, yin2019rademacher}.  The successful characterization of such performance-robustness tradeoffs in machine learning motivates the control objective we consider. However, the theoretical results from this area are largely intended for the supervised learning setting and do not immediately apply to our setting. The existence of  performance-robustness tradeoffs in control is shown in \cite{al2020accuracy},  but they are not characterized quantitatively.  Our prior work studies performance robustness tradeoffs in the setting of state feedback control \citep{lee2022performancerobustness} and finite horizon state estimation \citep{zhang2021adversarial}. This paper expands upon these works to consider performance robustness tradeoffs of both control and state estimation in a unified framework by considering them both as special cases of adversarially robust output feedback control. We end by noting that the extension of adversarial robustness results in machine learning to various control problems has recently received attention \citep{zhang2022adversarially, havens2022revisiting, pattanaik2017robust, tan2020robustifying, mandlekar2017adversarially, kuutti2021arc}.

\textbf{Notation:} The Euclidean norm of a vector $x$ is denoted by $\norm{x}$. For a matrix $A$, the spectral norm is denoted $\norm{A}$ and the Frobenius norm is denoted $\norm{A}_F$. The spectral radius of a square matrix $A$ is denoted $\rho(A)$. A symmetric, positive semidefinite (psd) matrix $A = A^\top$ is denoted $A \succeq 0$, and a symmetric positive definite (pd) matrix is denoted $A \succ 0$. Similarly $A \succeq B$ denotes that $A-B$ is positive semidefinite.
A sequence of vectors $x_t$ defined for $t \geq 0$ will be denoted by $\mathbf{x} = \curly{x_t}_{t \geq 0}$. The $\ell^2$ signal-norm of a sequence is denoted by $\norm{\mathbf{x}}_{\ell^2} := (\sum_{t\geq 0}\norm{x_t}^2)^{1/2}$.
For an autonomous system $x_{t+1} = Ax_t$ and symmetric matrix $Q$, we denote the solution $P$ to the discrete Lyapunov equation $$A^\top P A - P + Q = 0$$ by $\dlyap(A, Q)$.  Similarly, for a controlled system $x_{t+1} = Ax_t + Bu_t$ and symmetric matrices $Q, R$ of compatible size, we denote the solution $P$ to the discrete algebraic Riccati equation $$Q + A^\top P A - A^\top P B(B^\top P B + R)^{-1}B^\top P A = P$$ by $\dare(A, B, Q, R)$.
\section{Adversarially Robust Linear-Quadratic Control} \label{sec: advrobust lq control}

Consider a partially observed discrete-time linear-time-invariant (LTI) system with state  and measurement disturbances composed of both stochastic and adversarial components: let $x_t \in \R^{n_x}$ be the system state, $u_t \in \R^{n_u}$ the input, $y_t \in \R^{n_y}$ the measurement, $w_t \in \R^{n_w}$ and $\delta_t \in \R^{n_\delta}$ the stochastic and adversarial components of the process disturbance, respectively. The initial condition $x_0$ and stochastic component of the process noise $w_t$ are assumed to be i.i.d.\ zero-mean with covariance matrices $\Sigma_0,\;I$, respectively, and $\Ex\brac{x_0 w_t^\top}= 0$ for all $t \geq 0$. The performance will be measured by an output signal $z_t \in \R^{n_z}$ which depends on the current state and input. The LTI system is then defined by the following equations:
\begin{equation}\label{eq:ss}
    \begin{aligned}
     x_{t+1} &= Ax_t + B_0 w_t + B_1 \delta_t + B_2 u_t  \\
     z_t &= C_1 x_t + D_{12} u_t \\
     y_t &= C_2 x_t + D_{20} w_t + D_{21} \delta_t.
\end{aligned}
\end{equation}
We denote this system compactly by
\begin{align*}
        G = \left[\begin{array}{c|ccc}
           A & B_0 & B_1 & B_2 \\ 
           \hline 
           C_1 & D_{10} & 0 & D_{12}  \\ 
           C_2 & D_{20} & D_{21} & 0 
        \end{array}\right],
        \end{align*}
and adopt the notation $\mathcal{F}_l(G, K)$ to denote its feedback interconnection with a controller $K$ that takes the signal $y_t$ as input, and outputs control action $u_t$.

We assume that the adversarial perturbation sequence $\boldsymbol{\delta}$ is causal, i.e.,\ it can depend only on the states, inputs, and stochastic noise up to the current timestep; in particular, $\delta_t$ must be a measurable function of the randomness $x_0, w_{0:t}$.
We consider the infinite horizon objective defined in terms of the performance signal 
\begin{align}
    \label{eq: objective}
    \limsup_{T\to\infty} \frac{1}{T} \Ex_{\mathbf{w}, x_0} \brac{\sum_{t=0}^{T-1} \norm{z_t}^2},
\end{align}
subject to the dynamics \eqref{eq:ss}. We assume that $C_1 = \bmat{Q^{1/2} \\ 0}$ and $D_{12} = \bmat{0 \\ R^{1/2}}$ for $Q \succeq 0$, and $R \succ 0$, so that \eqref{eq: objective} reduces to the linear quadratic regulation cost. 
Therefore, if the adversarial perturbations $\boldsymbol \delta$ are set to zero, then the system \eqref{eq:ss} with objective \eqref{eq: objective} forms the nominal Linear Quadratic Gaussian (LQG) problem. If the stochasticity is set to zero ($\boldsymbol{w}, x_0 =0$) and $\boldsymbol{\delta}$ are worst-case perturbations with average power bounded by $\varepsilon$, then we instead recover the $\calH_\infty$ problem. When both stochastic noise and worst-case perturbations are present, we define the resulting control task as the \textit{adversarially robust LQG problem}.
We denote the three corresponding objectives by nominal cost ($\NC$), robust cost ($\RC$), and adversarial cost ($\AC$) respectively: 
\begin{align}
     \label{eq:nominal_cost}
    \NC(K) &= \limsup_{T \to \infty} 
    \frac{1}{T} \Ex_{\mathbf{w}, x_0}  \brac{ \sum_{t=0}^{T-1} \norm{z_t}^2}, \;\; \boldsymbol\delta = 0, 
    \\
    \label{eq:robustcost}
    \RC(K) &= \limsup_{T \to \infty} \frac{1}{T} \max_{\substack{\boldsymbol{\delta} \mbox{ causal} \\ \norm{\boldsymbol{\delta}}_{\ell_2}^2 \leq T \varepsilon}} \sum_{t=0}^{T-1} \norm{z_t}^2, \;\;  \bmat{x_0 \\ \boldsymbol w} = 0, 
    \\
    \label{eq:hardadversarialLQRobjective} 
    \AC(K)  &= \limsup_{T \to \infty} \frac{1}{T} \Ex_{\mathbf{w}, x_0}  \brac{\max_{\substack{\boldsymbol{\delta} \mbox{ causal} \\ \norm{\boldsymbol{\delta}}_{\ell_2}^2 \leq T \varepsilon}} \sum_{t=0}^{T-1} \norm{z_t}^2}. 
\end{align}

The constraint $\norm{\boldsymbol{\delta}}_{\ell^2}^2 \leq T\varepsilon$ is chosen such that the instance-wise adversarial budget satisfies $\norm{\delta_t}^2 \leq \varepsilon$ on average.
In order to ensure that there exists a stabilizing controller, and that minimizing either $\NC$ or $\RC$ provides a stabilizing controller, we make the following standard assumption \citep{zhou1996robust}.
\begin{assumption} $\;$
\label{as: system assumptions}
\begin{itemize}
    \item $(A,B_2,C_2)$ is stabilizable and detectable.
    \item $(A, Q^{1/2})$ is detectable.
    \item $R \succ 0$, $B_0 B_0^\top \succ 0$, $D_{20} D_{20}^\top \succ 0$,  $B_0 D_{20}^
    \top = 0$.
\end{itemize}
\end{assumption}
Under this assumption, there exists a stabilizing controller minimizing $\NC$, of the form
\begin{align*}
    \hat x_t &= (A+B_2 F_\star)(I - L_\star C_2) \hat x_t + (A+B_2 F_\star)L_\star y_t \\
    u_t & = F_\star(I-L_\star C_2)\hat x_t + F_\star L_\star y_t,
\end{align*}where
\begin{align}
    \label{eq:lqr}
    F_\star &= -(R + B_2^\top P_\star B_2)^{-1} B_2^\top P_\star A \\
    L_\star &= \Sigma_\star C_2^\top (C_2 \Sigma_\star C_2^\top + D_{20} D_{20})^{-1},
\end{align}
and $P_\star$ and $\Sigma_\star$ solve the following two Discrete Algebraic Riccati Equations (DARE):
\begin{align}
    \label{eq: nominal control riccati}
    P_\star &= \dare(A,B,Q,R) \\
    \label{eq: nominal estimation riccati}
    \Sigma_\star &= \dare(A^\top, C^\top, B_0 B_0^\top, D_{20} D_{20}^\top). 
\end{align}
The above solution is called the LQG controller. A solution minimizing $\RC$ is known as an $\calH_\infty$ controller, and may be expressed in terms of the solution to two modified DAREs \cite{basar1991hinf, hassibi1999}.

The remainder of this section is devoted to finding a controller minimizing $\AC$.  
Inspired by minimax dynamic games \citep{basar1991hinf}, we first find a controller which minimizes a soft-constrained version of the adversarial cost \eqref{eq:hardadversarialLQRobjective}:

\begin{align}
    \label{eq:adversarialLQRobjective}
     \limsup_{T \to \infty} 
    \frac{1}{T} \Ex \brac{\max_{\boldsymbol{\delta} \mbox{ causal}} \sum_{t=0}^{T-1} \norm{z_t}^2 - \gamma^2 \norm{ \delta_t}^2}.
\end{align}

The following lemma provides necessary and sufficient conditions for the existence of a stabilizing controller which minimizes the above objective in the state feedback setting ($C_2=I$, $D_{20}=0$, and $D_{21}=0$.) Similar statements in continuous time may be extracted from a result found in \cite{doyle1989optimal}.

\begin{lemma}
    \label{lem:adversarial LQR sol} 
    Suppose there exists a solution to the following DARE
    \begin{align}
         \label{eq:modified DARE}
        P_\gamma = \dare\paren{A,\bmat{B_1 & B_2}, Q, \bmat{-\gamma^2 I \\ & R}} 
    \end{align}
    satisfying $P_\gamma \succeq 0$ and $ B_1^\top P_\gamma B_1 \prec \gamma^2 I$. When the above condition holds, 
    \begin{enumerate}
        \item Let 
        \begin{align}
            \label{eq:shorthand psi}
            \Psi = \bmat{B_1 & B_2}^\top P_\gamma \bmat{B_1 & B_2} +\bmat{-\gamma^2 I  & 0\\ 0 & R}.
        \end{align} 
        Then the controller $u_t = F_\gamma x_t$ with $F_\gamma$ given by
        \begin{equation}
        \label{eq:suboptimal adversarial controller}
        \begin{aligned}
             \bmat{E_\gamma \\ F_\gamma} := -\Psi^{-1} \bmat{B_1 & B_2}^\top P A
        \end{aligned}
        \end{equation}
        satisfies $\rho(A+B_2 K) < 1$, and minimizes objective  \eqref{eq:adversarialLQRobjective}. \label{num:1}
        \item \label{num:2} The optimal adversarial perturbation under the controller $u_t=F_\gamma x_t$ is given by
        \begin{equation}
        \label{eq:adversarial perturbation}
        \begin{aligned}
            \Delta_\gamma &= (\gamma^2 I - B_1^\top PB_1 )^{-1} B_1^\top P B_0, \\
            \delta_t &=  E_\gamma x_t + \Delta_\gamma w_t.
        \end{aligned} 
        \end{equation}
        \item  The objective value \eqref{eq:adversarialLQRobjective} achieved under controller \eqref{eq:suboptimal adversarial controller} and adversary \eqref{eq:adversarial perturbation} is  $\trace(M_\gamma B_0 B_0^\top)$,
        where $M_\gamma = P_\gamma + P_\gamma B_1 (\gamma^2 I - B_1^\top P_\gamma B_1)^{-1} B_1^\top P_\gamma$. 
    \end{enumerate}
\end{lemma}

The solution approach for the above problem follows that in \cite{basar1991hinf} for minimax games. 
The finite horizon version of the problem is solved by defining a saddle point cost-to-go, then recursing backwards in time, solving both an unconstrained concave maximization problem and convex minimization problem at each time step to determine the optimal adversarial perturbation and control input. The causality of $\boldsymbol \delta$ is necessary in the recursion
to pull $\delta_t$ out of an expectation over future noise terms. Taking the limit as the horizon tends to infinity provides the steady state controller and adversary in the above lemma statement. See \ifnum\value{tac}>0{\cite{lee2023performance} }\else{Appendix~\ref{ap: soft-constrained state feedback sol} }\fi for the proof. It should be noted that in contrast to most adversarially robust machine learning problems, adversarially robust LQR provides a closed-form expression for the adversarial perturbation. 

We now leverage the state feedback solution to solve the soft-constrained output feedback problem. 

\begin{lemma}
    \label{lem:adversarial LQG sol} Suppose that 
    \begin{enumerate}
        \item There exists a solution to the following DARE
    \begin{align*}
        P_\gamma = \dare\paren{A,\bmat{B_1 & B_2}, Q, \bmat{-\gamma^2 I \\ & R}}
    \end{align*}
    satisfying $P_\gamma \succeq 0$ and $B_1^\top P_\gamma B_1 \prec \gamma^2 I$. Let $F_\gamma$, $E_\gamma$ and $\Delta_\gamma$ be the corresponding gains from Lemma~\ref{lem:adversarial LQR sol}. Also let $\Psi$ be as in \eqref{eq:shorthand psi}.
        \item There exists a nonsingular matrix $\bmat{\hat T_{11} & 0 \\ \hat T_{21} & \hat T_{22}}$ with $\hat T_{11} \in \mathbb{R}^{n_\delta \times n_\delta}$, and $\hat T_{22} \in \mathbb{R}^{n_u \times n_u}$ such that 
        \begin{align}
            \label{eq: decompose psi}
            \Psi  = \bmat{\hat T_{11} & 0 \\ \hat T_{21} & \hat T_{22} }^\top \bmat{-I \\ & I} \bmat{\hat T_{11} & 0 \\ \hat T_{21} & \hat T_{22} }.
        \end{align}
        \item There exists real matrices $L_\gamma, N_\gamma, \Sigma_\gamma, Y_\gamma$, where $\Sigma_\gamma, Y_\gamma \succ 0$, that satisfy
        \ifnum\value{tac}>0{
        \begin{equation}
        \label{eq: full control conditions}
        \begin{aligned}
            \tilde A &= \hat A+L_\gamma \hat C_2,  \quad \quad\,\,
            \tilde B_0 = \hat B_0 + L_\gamma \hat D_{20}, \\
            \tilde B_1 &= \hat B_1 + L_\gamma\hat D_{21}, \quad \quad
            \tilde C_1 = \hat C_1 + N_\gamma \hat C_2 \\
            \tilde D_{11} &= \hat D_{11} + N_\gamma \hat D_{21} \quad \tilde D_{10} = N_\gamma \hat D_{20} \\
            \Phi &=  I - \tilde D_{11}^\top \tilde D_{11} - \tilde B_1^\top Y_\gamma \tilde B_1 \succ 0 \\
            Y_\gamma &= \tilde C_1^\top \tilde C_1 + \tilde A^\top Y_\gamma \tilde A + \paren{\tilde D_{11}^\top \tilde C_1 + \tilde B_1^\top Y_\gamma \tilde A}^\top \Phi^{-1} \\
            &\qquad \times \paren{\tilde D_{11}^\top \tilde C_1 + \tilde B_1^\top Y_\gamma \tilde A} \\
            \Sigma_\gamma &= \brac{(I+\tilde B_1 \Phi^{-1} \tilde B_1^\top Y_\gamma) \tilde A + \tilde B_1 \Phi^{-1} \tilde D_{11}^\top \tilde C_1} \Sigma_\gamma \\
            &\quad \times \brac{(I+\tilde B_1 \Phi^{-1} \tilde B_1^\top Y_\gamma) \tilde A + \tilde B_1 \Phi^{-1} \tilde D_{11}^\top \tilde C_1}^\top \\
            &\quad + \brac{(I+\tilde B_1 \Phi^{-1} \tilde B_1^\top Y_\gamma) \tilde B_0 + \tilde B_1^\top \Phi^{-1} \tilde D_{11}^\top \tilde D_{10} } \\
            &\quad \times \brac{(I+\tilde B_1 \Phi^{-1} \tilde B_1^\top Y_\gamma) \tilde B_0 + \tilde B_1^\top \Phi^{-1} \tilde D_{11}^\top \tilde D_{10} }^\top \\
            L_\gamma, &N_\gamma \in \argmin_{L, N}  \trace \bigg(\tilde D_{10}^\top \tilde D_{10} +\tilde B_0^\top Y_\gamma \tilde B_0 +(\tilde D_{10}^\top \tilde D_{11} \\
            &\qquad + \tilde B_0^\top Y_\gamma \tilde B_1)  \Phi^{-1} (\tilde D_{10}^\top \tilde D_{11} + \tilde B_0^\top Y_\gamma \tilde B_1)^\top \bigg) \\ 
        &\qquad + \trace\bigg(\Sigma_\gamma \big[\tilde C_1^\top \tilde C_1 + \tilde A^\top Y_\gamma \tilde A + (\tilde D_{11}^\top \tilde C_1 \\
        &\qquad + \tilde B_1^\top Y_\gamma \tilde A )^\top\Phi^{-1}(\tilde D_{11}^\top \tilde C_1 + \tilde B_1^\top Y_\gamma \tilde A )  \big] \bigg),
        \end{aligned}
        \end{equation}
        }\else{
        \begin{equation}
        \label{eq: full control conditions}
        \begin{aligned}
            \tilde A &= \hat A+L_\gamma \hat C_2,  \\
            \tilde B_0 &= \hat B_0 + L_\gamma \hat D_{20}, \\
            \tilde B_1 &= \hat B_1 + L_\gamma\hat D_{21} \\
            \tilde C_1 &= \hat C_1 + N_\gamma \hat C_2 \\
            \tilde D_{11} &= \hat D_{11} + N_\gamma \hat D_{21} \\
            \tilde D_{10} &= N_\gamma \hat D_{20} \\
            \Phi &=  I - \tilde D_{11}^\top \tilde D_{11} - \tilde B_1^\top Y_\gamma \tilde B_1 \succ 0 \\
            Y_\gamma &= \tilde C_1^\top \tilde C_1 + \tilde A^\top Y_\gamma \tilde A + \paren{\tilde D_{11}^\top \tilde C_1 + \tilde B_1^\top Y_\gamma \tilde A}^\top \Phi^{-1} \paren{\tilde D_{11}^\top \tilde C_1 + \tilde B_1^\top Y_\gamma \tilde A} \\
            \Sigma_\gamma &= \brac{(I+\tilde B_1 \Phi^{-1} \tilde B_1^\top Y_\gamma) \tilde A + \tilde B_1 \Phi^{-1} \tilde D_{11}^\top \tilde C_1} \Sigma_\gamma \brac{(I+\tilde B_1 \Phi^{-1} \tilde B_1^\top Y_\gamma) \tilde A + \tilde B_1 \Phi^{-1} \tilde D_{11}^\top \tilde C_1}^\top \\
            &\qquad + \brac{(I+\tilde B_1 \Phi^{-1} \tilde B_1^\top Y_\gamma) \tilde B_0 + \tilde B_1^\top \Phi^{-1} \tilde D_{11}^\top \tilde D_{10} } \brac{(I+\tilde B_1 \Phi^{-1} \tilde B_1^\top Y_\gamma) \tilde B_0 + \tilde B_1^\top \Phi^{-1} \tilde D_{11}^\top \tilde D_{10} }^\top \\
            L_\gamma, &N_\gamma \in \argmin_{L, N}  \trace\paren{\tilde D_{10}^\top \tilde D_{10} +\tilde B_0^\top Y_\gamma \tilde B_0 +(\tilde D_{10}^\top \tilde D_{11} + \tilde B_0^\top Y_\gamma \tilde B_1)  \Phi^{-1} (\tilde D_{10}^\top \tilde D_{11} + \tilde B_0^\top Y_\gamma \tilde B_1)^\top } \\ 
        &\qquad\qquad + \trace\paren{\Sigma_\gamma \brac{\tilde C_1^\top \tilde C_1 + \tilde A^\top Y_\gamma \tilde A + (\tilde D_{11}^\top \tilde C_1 + \tilde B_1^\top Y_\gamma \tilde A )^\top\Phi^{-1}(\tilde D_{11}^\top \tilde C_1 + \tilde B_1^\top Y_\gamma \tilde A )  } },
        \end{aligned}
        \end{equation}}\fi
        where $Y_\gamma$ and $\Sigma_\gamma$ are treated as fixed parameters in the minimization problem in the final lines, and 
        \ifnum\value{tac}>0{
        \begin{equation}
        \begin{aligned}
            \label{eq: system for fc}
           & \hat G = \left[\begin{array}{c|ccc}
                \hat A & \hat B_0 & \hat B_1 & \hat B_2 \\ 
               \hline 
               \hat C_1 & 0 & \hat D_{11} & I  \\ 
               \hat C_2 & \hat D_{20} & \hat D_{21} & 0 
            \end{array}\right] :=\\
            & \left[\begin{array}{c|ccc}
               A + B_1 E_\gamma & B_0 + B_1 \Delta_\gamma &  B_1 \hat T_{11}^{-1} &  B_2 \hat T_{22}^{-1} \\ 
               \hline 
               -\hat T_{22} F_\gamma & 0 & \hat T_{21} \hat T_{11}^{-1} & I  \\ 
               C_2 + D_{21} E_\gamma & D_{20} + D_{21} \Delta_\gamma & D_{21} \hat T_{11}^{-1} & 0 
            \end{array}\right].
        \end{aligned}
        \end{equation}
        }\else{
        \begin{align}
            \label{eq: system for fc}
           \hat G = \left[\begin{array}{c|ccc}
                \hat A & \hat B_0 & \hat B_1 & \hat B_2 \\ 
               \hline 
               \hat C_1 & 0 & \hat D_{11} & I  \\ 
               \hat C_2 & \hat D_{20} & \hat D_{21} & 0 
            \end{array}\right] := \left[\begin{array}{c|ccc}
               A + B_1 E_\gamma & B_0 + B_1 \Delta_\gamma &  B_1 \hat T_{11}^{-1} &  B_2 \hat T_{22}^{-1} \\ 
               \hline 
               -\hat T_{22} F_\gamma & 0 & \hat T_{21} \hat T_{11}^{-1} & I  \\ 
               C_2 + D_{21} E_\gamma & D_{20} + D_{21} \Delta_\gamma & D_{21} \hat T_{11}^{-1} & 0 
            \end{array}\right].
        \end{align}}\fi
    \end{enumerate}
    
    When the above conditions hold, the controller
    \begin{align*}
        K =  
        \left[\begin{array}{c| cc}    
        \hat A-\hat B_2 (\hat C_1 + N_\gamma \hat C_2)+ L_\gamma \hat C_2 & L_\gamma - \hat B_2 N_\gamma \\
        \hline \hat T_{22}^{-1} (\hat C_1 + N_\gamma \hat C_2) &  \hat T_{22}^{-1} N_\gamma
        \end{array}\right]
    \end{align*}
    internally stabilizes the system \eqref{eq:ss} under the optimal adversarial perturbation and minimizes the objective \eqref{eq:adversarialLQRobjective}.
\end{lemma}

  The solution approach for the soft-constrained output feedback control synthesis above follows the separation argument outlined in e.g. \cite{doyle1989state, zhou1996robust, iglesias1991state} to reduce the output feedback control problem to that of estimating the optimal state feedback control input from noisy observations of the state. This can in turn be solved via another appeal to the dynamic programming argument from \cite{basar1991hinf}. The main challenge in doing so is that the optimal adversary causes a coupling between the future cost-to-go, and the estimation error-to-arrive. We do not know of existing numerical schemes which provably converge to the optimal $L_\gamma, N_\gamma, \Sigma_\gamma$ and $Y_\gamma$. However, an alternating solution which works well in practice is proposed in Algorithm~\ref{alg: soft-constrained}. The proof of Lemma~\ref{lem:adversarial LQG sol} can be found in \ifnum\value{tac}>0{\cite{lee2023performance}}\else{Appendix~\ref{ap: soft constrained output feedback sol}}\fi.
  
\begin{algorithm}
\caption{Computing Soft-Constrained Adversarially Robust Controller: \textbf{\texttt{SoftAdvController}}$(\hat G, \gamma, k)$} 
\label{alg: soft-constrained}
\begin{algorithmic}[1]
\State \textbf{Input:} System $\hat G$ from \eqref{eq: system for fc}, soft adversarial penalty $\gamma$, number of iterations $k$. 
\State Initialize $L_\gamma$ and $N_\gamma$ as the Kalman Filter gains for $\hat G$
\For{$k$ times}
  \State Given $L_\gamma$ and $N_\gamma$, compute $Y_\gamma$ according to \eqref{eq: full control conditions}, neglecting the final two conditions that $L_\gamma, N_\gamma$ minimize the provided function, and that $\Sigma_\gamma$ solves the provided Lyapunov equation.
  \State Given $L_\gamma$, $N_\gamma$, and $Y_\gamma$, compute $\Sigma_\gamma$ according to \eqref{eq: full control conditions}, neglecting the final condition that $L_\gamma, N_\gamma$ minimize the provided function.
  \State Given $Y_\gamma$ and $\Sigma_\gamma$, compute the gains $L_\gamma$ and $N_\gamma$ according to the final condition in \eqref{eq: full control conditions}, neglecting the Riccati equation for $Y_\gamma$ and the Lyapunov equation for $\Sigma_\gamma$. 
\EndFor
\State \textbf{Output: } Gains $L_\gamma$, $N_\gamma$
\end{algorithmic}
\end{algorithm}

We now return our attention to objective \eqref{eq:hardadversarialLQRobjective}.  It can be shown via strong duality that the hard-constrained problem may be solved by sequentially solving the soft-constrained problem using Lemma~\ref{lem:adversarial LQG sol}. Note that in contrast to the solution approach to minimize $\RC$, dualizing the constraint in $\AC$ results in an optimal dual variable $\gamma(\varepsilon)$ that is a random variable. Therefore, it is nontrivial to exchange the order of the minimization over the dual variable with the expectation (see \ifnum\value{tac}>0{\cite{lee2023performance} }\else{Appendix~\ref{appendix: advlqr thm proof} }\fi for details).  We propose Algorithm~\ref{alg:advControl} to minimize the adversarial cost \eqref{eq:hardadversarialLQRobjective}.  

\begin{algorithm}
\caption{Computing Adversarially Robust Controller: \textbf{\texttt{AdvControl}}$(G,\varepsilon, \gamma_{LB}, \gamma_{UB}, \texttt{tol})$} 
\label{alg:advControl}

\begin{algorithmic}[1]
\State \textbf{Input:} System $G$, adversary budget $\varepsilon > 0$, binary search bounds $\gamma_{LB}<\gamma_{UB}$, tolerance $\texttt{tol}$.\;
\State // \emph{Do binary search on $\gamma \in [\gamma_{LB}, \gamma_{UB}]$ to find optimal adversary with average power $\varepsilon > 0$}
\While{$\gamma_{UB}-\gamma_{LB}\geq \texttt{tol}$}
\State $\gamma = (\gamma_{UB} +\gamma_{LB})/2$
\If{Conditions $1-3$ of Lemma~\ref{lem:adversarial LQG sol} are satsified at level $\gamma$}
\State Compute the optimal controller $K$ for system $G$ at level $\gamma$ according to Lemma~\ref{lem:adversarial LQG sol}.
\If{$\texttt{AdvPower}(\mathcal{F}_l(G,K), \gamma) < \varepsilon$ (computed via \Cref{alg:advPower})}
\State $\gamma_{UB} = \gamma$
\Else 
\State $\gamma_{LB} = \gamma$
\EndIf
\Else
\State $\gamma_{LB}=\gamma$
\EndIf
\EndWhile
\State \textbf{Output:} Adversarially robust controller $K$, closed-loop system $G_{cl} = \mathcal{F}_{l}(G,K)$
\end{algorithmic}
\end{algorithm}

\begin{theorem}
    \label{thm:adversarial controller solution}
    Let $\gamma_{LB}$ be a small positive number such that the conditions of Lemma~\ref{lem:adversarial LQG sol} are satisfied for all $\gamma \geq \gamma_{LB}$, and $K_{\gamma_{LB}}$ be the corresponding optimal controller. For any $\varepsilon > 0$ such that $\texttt{AdvPower}(\mathcal{F}_l(G, K_{\gamma_{LB}}), \gamma_{LB}) > \varepsilon$, let $\gamma_{UB}$ be sufficiently large such that the controller $K_{\gamma_{UB}}$ at level $\gamma_{UB}$ satisfies $\texttt{AdvPower}(\mathcal{F}_l(G, K_{\gamma_{UB}}), \gamma_{UB}) < \varepsilon$. 
    Then the output of Algorithm~\ref{alg:advControl},  $[K, G_{cl}]
$  = \textnormal{\texttt{AdvControl}}$(G,\varepsilon, \gamma_{LB}, \gamma_{UB}, \textnormal{\texttt{tol}})$ satisfies (up to the numerical tolerance)
 \begin{enumerate}
        \item The controller $K$ minimizes $\AC(K)$. 
        \item The minimum value for the adversarial cost \eqref{eq:hardadversarialLQRobjective} is given by $\trace(J_\gamma)  + \gamma^2 \varepsilon$, where 
        \ifnum\value{tac}>0{
         \begin{equation}
        \label{eq: adversarial cost evaluation}
        \begin{aligned}
            J_\gamma &= D_{0, cl}^\top D_{0, cl} + B_{0, cl}^\top P_\gamma B_{0, cl} + (D_{1,cl}^\top D_{0, cl} \\
            &+ B_{1,cl}^\top P_\gamma B_{0,cl})^\top \Phi^{-1} (D_{1,cl}^\top D_{0, cl} + B_{1,cl}^\top P_\gamma B_{0,cl}) \\
            P_\gamma & = \texttt{DARE}(A_{cl}, B_{1,cl}, C_{cl}^\top C_{cl}, -\gamma^2 I, S= C_{cl}^\top D_{1,cl}) \\
            \Phi &= \gamma^2 I - D_{1,cl}^\top D_{1,cl} - B_{1,cl}^\top P_\gamma B_{1,cl} \\
           G_{cl} &= \left[\begin{array}{c|cc}
               A_{cl} & B_{0, cl} & B_{1,cl}  \\ 
               \hline 
               C_{cl} & D_{0, cl} & D_{1, cl} 
            \end{array}\right].
        \end{aligned}
        \end{equation}
        }\else{
        \begin{equation}
        \label{eq: adversarial cost evaluation}
        \begin{aligned}
            J_\gamma &= D_{0, cl}^\top D_{0, cl} + B_{0, cl}^\top P_\gamma B_{0, cl} + (D_{1,cl}^\top D_{0, cl} + B_{1,cl}^\top P_\gamma B_{0,cl})^\top \Phi^{-1} (D_{1,cl}^\top D_{0, cl} + B_{1,cl}^\top P_\gamma B_{0,cl}) \\
            P_\gamma & = \texttt{DARE}(A_{cl}, B_{1,cl}, C_{cl}^\top C_{cl}, -\gamma^2 I, S= C_{cl}^\top D_{1,cl}) \\
            \Phi &= \gamma^2 I - D_{1,cl}^\top D_{1,cl} - B_{1,cl}^\top P_\gamma B_{1,cl} \\
           G_{cl} &= \left[\begin{array}{c|cc}
               A_{cl} & B_{0, cl} & B_{1,cl}  \\ 
               \hline 
               C_{cl} & D_{0, cl} & D_{1, cl} 
            \end{array}\right].
        \end{aligned}
        \end{equation}}\fi
        Here, $\dare(A,B,Q,R,S)$ is overloaded to denote the solution $P$ to the generalized $\dare$, 
        \ifnum\value{tac}>0{
        \begin{align*}
            P &= Q +  A^\top P A \\
            &+ (A^\top P B + S)(B^\top P B+R)^{-1} (B^\top P A +S^\top).
        \end{align*}}\else{
         \begin{align*}
            P &= Q +  A^\top P A + (A^\top P B + S)(B^\top P B+R)^{-1} (B^\top P A +S^\top).
        \end{align*}
        }\fi
    \end{enumerate}
\end{theorem}

We observe from Theorem~\ref{thm:adversarial controller solution} that while the LQG controller is independent of the noise statistics, the adversarially robust controller output by Algorithm~\ref{alg:advControl} depends on the noise statistics through the optimal choice of $\gamma$.

In the subsequent sections, we examine the tradeoff in the objective values from minimizing either the nominal objective or the adversarially robust objective. This motivates the following two results, which allow us to compute the nominal and adversarial costs for an arbitrary LTI controller $K$. 

\begin{proposition} 
    For an LTI controller $K$, let $A_{cl}$, $B_{cl}$, $C_{cl}$ and $D_{cl}$ be a state space realization for $\mathcal{F}_l(G,K)$. If $\rho(A_{cl}) < 1$, then the nominal cost may be expressed as
\ifnum\value{tac}<1{
\begin{align*}
    \NC(K) = \limsup_T \frac{1}{T} \Ex_w \brac{\sum_{t=0}^{T-1} \norm{z_t}^2} =  \trace \paren{C_{cl} \Sigma_{cl} C_{cl}^\top + D_{cl} D_{cl}^\top},
\end{align*}}
\else{
\begin{align*}
    \NC(K) &= \limsup_T \frac{1}{T} \Ex_w \brac{\sum_{t=0}^{T-1} \norm{z_t}^2} \\
    &=  \trace \paren{C_{cl} \Sigma_{cl} C_{cl}^\top + D_{cl} D_{cl}^\top},
\end{align*}
}\fi
where
    $\Sigma_{cl} = A_{cl} \Sigma_{cl} A_{cl}^\top + B_{cl} 
B_{cl}^\top.$
\end{proposition}

\begin{algorithm}
\caption{Computing Adversarially Power: \textbf{\texttt{AdvPower}}$(G_{cl},\gamma)$} 
\label{alg:advPower}

\begin{algorithmic}[1]
\State \textbf{Input:} Closed-loop system $G_{cl}$, adversary penalty $\gamma$. 
\State Determine a state space realization for the closed-loop system $
    \left[\begin{array}{c|cc}
       A_{cl} & B_{0, cl} & B_{1,cl}  \\ 
       \hline 
       C_{cl} & D_{0, cl} & D_{1, cl} 
    \end{array}\right] = G_{cl}$.
\State Set $P_\gamma = \texttt{DARE}(A_{cl}, B_{1,cl}, C_{cl}^\top C_{cl}, -\gamma^2 I, S= C_{cl}^\top D_{1,cl})$. 
\State Let $\Phi = \gamma^2 I - B_{1, cl}^\top P_\gamma B_{1,cl}$.
\State Let $\Gamma_x = D_{1,cl}^\top C_{cl} + B_{1,cl}^\top P_\gamma A_{cl}$, $\Gamma_w =  D_{1,cl}^\top D_{0,cl} + B_{1,cl}^\top P_\gamma B_{0, cl}$.
\State Assign $\Sigma_\gamma = \texttt{dlyap}(A_{cl} + B_{1,cl}\Phi^{-1} \Gamma_x, (B_{0,cl} + B_{1,cl}\Phi^{-1} \Gamma_w)(B_{0,cl} + B_{1,cl}\Phi^{-1} \Gamma_w)^\top)$.
\State \textbf{Output: } Adversarial power $\trace(\Gamma_w^\top \Phi^{-2} \Gamma_w) + \trace(\Gamma_x^\top \Phi^{-2} \Gamma_x \Sigma_\gamma)$
\end{algorithmic}
\end{algorithm}

\begin{proposition}
    For an LTI controller $K$, the same results used to solve for the optimal controller minimizing $\AC(\cdot)$ may be adapted to evaluate $\AC(K)$. In particular, for a closed-loop system $G_{cl}=\mathcal{F}_{l}(G,K)$ and under any value of $\gamma$ such that the Riccati equation for $P_\gamma$ in \eqref{eq: adversarial cost evaluation} yields a positive semidefinite solution,
    the soft-penalized objective \eqref{eq:adversarialLQRobjective} evaluates to $\trace(J_\gamma)$, where $J_\gamma$ is given by \eqref{eq: adversarial cost evaluation}. Similarly, for any $\varepsilon > 0$, the adversarial cost satisfies $\AC(K)=\trace(J_\gamma)+\gamma^2 \varepsilon$, where $
    \gamma$ is the optimal value obtained from \Cref{alg:advControl} applied to $G_{cl}$ with appropriately chosen $\gamma_{LB}$, $\gamma_{UB}$ and $\texttt{tol}$, and the matrix $J_\gamma$ results from \eqref{eq: adversarial cost evaluation}.  
\end{proposition}


\section{Performance-Robustness Tradeoff Bounds: State Feedback}
\label{s: state feedback performance robustness tradeoffs}

In this section, we summarize the results of our prior work \citep{lee2022performancerobustness} to study the tradeoffs that arise in adversarial control by investigating the interplay between the two objectives \eqref{eq:nominal_cost} and \eqref{eq:adversarialLQRobjective}\footnote{We note that the original problem of interest is the hard-constrained problem \eqref{eq:hardadversarialLQRobjective}. The tradeoffs between the nominal controller and the soft-constrained robust controller are different than those between the nominal controller and hard-constrained robust controller. However, we conjecture that the dependencies of these tradeoffs on the underlying system-theoretic quantities are similar. This conjecture is supported by the numerical experiments.} in the state feedback setting. In particular, we assume that $C_2 = I$, $D_{20} = 0$ and $D_{21}=0$. For ease of exposition, we additionally assume that $B_0 = \Sigma_w^{1/2}=\sigma_w I$ and $B_1 = I$. As such, we drop the subscript on $B_2$ such that $B_2 \equiv B$ for the remainder of the section. 

We consider the gap between the nominal and $\gamma$-adversarially robust controllers when  applied in the nominal setting, i.e.,\ we seek to bound the gap $\NC(F_\gamma) - \NC(F_\star)$, where $F_\gamma$ is the $\gamma$-adversarially robust controller given by Lemma~\ref{lem:adversarial LQR sol}, and $F_\star$ is the LQR controller given by equation \eqref{eq:lqr}. Let $P_\star$ be the nominal DARE solution given by the solution to equation \eqref{eq: nominal control riccati}, let $P_\gamma$ be the solution to equation \eqref{eq:modified DARE} and let 
\begin{align}
\label{eq: Mgamma def}
M_\gamma =  P_\gamma + P_\gamma (\gamma^2 I - P_\gamma)^{-1}  P_\gamma.
\end{align}

Given an arbitrary stabilizing linear state feedback controller $F$, 
Lemma 12 of  \cite{fazel2018global} allows us to characterize the gap in the cost between $F$ and the optimal LQR controller $F_\star$ as
\ifnum\value{tac}>0{
\begin{equation}
    \begin{aligned}
    \label{eq: fazel result}
    \NC&(F)  - \NC(F_\star) \\
    &=\trace(\Sigma(F) (F-F_\star)^\top (R + B^\top P_\star B) (F- F_\star)),
\end{aligned}
\end{equation}
}\else{
\begin{align}
    \label{eq: fazel result}
    \NC(F) - \NC(F_\star) = \trace(\Sigma(F) (F-F_\star)^\top (R + B^\top P_\star B)(F- F_\star)),
\end{align}
}\fi

where $\Sigma(F) := \dlyap(A+BF, \Sigma_w)$ is the steady state covariance of the closed-loop system under controller $F$. The following bounds on the gap $\NC(F_\gamma) - \NC(F_\star)$ then follow immediately:
\begin{align}
    \label{eq:cost gap lower bound}
    \NC(F_\gamma) \!- \!\NC(F_\star) \!&\geq \! \sigma_w^2 \sigma_{\min}(R \!+\!B^\top \! P_\star B) \! \norm{F_\gamma \!-\! F_\star}_F^2 \\
    \label{eq:cost gap upper bound}
    \NC(F_\gamma) \!-\! \NC(F_\star) \!&\leq\! \norm{\Sigma(F_\gamma)}\! \norm{R \!+\! B^\top \!P_\star B} \! \norm{F_\gamma \!-\! F_\star}_F^2.
\end{align}
We have therefore reduced the task of upper and lower bounding the cost gap between the $\gamma$-adversarially robust controller and the nominal LQR controller in the nominal setting to directly bounding the gap between the two controller gains. Recalling that $\norm{F_\gamma - F_\star}^2_F \leq \min\curly{n_u, n_x} \norm{F_\gamma - F_\star}^2$, we use  
the following lemma to bound the difference $\norm{F_\gamma - F_\star}$ in terms of the difference between the solutions to the  corresponding adversarial and nominal DAREs. 

\begin{lemma}(Adapted from Lemma 2 of \cite{mania2019certainty})
    \label{lem: controller gap to DARE gap}
    Suppose that $f_1(u; x) = \frac{1}{2} u^\top R u + \frac{1}{2}(Ax + Bu)^\top M (Ax+Bu)$ and $f_2(u; x) = \frac{1}{2} u^\top R u + \frac{1}{2}(Ax + Bu)^\top P (Ax+Bu)$ with $M \succeq P$. Furthermore, for $i \in [2]$ and any $x$, let $u_i = F_i x = \argmin_u f_i(u,x)$. Then 
    \ifnum\value{tac}>0{
    \begin{align*}
        \norm{F_1 - F_2} &\geq \frac{\norm{B^\top (M - P) (A+BF_2)}}{\norm{R + B^\top M B)}}, \textrm{ and }   \\
        \norm{F_1 - F_2} &\leq \frac{\norm{B^\top (M - P) (A+BF_2)}}{\sigma_{\min}(R + B^\top P B)}.
    \end{align*}
    }\else{
    \[
        \frac{\norm{B^\top (M - P) (A+BF_2)}}{\norm{R + B^\top M B)}} \leq \norm{F_1 - F_2} \leq \frac{\norm{B^\top (M - P) (A+BF_2)}}{\sigma_{\min}(R + B^\top P B)}.
    \]    
    }\fi
\end{lemma}

 We also define $\gamma_\infty$ as the minimum $\calH_\infty$ norm for the closed loop system, i.e., the smallest value of $\gamma$ for which the conditions of Lemma~\ref{lem:adversarial LQR sol} hold. Similarly, we define $\tilde \gamma_\infty$ as the $\calH_\infty$ norm of the closed loop system under the nominal LQR controller. Additionally, we define the $\ell$-step controllability gramian as $W_{\ell}(A,B): = \sum_{t=0}^\ell A^t BB^\top (A^t)^\top$. If $\rho(A) < 1$ we define the controllability gramian as $W_\infty(A,B):=\lim_{\ell \to \infty} W_\ell(A,B)$.

\subsection{Upper Bound}\label{sec: upper bound}

Applying Lemma~\ref{lem: controller gap to DARE gap} with $P =P_\star$, the nominal DARE solution \eqref{eq: nominal control riccati}, and $M = M_\gamma$, the modified robust DARE solution in \Cref{lem:adversarial LQR sol}, reduces our goal to bounding the spectrum of $M_\gamma - P_\star$. From the definition of $M_\gamma$ \eqref{eq: Mgamma def}, we can write $
    \norm{P_\star - M_\gamma} \leq \norm{P_\star - P_\gamma} + \frac{\norm{P_\gamma}^2}{\gamma^2 - \norm{P_\gamma}}.$
For $\gamma > \gamma_\infty$ we have $P_\gamma \prec P_{\gamma_\infty} \prec \gamma_\infty^2 I$\ifnum\value{tac}>0{}\else{ (Lemma~\ref{lem: gamma ordering})}\fi, and thus $
    \norm{P_\star - M_\gamma} \leq \norm{P_\star - P_\gamma} + \frac{\gamma_\infty^4}{\gamma^2 - \gamma_\infty^2},$
reducing our task to bounding $\norm{P_\star-P_\gamma}$, the gap between solutions to the $\gamma$-adversarial and nominal DAREs.

Toward bounding the norm difference of solutions to DAREs, we show that the closed-loop dynamics under the adversary $\delta_t$ can be expressed as perturbations of the nominal system matrices. In particular, recall that for a noiseless adversarial LQR instance at level $\gamma > 0$, the adversary can be represented as $
\delta_t =  \paren{\gamma^2 I - P_\gamma }^{-1} P_\gamma \paren{Ax_t + Bu_t},$
such that we may write $x_{t+1} = Ax_t + Bu_t + \delta_t = \tilde{A}x_t + \tilde{B}u_t$, for $\tilde A := \paren{I + (\gamma^2 I -P_\gamma)^{-1}P_\gamma}A$ and $\tilde B :=\paren{I + (\gamma^2 I -P_\gamma)^{-1}P_\gamma}B $. This allows us to bound the gaps $||\tilde{A} - A||,\; ||\tilde{B}-B||$ in terms of $\gamma$. This is formalized in \ifnum\value{tac}>0{\cite{lee2023performance}}\else{Lemma~\ref{lem: eps to gamma conversion} in Appendix~\ref{appendix: upper bound}}\fi.

By bounding the gap between system matrices in the adversarial setting and in the nominal setting, we derive bounds on the gap %
$\norm{P_\gamma-P_\star}$ between the adversarial and nominal DARE solutions, which ultimately leads to the following upper bound on $\NC(F_\gamma) - \NC(F_\star)$.

\begin{theorem}
\label{thm: nominal cost gap upper bound}
Suppose $(A,B, Q^{1/2})$ is controllable and detectable. Define the condition number $\kappa(Q, R) := \frac{\max\curly{\sigma_{\max}(Q), \sigma_{\max}(R)}}{\min\curly{\sigma_{\min}(Q), \sigma_{\min}(R)}}$, 

$\tau(A, \rho) := \sup\curly{\norm{A^k}\rho^{-k}: k \geq 0}$,
and $\beta := \max\curly{1, \frac{\gamma_{\infty}^2\max\curly{\norm{A}, \norm{B}}}{\gamma^2 - \gamma_\infty^2} \tau\paren{A, \rho} + \rho}$, where $\rho > \rho(A)$. Furthermore, 
let $\ell$ be any natural number $1 \leq \ell \leq n_x$. For $\gamma > 0$ satisfying

\ifnum\value{tac}>0{
\begin{align*}
    \gamma^2 &\geq \gamma_\infty^2 + \frac{3}{2} \ell^{3/2} \beta^{\ell - 1} \sigma_{\min}(W_\ell(A,B))^{-1/2} \tau(A, \rho)^2 (\norm{B} \\
    &\qquad + 1) \max\curly{\norm{A}, \norm{B}}  \gamma_\infty^2,
\end{align*}
the following upper bound holds:
\begin{align*}
    &\NC(F_\gamma) - \NC(F_\star) \leq O(1) \sigma_w^2  \\
    &\quad \times \Big(\frac{\gamma_\infty^2}{\gamma^2 - \gamma_\infty^2}\Big)^2 n_u \ell^{5} \beta^{4(\ell - 1)} \paren{1 + \sigma_{\min}(W_\ell(A,B))^{-1/2}}^2  \\
    &\quad \times \norm{A + BK_\star}^2 \norm{W_\infty (A + BF_\gamma, I)} \tau(A, \rho)^6 \\
    &\quad \times \frac{\norm{R + B^\top P_\star B}}{\sigma_{\min}\paren{R + B^\top P_\star B}^2} \kappa(Q,R)^2 \norm{B}^2\paren{\norm{B} + 1}^4  \\ 
    &\quad \times \paren{\max\curly{\norm{A}, \norm{B}}^2 \norm{P_\star}^2 + \gamma_\infty^2}.
\end{align*}
}\else{
\begin{align*}
    \gamma^2 \geq \gamma_\infty^2 + \frac{3}{2} \ell^{3/2} \beta^{\ell - 1} \sigma_{\min}(W_\ell(A,B))^{-1/2} \tau(A, \rho)^2 \paren{\norm{B} + 1} \max\curly{\norm{A}, \norm{B}}  \gamma_\infty^2,
\end{align*}
the following upper bound holds:
\begin{align*}
    \NC(F_\gamma) - \NC(F_\star) &\leq O(1) \sigma_w^2 \Big(\frac{\gamma_\infty^2}{\gamma^2 - \gamma_\infty^2}\Big)^2 n_u \ell^{5} \beta^{4(\ell - 1)} \paren{1 + \sigma_{\min}(W_\ell(A,B))^{-1/2}}^2  \\
    &\quad \times \norm{A + BK_\star}^2 \norm{W_\infty (A + BF_\gamma, I)} \tau(A, \rho)^6 \\
    &\quad \times \frac{\norm{R + B^\top P_\star B}}{\sigma_{\min}\paren{R + B^\top P_\star B}^2} \kappa(Q,R)^2 \norm{B}^2\paren{\norm{B} + 1}^4  \\ 
    &\quad \times \paren{\max\curly{\norm{A}, \norm{B}}^2 \norm{P_\star}^2 + \gamma_\infty^2}.
\end{align*}
}\fi

\end{theorem}
As $\gamma \to \infty$, our upper bound decays to $0$ as expected, since the adversarial controller converges to the nominal controller in the limit. However, the steepness of this cost gap is affected by system properties such as the minimum singular value of the $\ell$-step controllability gramian. Specifically, poor controllability causes the upper bound to increase, as captured by the minimum singular value of the controllability gramian $\sigma_{\min}(W_{\ell}(A,B))$ in the bound above. We note that in contrast to the perturbation gap requirements on $\|\tilde A - A|, \|\tilde B - B\|$ in \cite{mania2019certainty}, our condition on the perturbation gap via lower bounds on $\gamma$ are much less conservative. We only require a lower bound on $\gamma$ to guarantee that the controllability of the adversarially perturbed system $(\tilde{A}, \tilde{B})$ is on the same order as that of the nominal system $(A,B)$.

\subsection{Lower Bound}\label{s: p-r tradeoffs lower bounds}

Applying Lemma \ref{lem: controller gap to DARE gap} with $M = M_\gamma$ and $P = P_\star$, we conclude that

\ifnum\value{tac}>0{
\begin{equation}
\begin{aligned}
    \label{eq: general controller lb}
    \norm{F_\gamma - F_\star} &\geq \frac{\norm{B^\top (M_\gamma-P_\star) (A+BF_\star) }}{\norm{R + B M_\gamma B}} \\
    &\geq \frac{\norm{B^\top (M_\gamma-P_\star) } \sigma_{\min}(A+BF_\star) }{\norm{R + B M_\gamma B}}.
\end{aligned}
\end{equation}
}\else{
\begin{align}
    \label{eq: general controller lb}
    \norm{F_\gamma - F_\star} &\geq \frac{\norm{B^\top (M_\gamma-P_\star) (A+BF_\star) }}{\norm{R + B M_\gamma B}} \geq \frac{\norm{B^\top (M_\gamma-P_\star) } \sigma_{\min}(A+BF_\star) }{\norm{R + B M_\gamma B}}.
\end{align}}\fi
Next, we add and subtract a particular DARE solution to the $M_\gamma-P_\star$ term in the above bound. Specifically, for $\gamma \geq \tilde \gamma_\infty$, we let $\tilde P_\gamma = \dare(A+BF_\star, I, Q, -\gamma^2 I)$, and note that $x^\top \tilde P_\gamma x$ represents the cost of applying controller $F_\star$ in the adversarial setting at level $\gamma$ starting from state $x$ with $w_t = 0$ for all $t \geq 0$. Adding and subtracting $\tilde P_\gamma$ in the lower bound in \eqref{eq: general controller lb} yields
\ifnum\value{tac}>0{
\begin{equation}
\begin{aligned}
    &\norm{F_\gamma - F_\star}
    \geq \frac{\norm{B^\top (M_\gamma-\tilde P_\gamma + \tilde P_\gamma - P_\star) }\sigma_{\min}(A+BF_\star) }{\norm{R + B M_\gamma B}} \\
    &\geq  \frac{\sigma_{\min}(A+BF_\star)}{\norm{R + B M_\gamma B}} \bigg(\norm{B^\top (P_\gamma(\gamma^2 I - P_\gamma)^{-1} P_\gamma + \tilde P_\gamma - P_\star)  }\\
    &\qquad -\norm{B^\top (\tilde P_\gamma - P_\gamma)}\bigg).
    \label{eq:lb on controller gap}
\end{aligned}
\end{equation}
}\else{
\begin{equation}
\begin{aligned}
    \norm{F_\gamma - F_\star}
    &\geq \frac{\norm{B^\top (M_\gamma-\tilde P_\gamma + \tilde P_\gamma - P_\star) }\sigma_{\min}(A+BF_\star) }{\norm{R + B M_\gamma B}} \\
    &\geq  \frac{\sigma_{\min}(A+BF_\star)}{\norm{R + B M_\gamma B}} \bigg(\norm{B^\top (P_\gamma(\gamma^2 I - P_\gamma)^{-1} P_\gamma + \tilde P_\gamma - P_\star)  } -\norm{B^\top (\tilde P_\gamma - P_\gamma)}\bigg).
    \label{eq:lb on controller gap}
\end{aligned}
\end{equation}
}\fi
To obtain a lower bound on the above expression, we can upper and lower bound the spectra of $\tilde P_\gamma - P_\gamma$ and $\tilde P_\gamma - P_\star$, respectively. To upper bound the spectrum of $\tilde P_\gamma - P_\gamma$, we apply a result similar to equation \eqref{eq: fazel result} for the noiseless adversarial setting. We may lower bound the spectrum of $\tilde P_\gamma - P_\star$ by writing each as the solution to a Lyapunov equation and observing that their difference may also be written as the solution to a Lyapunov equation. \ifnum\value{tac}>0{}\else{This is stated formally and proven in Lemma~\ref{lem: spectral UB+LB}. }\fi Combining the above leads to the following theorem.

\begin{theorem}
\label{thm:lower bound}
Suppose 
\ifnum\value{tac}>0{
\begin{align*}
    \gamma^2 &\geq \sigma_{\min}(P_\star) + \frac{1}{2}\sigma_{\min}(P_\star)^2\frac{\norm{B^\top W_\infty(A+BF_\star, I) }}{\norm{R + B^\top M_\gamma B}} \\
    &\times\norm{B^\top  W_\infty\paren{ \paren{I+(\gamma^2 I - \tilde P_\gamma)^{-1} \tilde P_\gamma} (A +B F_\star), I}} \\
    &\times \sigma_{\min}(A+BF_\star)^2.
\end{align*}
Then the following lower bound holds:
\begin{align*}
    \NC&(F_\gamma) - \NC(F_\star) \geq \frac{\sigma_w^2}{2} \bigg( \frac{\sigma_{\min}(P_\star)^2}{\gamma^2 - \sigma_{\min}(P_\star)} \bigg)^2 \\
    &\times \frac{\sigma_{\min}(R + B^\top P_\star B)}{\norm{R + B^\top M_\gamma B}^2} \norm{B^\top W_\infty(A+BF_\star, I) }^2 \\
    & \times \sigma_{\min}(A+BF_\star)^2.
\end{align*}
}\else{
\begin{align*}
    \gamma^2 &\geq \sigma_{\min}(P_\star) \\
    &+ \frac{1}{2}\sigma_{\min}(P_\star)^2\frac{\norm{B^\top W_\infty(A+BF_\star, I) }}{\norm{R + B^\top M_\gamma B}} \norm{B^\top  W_\infty\paren{ \paren{I+(\gamma^2 I - \tilde P_\gamma)^{-1} \tilde P_\gamma} (A +B F_\star), I}} \sigma_{\min}(A+BF_\star)^2.
\end{align*}
Then the following lower bound holds:
\begin{align*}
    \NC(F_\gamma) - \NC(F_\star) &\geq \frac{\sigma_w^2}{2} \bigg( \frac{\sigma_{\min}(P_\star)^2}{\gamma^2 - \sigma_{\min}(P_\star)} \bigg)^2 \frac{\sigma_{\min}(R + B^\top P_\star B)}{\norm{R + B^\top M_\gamma B}^2} \norm{B^\top W_\infty(A+BF_\star, I) }^2 \sigma_{\min}(A+BF_\star)^2.
\end{align*}
}\fi

\end{theorem}
Keeping the nominal system fixed, both the upper and lower bounds decay at a rate $\gamma^{-4}$. We note that instead of the $\ell$-step controllability gramian that manifests in the upper bound, we have instead the system parameter $W_{\infty}(A+BF_\star, I)$, which is the controllability gramian of the closed-loop system under controller $F_\star$ with disturbances as inputs. That is, a large $\norm{B^\top W_\infty(A+BF_\star, I) }$ implies that the nominal closed-loop system is quantifiably more controllable by the disturbance input, hence more susceptible to adversarial disturbances of fixed energy.


\section{Performance Robustness Tradeoff Bounds: State Prediction}
\label{s: state prediction}

We now turn to the problem of state prediction, where the goal is to use the history of observations $y_{-\infty:t}$ to propose a state estimate $\hat x_{t+1}$ such that $\norm{\hat x_{t+1} - x_{t+1}}^2$ is small. By denoting $e_t = x_t - \hat x_t$, we represent the state prediction error dynamics as
\begin{align*}
    e_{t+1} &= x_{t+1}-\hat x_{t+1} = A x_t + B_0 w_t + B_1 \delta_t - \hat x_{t+1}.
\end{align*}
We may perform a change of variables $\hat x_{t+1} = A \hat x_t - u_t$, where $u_t$ is an estimate of $ A \hat x_t - x_{t+1}$ given the history of measurements. Doing so allows us to express
\begin{align*}
    e_{t+1} &= A e_t + B_0 w_t + B_1 \delta_t + u_t \\
    e_{y_t} &:= y_t - C_2 \hat x_t =  C_2 e_t + D_{20} w_t + D_{21} \delta_t. 
\end{align*}
With this representation, the state prediction problem becomes a control problem in the language of \eqref{eq:ss} and \eqref{eq: objective} by letting $B_2 = I$, $C_1 = I$, and $D_{12} = 0$. In particular, the state $x_t$ in \eqref{eq:ss} is replaced by the state prediction error, $e_t$. The performance signal $z_t$ also becomes $e_t$, as the goal is to make the state estimation error small. The measurement $y_t$ becomes the innovation $e_{y_t}$. 

The optimal solution to the nominal state prediction problem (no adversarial perturations) is given by the Kalman Filter, which takes the form $u_t = -A\Sigma_\star C_2^\top(C_2\Sigma_\star C_2^\top + D_{20}^\top D_{20})^{-1} e_{y_t} =: L_\star e_{y_t}$\footnote{Note that $L_\star$ is slightly different from the Kalman Filter defined in \eqref{eq:lqr}. This is to account for the fact that the problem of interest in this section is state prediction rather than current state estimation.}, where $\Sigma_\star$ is given by \eqref{eq: nominal estimation riccati}. Meanwhile, the solution to the soft-constrained adversarially robust state prediction problem at level $\gamma$ is given by $u_t = L_\gamma e_{y_t}$, as long as there exists real matrices $L_\gamma, \Sigma_\gamma, Y_\gamma$ with $\Sigma_\gamma$ and $Y_\gamma$ symmetric positive definite that satisfy
\ifnum\value{tac}>0{
        \begin{equation}
        \label{eq: adv rob observer gain}
        \begin{aligned}
            \tilde A &= A+L_\gamma C_2 \\
            \tilde B_0 &= B_0 + L_\gamma D_{20} \\
            \tilde B_1 &=  B_1 + L_\gamma D_{21} \\
            \Phi &=  \gamma^2 I - \tilde B_1^\top Y_\gamma \tilde B_1 \succ 0\\
            Y_\gamma &= I + \tilde A^\top (Y_\gamma + Y_\gamma \tilde B_1 \Phi^{-1} \tilde B_1 Y_\gamma) \tilde A \\
            \Sigma_\gamma &= (I+\tilde B_1 \Phi^{-1} \tilde B_1^\top Y_\gamma) \tilde A  \Sigma_\gamma \tilde A^\top (I+\tilde B_1 \Phi^{-1} \tilde B_1^\top Y_\gamma)^\top \\ 
            &\qquad+ (I+\tilde B_1 \Phi^{-1} \tilde B_1^\top Y) \tilde B_0 \tilde B_0^\top (I+\tilde B_1 \Phi^{-1} \tilde B_1^\top Y)^\top \\
            L_\gamma  &\in \argmin_{L}  \trace\paren{\tilde B_0^\top Y_\gamma \tilde B_0 +\tilde B_0^\top Y_\gamma \tilde B_1  \Phi^{-1}  \tilde B_1^\top Y_\gamma \tilde B_0 } \\
            &\qquad+ \trace\paren{\Sigma_\gamma \brac{\tilde A^\top Y_\gamma \tilde A +  \tilde A^\top  Y_\gamma \tilde B_1 \Phi^{-1} \tilde B_1^\top Y_\gamma \tilde A   } },
    \end{aligned}
\end{equation}
}\else{
        \begin{equation}
        \label{eq: adv rob observer gain}
        \begin{aligned}
            \tilde A &= A+L_\gamma C_2 \\
            \tilde B_0 &= B_0 + L_\gamma D_{20} \\
            \tilde B_1 &=  B_1 + L_\gamma D_{21} \\
            \Phi &=  \gamma^2 I - \tilde B_1^\top Y_\gamma \tilde B_1 \succ 0\\
            Y_\gamma &= I + \tilde A^\top (Y_\gamma + Y_\gamma \tilde B_1 \Phi^{-1} \tilde B_1 Y_\gamma) \tilde A \\
            \Sigma_\gamma &= (I+\tilde B_1 \Phi^{-1} \tilde B_1^\top Y_\gamma) \tilde A  \Sigma_\gamma \tilde A^\top (I+\tilde B_1 \Phi^{-1} \tilde B_1^\top Y_\gamma)^\top + (I+\tilde B_1 \Phi^{-1} \tilde B_1^\top Y) \tilde B_0 \tilde B_0^\top (I+\tilde B_1 \Phi^{-1} \tilde B_1^\top Y)^\top \\
            L_\gamma  &\in \argmin_{L}  \trace\paren{\tilde B_0^\top Y_\gamma \tilde B_0 +\tilde B_0^\top Y_\gamma \tilde B_1  \Phi^{-1}  \tilde B_1^\top Y_\gamma \tilde B_0 }  + \trace\paren{\Sigma_\gamma \brac{\tilde A^\top Y_\gamma \tilde A +  \tilde A^\top  Y_\gamma \tilde B_1 \Phi^{-1} \tilde B_1^\top Y_\gamma \tilde A   } },
    \end{aligned}
\end{equation}
}\fi
where $\Sigma_\gamma$ and $Y_\gamma$ are treated as fixed parameters in the minimization on the final line above. 

As in the previous section, we bound the nominal performance gap between the nominal state estimator $L_\star$ and the adversarially robust state estimator $L_\gamma$. In this setting, the nominal performance metric reduces to the trace of the covariance of the state estimation error. Under an arbitrary linear stabilizing state estimator $L$, the state estimation error covariance becomes $\Sigma_L = \dlyap((A+L C_2)^\top, (B_0+L D_{20})(B_0+L D_{20})^\top)$.  In the nominal setting, the state estimation error covariance induced by the Kalman Filter is $\Sigma_{L_\star} = \Sigma_\star$, while the covariance induced by the adversarially robust state estimator $L_\gamma$ is given by $\Sigma_{L_\gamma}$. Then our objective is to upper and lower bound the state estimation error cost gap defined by 
\[
    \trace(\Sigma_{L_\gamma}) -\trace(\Sigma_{\star}).
\]
We may leverage Lemma 12 in  \cite{fazel2018global} to show the following reduction.
\begin{lemma}   
    For any observer gain $L$ such that $\rho(A+LC_2)<1$, 
    \ifnum\value{tac}>0{
    \begin{align*}
        \trace(\Sigma_{L}) &-\trace(\Sigma_{\star}) = \trace(\dlyap(A+L C_2, I) \\
        &\times (L- L_\star)(D_{20} D_{20}^\top + C_2 \Sigma C_2^\top) (L - L_\star)^\top).
    \end{align*}}
    \else{
    \begin{align*}
        \trace(\Sigma_{L}) &-\trace(\Sigma_{\star}) = \trace(\dlyap(A+L C_2, I)  (L- L_\star)(D_{20} D_{20}^\top + C_2 \Sigma C_2^\top) (L - L_\star)^\top).
    \end{align*}
    }\fi
\end{lemma}
\textit{Proof:}
    We have that
    \begin{align*}
        \Sigma_L &= \dlyap((A + LC_2)^\top, B_0 B_0^\top +  LD_{20} D_{20}^\top L^\top) \\
        \Sigma_{\star} &= \dare(A^\top, C_2^\top, B_0B_0^\top, D_{20}, D_{20}^\top).
    \end{align*}
    By interpreting this as a control problem for the system $x_{t+1} = A^\top x_t + C_2^\top u_t$ with policies $u_t = L^\top x_t$, and $u_t = L_\star^\top x_t$, $\trace(\Sigma_L) - \trace(\Sigma_{L_{\star}})$ may be interpreted as the expected gap in control costs under initial state distribution $x_0 \sim \calN(0,I)$. Therefore Lemma 12 in \cite{fazel2018global} applies immediately, yielding the result. 
    \hfill $\blacksquare$

Using the above result and pulling largest or smallest singular values from the trace, we have the following lower and upper bounds on the quantity of interest:
\ifnum\value{tac}>0{
\begin{align*}
    &\sigma_{\min}\paren{\dlyap((A+L_\gamma C_2)^\top, I)} \sigma_{\min}\paren{D_{20} D_{20}^\top + C_2 \Sigma_{\star} C_2^\top} \\
    &\times \norm{L_\gamma - L_\star}_F^2 \le \trace(\Sigma_{L_\gamma}) -\trace(\Sigma_{\star}) \le  \norm{L_\gamma - L_\star}_F^2\\
    &\times \norm{\dlyap((A+L_\gamma C_2)^\top, I)} \norm{D_{20} D_{20}^\top + C_2 \Sigma_{\star} C_2^\top} . 
\end{align*}
}\else{
\begin{align*}
    \trace(\Sigma_{L_\gamma}) -\trace(\Sigma_{\star}) &\geq \sigma_{\min}\paren{\dlyap((A+L_\gamma C_2)^\top, I)} \sigma_{\min}\paren{D_{20} D_{20}^\top + C_2 \Sigma_{\star} C_2^\top} \norm{L_\gamma - L_\star}_F^2, \quad \textrm{ and }  \\
    \trace(\Sigma_{L_\gamma}) -\trace(\Sigma_{\star}) &\le    \norm{\dlyap((A+L_\gamma C_2)^\top, I)} \norm{D_{20} D_{20}^\top + C_2 \Sigma_{\star} C_2^\top} \norm{L_\gamma - L_\star}_F^2.
\end{align*}
}\fi

We have therefore reduced the problem to upper and lower bounding the Frobenius norm of the gap between the adversarially robust observer gain $L_\gamma$, and the nominal observer gain $L_\star$. The frobenius norm can in turn be upper and lower bounded as $\norm{L_\gamma  - L_\star}^2 \leq \norm{L_\gamma - L_\star}_F^2 \leq \min\curly{n_y, n_u} \norm{L_\gamma - L_\star}^2$. Unlike the adversarially robust state feedback controller gain, the adversarially robust filter gain does not have a closed form expression, and instead requires solving the collection of equations in \eqref{eq: adv rob observer gain}. Therefore, rather than presenting a bound on the gap in general, we focus on two special cases where the equations simplify substantially. The first is when the adversary perturbs only the state, and not the measurement (i.e. $D_{21} = 0$.) The second is when both the state and the measurement are scalar. These two simplifications and the resulting bounds on the nominal performance gap are presented in the subsequent sections. 





\subsection{Adversarial Perturbations Entering The State}
\label{s: adv state only}
In this setting, we assume that $D_{21} = 0$, i.e.,\ the adversary impacts only the state of the system. This assumption is reasonable in settings where the sensor noise is stochastic, but the process noise consists of a mix of both stochastic and adversaial components to model environment and model uncertainty. 

With this simplification, the optimization problem that must be solved for $L_\gamma$ reduces to
\[
    L_\gamma \in \argmin f(L)
\]
where 
\ifnum\value{tac}>0{
\begin{align*}
    f(L) &= \trace((B_0 + L D_{20})^\top M_\gamma (B_0 + L D_{20})) \\
    &\qquad + \trace(\Sigma_\gamma (A+L C_2)^\top M_\gamma (A+ L C_2)),
\end{align*}
}\else{
\[
    f(L) = \trace((B_0 + L D_{20})^\top M_\gamma (B_0 + L D_{20})) + \trace(\Sigma_\gamma (A+L C_2)^\top M_\gamma (A+ L C_2)),
\]
}\fi
and 
\[
    M_\gamma = Y_\gamma + Y_\gamma  B_1 (\gamma^2 I - B_1^\top Y_\gamma  B_1)^{-1} B_1^\top Y_\gamma.
\]
As $M_\gamma$ is full rank, the solution reduces to $L_\gamma = -A \Sigma_\gamma C_2^\top (D_{20} D_{20}^\top + C_2 \Sigma_\gamma C_2^\top)^{-1}$. In particular, it has a form identical to the Kalman Filter but using the solution $\Sigma_\gamma$ to a different Riccati equation. Applying Lemma~\ref{lem: controller gap to DARE gap} to the transposes of $L_\star$ and $L_\gamma$, we find that 
\ifnum\value{tac}>0{
\begin{align*}
    \norm{L_\gamma - L_\star} &\geq \frac{\norm{(A+L_\star C_2)^\top (\Sigma_\gamma - \Sigma_\star) C_2}}{\norm{D_{20} D_{20}^\top + C_2 \Sigma_\gamma C_2^\top}}, \quad\textrm{ and }\\
    \norm{L_\gamma - L_\star} &\leq \frac{\norm{(A+L_\star C_2)^\top (\Sigma_\gamma - \Sigma_\star) C_2}}{\sigma_{\min}(D_{20} D_{20}^\top + C_2 \Sigma_\star C_2^\top)}.
\end{align*}
}\else{
\begin{align*}
   \frac{\norm{(A+L_\star C_2)^\top (\Sigma_\gamma - \Sigma_\star) C_2)}}{\norm{D_{20} D_{20}^\top + C_2 \Sigma_\gamma C^\top}} \leq  \norm{L_\gamma - L_\star} \leq \frac{\norm{(A+L_\star C_2)^\top (\Sigma_\gamma - \Sigma_\star) C_2}}{\sigma_{\min}(D_{20} D_{20}^\top + C_2 \Sigma_\star C_2^\top)}.
\end{align*}
}\fi
It therefore remains to upper and lower bound the spectrum of $\Sigma_\gamma - \Sigma_\star$.

\subsubsection{Upper Bound}
\label{s: adv state only upper}
To obtain an upper bound on the cost gap, we must simply upper bound $
    \norm{\Sigma_\gamma - \Sigma_\star}.$
To do so, observe that if we denote $\Lambda_\gamma = B_1(\gamma^2 I - B_1^\top Y_\gamma B_1)^{-1} B_1^\top Y_\gamma$, then we may write
\ifnum\value{tac}>0{
\begin{align*}
    \Sigma_\gamma &= (I + \Lambda_\gamma)(A+ L_\gamma C_2) \Sigma_\gamma (A+L_\gamma C_2)^\top (I + \Lambda_\gamma)^\top \\
    &\qquad + (I+\Lambda_\gamma)(B_0 + L D_{20})(B_0+L D_{20})^\top (I+\Lambda_\gamma)^\top.
\end{align*}
}\else{
\begin{align*}
    \Sigma_\gamma = (I + \Lambda_\gamma)(A+ L_\gamma C_2) \Sigma_\gamma (A+L_\gamma C_2)^\top (I + \Lambda_\gamma)^\top + (I+\Lambda_\gamma)(B_0 + L D_{20})(B_0+L D_{20})^\top (I+\Lambda_\gamma)^\top.
\end{align*}}\fi
Defining $\tilde \Sigma_\gamma = (I+\Lambda_\gamma)^{-1} \Sigma_\gamma \paren{(I+\Lambda_\gamma)^{-1}}^\top$, we have
\ifnum\value{tac}>0{
\begin{align*}
    \tilde \Sigma_\gamma &= (A+ L_\gamma C_2)(I + \Lambda_\gamma)\tilde  \Sigma_\gamma (I + \Lambda_\gamma)^\top(A+L_\gamma C_2)^\top  \\
    &\qquad + (B_0 + L D_{20})(B_0+L D_{20})^\top \\
    &= \dare( (A(I+\Lambda_\gamma))^\top, (C_2(I+\Lambda_\gamma))^\top, B_0  B_0^\top, D_{20} D_{20}^\top).
\end{align*}
}\else{
\begin{align*}
    \tilde \Sigma_\gamma &= (A+ L_\gamma C_2)(I + \Lambda_\gamma)\tilde  \Sigma_\gamma (I + \Lambda_\gamma)^\top(A+L_\gamma C_2)^\top  + (B_0 + L D_{20})(B_0+L D_{20})^\top \\
    &= \dare( (A(I+\Lambda_\gamma))^\top, (C_2(I+\Lambda_\gamma))^\top, B_0  B_0^\top, D_{20} D_{20}^\top).
\end{align*}
}\fi
By the triangle inequality,
\begin{align*}
    \norm{\Sigma_\gamma - \Sigma_\star} \leq \norm{\Sigma_\gamma - \tilde \Sigma_\gamma} + \norm{\tilde \Sigma_\gamma - \Sigma_\star}.
\end{align*}
As in \Cref{sec: upper bound}, we may bound $\norm{\tilde \Sigma_\gamma - \Sigma_\star}$ when both $A\Lambda_\gamma^\top$ and $C_2\Lambda_\gamma^\top$ are small. To bound $\norm{\Sigma_\gamma-\tilde \Sigma_\gamma}$, note that 
\[
    \Sigma_\gamma - \tilde \Sigma_\gamma = \Lambda_\gamma \tilde \Sigma_\gamma + \tilde \Sigma_\gamma \Lambda_\gamma^\top   + \Lambda_\gamma \tilde \Sigma_\gamma \Lambda_\gamma^\top.
\]
Therefore, 
\ifnum\value{tac}>0{
\begin{align*}
    \norm{\Sigma_\gamma - \tilde \Sigma_\gamma} &\leq \bigg(2\frac{\norm{Y_\gamma} \norm{B_1}^2}{\gamma^2 - \norm{Y_\gamma}\norm{B_1}^2} \\
    &\qquad+ \paren{\frac{\norm{Y_\gamma} \norm{B_1}^2}{\gamma^2 - \norm{Y_\gamma}\norm{B_1}^2}}^2\bigg)
    \norm{\tilde \Sigma_\gamma}.
\end{align*}
}\else{
\[
    \norm{\Sigma_\gamma - \tilde \Sigma_\gamma} \leq \paren{2\frac{\norm{Y_\gamma} \norm{B_1}^2}{\gamma^2 - \norm{Y_\gamma}\norm{B_1}^2} + \paren{\frac{\norm{Y_\gamma} \norm{B_1}^2}{\gamma^2 - \norm{Y_\gamma}\norm{B_1}^2}}^2} \norm{\tilde \Sigma_\gamma}.
\]}\fi
Combining these results leads to the following upper bound on the cost gap. 
\begin{theorem}
    \label{thm: state prediction upper bound}
    Define $\bar Y$ such that $\bar Y \geq \norm{Y_\gamma}$ for all $\gamma$ such that the conditions of \eqref{eq: adv rob observer gain} are satisifed. Suppose that $(A, C_2)$ is observable. Define $\kappa(B_0 B_0^\top, D_{20} D_{20}^\top ) = \frac{\max\curly{\sigma_{\max}(B_0B_0^\top), \sigma_{\max}(D_{20} D_{20}^\top)}}{\min\curly{\sigma_{\min}(B_0 B_0^\top), \sigma_{\min}(D_{20} D_{20}^\top)}}$,
    $\tau(A, \rho) := \sup\curly{\norm{A^k}\rho^{-k}: k \geq 0}$, 
    and $\beta = \max\curly{1, \frac{\norm{B_1}^2 \bar Y}{\gamma^2 - \norm{B_1}^2 \bar Y} \max\curly{\norm{A}, \norm{C_2}} \tau\paren{A, \rho} + \rho}$, where $\rho > \rho(A)$. Then for $1\leq \ell\leq n_x$ and
    \ifnum\value{tac}>0{
    \begin{align*}
        \gamma^2 &\geq  \norm{B_1}^2 \bar Y + \frac{3}{2} \ell^{3/2} \beta^{\ell - 1} \sigma_{\min}(W_\ell(A^\top,C_2^\top))^{-1/2} \tau(A, \rho)^2 \\
        &\qquad\times \paren{\norm{C_2} + 1} \max\curly{\norm{A}, \norm{C_2}}  \norm{B_1}^2 \bar Y,
    \end{align*}
    the following bound holds:
    \begin{align*}
        & \trace(\Sigma_{L_\gamma}) -\trace(\Sigma_{\star}) \leq \norm{\dlyap(A+L_\gamma C_2, I)} \\
        &\times \frac{\norm{D_{20} D_{20}^\top + C_2 \Sigma_\star C_2^\top }}{\sigma_{\min}(D_{20} D_{20}^\top + C_2 \Sigma_\star C_2^\top)^2} n_y \norm{A+L_\star C_2}^2 \norm{C_2}^2  \\
        & \times O(1) (\norm{\tilde \Sigma_{\gamma}}+ \max\curly{\norm{A}, \norm{C_2}} \norm{\Sigma_{\star}})^2 \\&\times\paren{\frac{\norm{B_1}^2 \bar Y}{\gamma^2 - \norm{B_1}^2 \bar Y}}^2  \paren{1 + \sigma_{\min}\paren{W_{l}(A^\top, C_2^\top)^{-1/2}}}^2  \\
        & \times \ell^5 \beta^{4 (\ell -1)} \tau(A,\rho)^6\paren{\norm{C_2} +1}^2 \kappa(B_0B_0^\top, D_{20} D_{20}^\top).
    \end{align*}
    }\else{
    \begin{align*}
        \gamma^2 \geq  \norm{B_1}^2 \bar Y + \frac{3}{2} \ell^{3/2} \beta^{\ell - 1} \sigma_{\min}(W_\ell(A^\top,C_2^\top))^{-1/2} \tau(A, \rho)^2 \paren{\norm{C_2} + 1} \max\curly{\norm{A}, \norm{C_2}}  \norm{B_1}^2 \bar Y,
    \end{align*}
    the following bound holds:
    \begin{align*}
        & \trace(\Sigma_{L_\gamma}) -\trace(\Sigma_{\star}) \leq \norm{\dlyap(A+L_\gamma C_2, I)} \frac{\norm{D_{20} D_{20}^\top + C_2 \Sigma_\star C_2^\top }}{\sigma_{\min}(D_{20} D_{20}^\top + C_2 \Sigma_\star C_2^\top)^2} n_y \norm{A+L_\star C_2}^2 \norm{C_2}^2  \\
        & \cdot O(1) (\norm{\tilde \Sigma_{\gamma}}+ \max\curly{\norm{A}, \norm{C_2}} \norm{\Sigma_{\star}})^2 \paren{\frac{\norm{B_1}^2 \bar Y}{\gamma^2 - \norm{B_1}^2 \bar Y}}^2 \ell^5 \beta^{4 (\ell -1)} \paren{1 + \sigma_{\min}\paren{W_{l}(A^\top, C_2^\top)^{-1/2}}}^2 \tau(A,\rho)^6 \\
        & \cdot \paren{\norm{C_2} +1}^2 \kappa(B_0B_0^\top, D_{20} D_{20}^\top).
    \end{align*}}\fi
\end{theorem}
\ifnum\value{tac}>0{}\else{The proof combines the results of this section with \Cref{prop: advrobust covariance DARE upper bound obsility} in \Cref{s: proofs for adv state upper}. }\fi The result is roughly a dual result to \Cref{thm: state prediction upper bound}. Instead of the bound decaying with the inverse of the minimum singular value of the $\ell$-step controllability gramian, it decays with the inverse of the minimum singular value of the $\ell$-step observability gramian, $W_\ell(A^\top, C_2^\top)$. Therefore, uniformly high observability causes the upper bound to decrease. 

\subsubsection{Lower Bound}
\label{s: adv state only lower}

To obtain our lower bound, we lower bound $\sigma_{\min}(\Sigma_\gamma - \Sigma_\star)$. To do so, note that 
\begin{align*}
    \Sigma_\gamma - \Sigma_\star &=  \Sigma_\gamma - \tilde \Sigma_\gamma + \tilde \Sigma_\gamma - \Sigma_{L_\gamma} + \Sigma_{L_\gamma} - \Sigma_\star,
\end{align*}
where $\tilde \Sigma_\gamma$ is as in \Cref{s: adv state only upper}. By the fact that $\Sigma_{L_\gamma}$ is the estimation error covariance under the controller $L_\gamma$ for the nominal setting, we have that $\Sigma_{L_\gamma} - \Sigma_\star \succeq 0$, and $\Sigma_\gamma - \Sigma_\star \succeq  \Sigma_\gamma - \tilde \Sigma_\gamma + \tilde \Sigma_\gamma -  \Sigma_{L_\gamma}$.
Writing 
\ifnum\value{tac}>0{
\begin{align*}
    \tilde \Sigma_\gamma &= \dlyap( (I+\Lambda_\gamma)^\top (A+L_\gamma C_2)^\top, \\
    &\qquad (B_0 + L D_{20}) (B_0 + L D_{20})^\top),
\end{align*}
}\else{
\begin{align*}
    \tilde \Sigma_\gamma &= \dlyap( (I+\Lambda_\gamma)^\top (A+L_\gamma C_2)^\top, (B_0 + L D_{20}) (B_0 + L D_{20})^\top),
\end{align*}}\fi
we may take the difference of the Lyapunov equations to find that 
\ifnum\value{tac}>0{
\begin{align*}
    &\tilde \Sigma_\gamma - \Sigma_{L_\gamma} =\\& \dlyap((A+L_\gamma C_2)^\top, (A+L_\gamma C_2) (\Sigma_\gamma - \tilde \Sigma_\gamma)(A+L_\gamma C_2)^\top ),
\end{align*}
}\else{
\begin{align*}
    \tilde \Sigma_\gamma - \Sigma_{L_\gamma} = \dlyap((A+L_\gamma C_2)^\top, (A+L_\gamma C_2) (\Sigma_\gamma - \tilde \Sigma_\gamma)(A+L_\gamma C_2)^\top ),
\end{align*}}\fi
and therefore
\begin{align*}
    \Sigma_\gamma -   \Sigma_{L_\gamma} = \dlyap((A+L_\gamma C_2)^\top, \Sigma_\gamma - \tilde \Sigma_\gamma).
\end{align*}
To conclude, we may lower bound $\sigma_{\min}(\Sigma_\gamma -  \tilde \Sigma_\gamma)$.
\begin{restatable}{lemma}{scalarLBlemma}
    We have
    \ifnum\value{tac}>0{
    \begin{align*}
        \sigma_{\min}(\Sigma_\gamma -  \tilde \Sigma_\gamma)
        &\geq \sigma_{\min}(Y_\gamma) \Bigg(\paren{\frac{\gamma^2}{\gamma^2 - \norm{B_1^\top Y_\gamma^{1/2}}}}^2 \\
        &\times \frac{1}{\norm{Y_\gamma}} \sigma_{\min}(\tilde \Sigma_\gamma) - \norm{\tilde \Sigma_\gamma}/\sigma_{\min}(Y_\gamma)\Bigg).
    \end{align*}    
    }\else{
    \begin{align*}
    \sigma_{\min}(\Sigma_\gamma -  \tilde \Sigma_\gamma) 
    &\geq \sigma_{\min}(Y_\gamma) \paren{\paren{\frac{\gamma^2}{\gamma^2 - \norm{B_1^\top Y_\gamma^{1/2}}}}^2 \frac{1}{\norm{Y_\gamma}} \sigma_{\min}(\tilde \Sigma_\gamma) - \norm{\tilde \Sigma_\gamma}/\sigma_{\min}(Y_\gamma)}.
\end{align*}}\fi
\end{restatable}

Combining results, we have the following theorem.
\begin{theorem}
     Define $\bar \kappa$ such that $\bar \kappa \geq \norm{Y_\gamma}/\sigma_{\min}(Y_\gamma)$ for all $\gamma$ such that the conditions of \eqref{eq: adv rob observer gain} are satisifed. Then 
     \ifnum\value{tac}>0{
        \begin{align*}
        \trace&(\Sigma_{L_\gamma})  - \trace(\Sigma_\star) \geq \sigma_{\min}(\dlyap((A+L_\gamma C_2)^\top, I))^3 \\
        &\times \frac{\sigma_{\min}(D_{20} D_{20}^\top + C_2 \Sigma_\star C_2^\top)}{\norm{D_{20} D_{20}^\top + C_2 \Sigma_{L_\gamma} C^\top }^2} \sigma_{\min}(A+L_\star C_2)^2 \norm{C_2}^2  \\
        & \times \paren{\paren{\frac{\gamma^2}{\gamma^2 - \norm{B_1 Y_\gamma^{1/2}}}}^2 \frac{\sigma_{\min}(\tilde \Sigma_\gamma)}{\bar \kappa}  - \norm{\tilde \Sigma_\gamma}}.
    \end{align*}
     }\else{
    \begin{align*}
        \trace(\Sigma_{L_\gamma}) - \trace(\Sigma_\star) &\geq \sigma_{\min}(\dlyap((A+L_\gamma C_2)^\top, I))^3 \frac{\sigma_{\min}(D_{20} D_{20}^\top + C_2 \Sigma_\star C_2^\top)}{\norm{D_{20} D_{20}^\top + C_2 \Sigma_{L_\gamma} C_2^\top }^2} \sigma_{\min}(A+L_\star C_2)^2 \norm{C_2}^2  \\
        & \qquad \times \paren{\paren{\frac{\gamma^2}{\gamma^2 - \norm{B_1 Y_\gamma^{1/2}}}}^2 \frac{\sigma_{\min}(\tilde \Sigma_\gamma)}{\bar \kappa}  - \norm{\tilde \Sigma_\gamma}}.
    \end{align*}
    }\fi
\end{theorem}
The lower bound above is loose when $\gamma$ is large. In particular, it becomes negative as $\gamma \to \infty$. However, for small $\gamma$, corresponding to a large adversarial power, the bound becomes positive. It scales with the minimum singular value of $\dlyap((A+L_\gamma C_2)^\top, I)$ which measures the controllability of the closed-loop state prediction error system under the adversarially robust observer by the external disturbances.

\subsection{Scalar State and Measurement Case}
\label{s: scalar state and measurement}
Here, we let $A = a$, $B_0 = B_1 = \bmat{1 & 0}$, $B_2=1$, $C_2 = c$, and $D_{20} = D_{21} = \bmat{0 & 1}$. The values of $1$ in $B_0$, $B_1$, $D_{20}$ and $D_{21}$ are for ease of exposition only, the derivations go through for arbitrary values. In this setting, we may solve for $L_\gamma$ explicitly in terms of the $a$, $c$, $Y_\gamma$, $\Sigma_\gamma$, and $\gamma$. In particular we find that $L_\gamma$ is given by one of the solutions to 
\ifnum\value{tac}>0{
\begin{equation}
\begin{aligned}
\label{eq: quadratic equation}
0 &= \gamma^2\paren{L(1+c^2 \Sigma_\gamma) + ac\Sigma_\gamma } + ((a^2 - c^2) L \\
&\qquad+ ac(L^2 -1)) \Sigma_\gamma Y_\gamma.
\end{aligned}
\end{equation}
}\else{
\begin{align}
\label{eq: quadratic equation}
0 = \gamma^2\paren{L(1+c^2 \Sigma_\gamma) + ac\Sigma_\gamma } + ((a^2 - c^2) L + ac(L^2 -1)) \Sigma_\gamma Y_\gamma.
\end{align}}\fi
Solving for the relevant root of this equation yields
\ifnum\value{tac}>0{
\begin{align*}
   &L_\gamma = \frac{1}{2 a c \Sigma_\gamma Y_\gamma}\Bigg(-\gamma^2 - c^2 \gamma^2 \Sigma_\gamma + (c^2 - a^2) \Sigma_\gamma Y_\gamma +\\
    &+ \bigg(-4 a c \Sigma_\gamma Y (a c \gamma^2 \Sigma_\gamma - a c \Sigma_\gamma Y_\gamma) \\
    & + (\gamma^2 + c^2 \gamma^2 \Sigma_\gamma +a^2 \Sigma_\gamma Y_\gamma - c^2 \Sigma_\gamma Y_\gamma)^2 \bigg)^{1/2}\Bigg).
\end{align*}
}\else{
\[
    L_\gamma = \frac{-\gamma^2 - c^2 \gamma^2 \Sigma_\gamma + (c^2 - a^2) \Sigma_\gamma Y_\gamma + \sqrt{-4 a c \Sigma_\gamma Y (a c \gamma^2 \Sigma_\gamma - a c \Sigma_\gamma Y_\gamma) + (\gamma^2 + c^2 \gamma^2 \Sigma_\gamma +
a^2 \Sigma_\gamma Y_\gamma - c^2 \Sigma_\gamma Y_\gamma)^2}}{2 a c \Sigma_\gamma Y_\gamma}. 
\]}\fi
If the second term in \eqref{eq: quadratic equation} were zero, then the solution would be $L_\gamma = -\frac{a \Sigma_\gamma c}{c^2 \Sigma_\gamma +1}$, which has a form identical to the Kalman Filter, $L_\star = -\frac{a\Sigma_\star c}{c^2 \Sigma_\star +1}$. This will approximately be true when $\gamma$ is large. This is made concrete in the following lemma. 


\begin{restatable}{lemma}{taylorFilterGain}\label{lem: taylor_filter_gain}
    Suppose 
    \begin{align}  
    \label{eq: gamma bounds taylor filter gain}
        \gamma^2 \geq \max\curly{2 \Sigma_\gamma Y_\gamma (a^2 - c^2), 32 \Sigma_\gamma^2 Y_\gamma a^2c^2, \frac{Y_\gamma}{2}}.
    \end{align}
    Then 
    \[
        L_\gamma = -\frac{a \Sigma_\gamma c}{c^2 \Sigma_\gamma +1} + \xi,
    \]
    where 
    \begin{align*}
        \abs{\xi} & \leq 64 \abs{a}\abs{c} \frac{ \Sigma_\gamma Y_\gamma}{\gamma^2}  \paren{\Sigma_\gamma \abs{a^2-c^2} + 1 + (ac \Sigma_\gamma)^2}.
\end{align*}
\end{restatable}

If we now consider the gap $L_\gamma - L_\star$, we find that if the condition on $\gamma$ from \Cref{lem: taylor_filter_gain} is satisfied, then by using \Cref{lem: controller gap to DARE gap},
\ifnum\value{tac}>0{
\begin{align*}
       \abs{L_\gamma - L_\star} &\geq \frac{\abs{c}\abs{a+L_\star c} \abs{\Sigma_\gamma -\Sigma_\star}}{1 + c^2 \Sigma_\gamma} - \abs{\xi}, \quad \textrm{ and }\\
       \abs{L_\gamma - L_\star} &\leq \frac{\abs{c}\abs{a+L_\star c} \abs{\Sigma_\gamma -\Sigma_\star}}{1 + c^2 \Sigma_\gamma} + \abs{\xi}.
\end{align*}
}\else{
\[
       \frac{\abs{c}\abs{a+L_\star c} \abs{\Sigma_\gamma -\Sigma_\star}}{1 + c^2 \Sigma_\gamma} - \abs{\xi} \leq  \abs{L_\gamma - L_\star} \leq \frac{\abs{c}\abs{a+L_\star c} \abs{\Sigma_\gamma -\Sigma_\star}}{1 + c^2 \Sigma_\gamma} + \abs{\xi}.
\]}\fi
To complete our bounds, we must bound $\abs{\Sigma_\gamma - \Sigma_\star}$ above and below. 

\subsubsection{Upper Bound}
\label{s: scalar upper}
To obtain an upper bound, we must upper bound $
    \abs{\Sigma_\gamma - \Sigma_\star}.$ The approach to do so is similar to that in \Cref{s: adv state only upper}. In particular, by again appealing to the Riccati perturbation bounds of \cite{mania2019certainty}, \ifnum\value{tac}>0{}\else{(specialized to this setting in \Cref{prop: scalar advrobust covariance DARE upper bound obsility})}\fi and combining with previous results, we obtain the following theorem. 

\begin{restatable}{theorem}{predUBscalar}
    \label{thm: state prediction upper bound scalar}
    Define $\bar Y$ such that $\bar Y \geq Y_\gamma$ for all $\gamma$ such that the conditions of \eqref{eq: adv rob observer gain} are satisfied and $\underline \gamma$ as the smallest $\gamma$ such that these conditions are satisfied. Suppose that $c\neq 0$. Suppose $\gamma$ is sufficiently large that the condition \eqref{eq: gamma bounds taylor filter gain} holds. Then if
    \ifnum\value{tac}>0{
     \begin{align*}
    \gamma^2 &\geq  \bar Y(1+((\abs{a} +1)/\abs{c})^2 + \frac{3}{2}  \abs{c}^{-1} \paren{\abs{c} + 1}  \\
    &\qquad \times \max\curly{\abs{a}, \abs{c}}\bar Y(1+((\abs{a} +1)/\abs{c})^2,
    \end{align*}
    the following bound holds:
    \begin{align*}
       & \Sigma_{L_\gamma} -\Sigma_\star  \leq \frac{1}{1 - (a + L_\gamma c)^2} \bigg(16 \frac{\abs{a+L_\star c}}{1 + c^2 \Sigma_\star} \\
       &\times \frac{ \bar Y (1 + ((\abs{a}+1)/\abs{c})^2)}{\gamma^2 -  \bar Y (1 + ((\abs{a}+1)/\abs{c})^2)}  \\ 
        &\times \bigg(\tilde \Sigma_\gamma + \max\curly{\abs{a}, \abs{c}} (\abs{c}+1)^3 \Sigma_\star\bigg) + \abs{\xi}\bigg)^2 (1 + c^2 \Sigma_\star),
    \end{align*}
    where $
        \abs{\xi} \leq 64 \abs{a}\abs{c} \frac{\Sigma_{\underline \gamma} \bar Y}{\gamma^2}\paren{\Sigma_{\underline \gamma} \abs{a^2 - c^2} + 1 + (ac \Sigma_{\underline \gamma})^2}.$
    }\else{
    \begin{align*}
    \gamma^2 \geq  \bar Y(1+((\abs{a} +1)/\abs{c})^2 + \frac{3}{2}  \abs{c}^{-1} \paren{\abs{c} + 1} \max\curly{\abs{a}, \abs{c}}  \bar Y(1+((\abs{a} +1)/\abs{c})^2,
    \end{align*}
    the following bound holds:
    \begin{align*}
        \Sigma_{L_\gamma} -\Sigma_\star & \leq \frac{1}{1 - (a + L_\gamma c)^2} \paren{16 \frac{\abs{a+L_\star c}}{1 + c^2 \Sigma_\star}\frac{\bar Y (1 + ((\abs{a}+1)/\abs{c})^2)}{\gamma^2 - \bar Y (1 + ((\abs{a}+1)/\abs{c})^2)} \bigg(\tilde \Sigma_\gamma + \max\curly{\abs{a}, \abs{c}} (\abs{c}+1)^3 \Sigma_\star} \\ 
        &\qquad\qquad + \abs{\xi}\bigg)^2 (1 + c^2 \Sigma_\star),
    \end{align*}
    where 
    \[
        \abs{\xi} \leq 64 \abs{a}\abs{c} \frac{\Sigma_{\underline \gamma} \bar Y}{\gamma^2}\paren{\Sigma_{\underline \gamma} \abs{a^2 - c^2} + 1 + (ac \Sigma_{\underline \gamma})^2}.
    \]}\fi
\end{restatable}

In the scalar setting, $\abs{c}$ is a measure of observability. Therefore, the lower bound is presented directly in terms of $\abs{a}$ and $\abs{c}$ rather than the observability gramian for the system. As with \Cref{thm: state prediction upper bound}, the bound decays to zero as $\gamma\to\infty$. Additionally, when observability becomes poor, as measured by $\abs{c}$ becoming small, the bound becomes large due to the appearance of $\abs{c}$ in the denominator of $\bar Y (1 + (\abs{a}+1)/\abs{c})$. In particular, $\gamma^2$ is constrained to be larger than this quantity for the bound to be valid. Thus as $\abs{c}$ becomes small, $\bar Y (1 + (\abs{a}+1)/\abs{c})$ becomes large, making the term $\frac{\bar Y (1 + (\abs{a}+1)/\abs{c})^2}{\gamma^2 - \bar Y (1 + (\abs{a}+1)/\abs{c})^2}$  large. The stability of the closed-loop system under the adversarially robust observer also makes an appearance in the bound. When this closed-loop system is near marginally stable, the bound becomes large due to the term $\frac{1}{1 - (a + L_\gamma c)^2}$.

\subsubsection{Lower Bound}
\label{s: scalar lower}

We have that 
\begin{align*}
    \Sigma_\gamma - \Sigma_\star &= \Sigma_\gamma - \tilde \Sigma_\gamma + \tilde \Sigma_\gamma - \Sigma_{L_\gamma} + \Sigma_{L_\gamma} - \Sigma_\star \\
    & \geq \Sigma_\gamma - \tilde \Sigma_\gamma + \tilde \Sigma_\gamma -  \Sigma_{L_\gamma},
\end{align*}
where $\tilde \Sigma_\gamma = \dare( (a(1+\Lambda_\gamma), c(1+\Lambda_\gamma), 1,  1)$ and $\Lambda_\gamma =  \frac{Y_\gamma(1+L_\gamma^2)}{\gamma^2 - Y_\gamma (1+L_\gamma^2)}$.

By expressing both $\tilde \Sigma_\gamma$ and $\Sigma_{L_\gamma}$ as the solution to a Lyapunov equation, we have that
\[
    \tilde \Sigma_\gamma - \Sigma_{L_\gamma} = \dlyap(a+L_\gamma c, (a+L_\gamma c)^2 \tilde \Sigma_\gamma (2 \Lambda_\gamma + \Lambda_\gamma^2)).
\]
Similarly,  $\Sigma_\gamma - \tilde \Sigma_\gamma=\tilde \Sigma_\gamma (2 \Lambda_\gamma + \Lambda_\gamma^2)$. Therefore, 
\ifnum\value{tac}>0{
\begin{align*}
    &\Sigma_\gamma - \Sigma_\star \geq (2 \Lambda_\gamma + \Lambda_\gamma^2) \tilde \Sigma_\gamma \sum_{t=0}^\infty (a+L_\gamma c)^{2t} \\
    &\qquad \geq 2 \frac{Y_\gamma(1+L_\gamma^2)}{\gamma^2 - Y_\gamma (1+L_\gamma^2)}  \tilde \Sigma_\gamma \sum_{t=0}^\infty (a+L_\gamma c)^{2t}. 
\end{align*}
}\else{
\begin{align*}
    \Sigma_\gamma - \Sigma_\star \geq (2 \Lambda_\gamma + \Lambda_\gamma^2) \tilde \Sigma_\gamma \sum_{t=0}^\infty (a+L_\gamma c)^{2t} \geq 2 \frac{Y_\gamma(1+L_\gamma^2)}{\gamma^2 - Y_\gamma (1+L_\gamma^2)}  \tilde \Sigma_\gamma \sum_{t=0}^\infty (a+L_\gamma c)^{2t}. 
\end{align*}
}\fi
We have $Y_\gamma \geq 1$ for all $\gamma$ and $\tilde \Sigma_\gamma \geq \Sigma_\star$. Therefore, 
\begin{align*}
    \Sigma_\gamma - \Sigma_\star  \geq 2 \frac{(1+L_\gamma^2)}{\gamma^2 - (1+L_\gamma^2)}   \Sigma_\star \sum_{t=0}^\infty (a+L_\gamma c)^{2t}. 
\end{align*}
Combining these results, we have the following theorem. 
\begin{restatable}{theorem}{predLBscalar}
    \label{thm: state prediction lower bound scalar}
     Define $\bar Y$ such that $\bar Y \geq Y_\gamma$ for all $\gamma$ such that the conditions of \eqref{eq: adv rob observer gain} are satisfied and $\underline \gamma$ as the smallest $\gamma$ such that these conditions are satisfied. Suppose $\gamma$ is sufficiently large that the condition \eqref{eq: gamma bounds taylor filter gain} holds. Then
     \ifnum\value{tac}>0{
     \begin{align*}
        \Sigma_{L_\gamma} - \Sigma_\star& \geq \frac{1}{1-(a + L_\gamma c)^2} (1 + c^2 \Sigma_\star) \Bigg(\frac{\abs{c}\abs{a+L_\star c}}{1 + c^2 \Sigma_\gamma} \\
        &\times 2  \frac{(1+L_\gamma^2)}{\gamma^2 -  (1+L_\gamma^2)}   \Sigma_\star \dlyap(a+L_\gamma c, 1) - \abs{\xi}\Bigg)^2, 
    \end{align*}
     }\else{
    \begin{align*}
        \Sigma_{L_\gamma} - \Sigma_\star \geq \frac{1}{1-(a + L_\gamma c)^2} (1 + c^2 \Sigma_\star)\paren{\frac{\abs{c}\abs{a+L_\star c}}{1 + c^2 \Sigma_\gamma}2 \frac{(1+L_\gamma^2)}{\gamma^2 -  (1+L_\gamma^2)}   \Sigma_\star \dlyap(a+L_\gamma c, 1) - \abs{\xi}}^2, 
    \end{align*}}\fi
    where \[
        \abs{\xi} \leq 64 \abs{a}\abs{c} \frac{\Sigma_{\underline \gamma} \bar Y}{\gamma^2}\paren{\Sigma_{\underline \gamma} \abs{a^2 - c^2} + 1 + (ac \Sigma_{\underline \gamma})^2}.
    \]
\end{restatable}
We can again observe the dependence of the above bound on $\abs{c}$ as a proxy for observability and $\abs{a}$ as the measure of stability. We see that when $\abs{c}$ becomes small, and $\abs{a}$ approaches $1$ both $L_\gamma$ and $\Sigma_\star$ become large. Indeed, if $\abs{a} = 1-\lambda$ and $\abs{c} = \lambda$, and then $\Sigma_\star$ grows with $1/\lambda^2$ as $\lambda$ becomes small, and the overall bound becomes large. The bound then grows as observability becomes poor, and the system approaches marginal stability. As with \Cref{thm: state prediction upper bound scalar}, the appearance of $\frac{1}{1-(a + L_\gamma c)^2}$ indicates that the bound grows when the closed-loop system under the adversarially robust observer approaches marginal stability.

\section{Numerical Experiments} \label{sec: numerical experiments}

We now empirically study the trends suggested by our  tradeoff bounds in the previous two sections.

\begin{figure} 
    \centering
    \begin{subfigure}[b]{0.45\textwidth}
    \includegraphics[width=\textwidth]{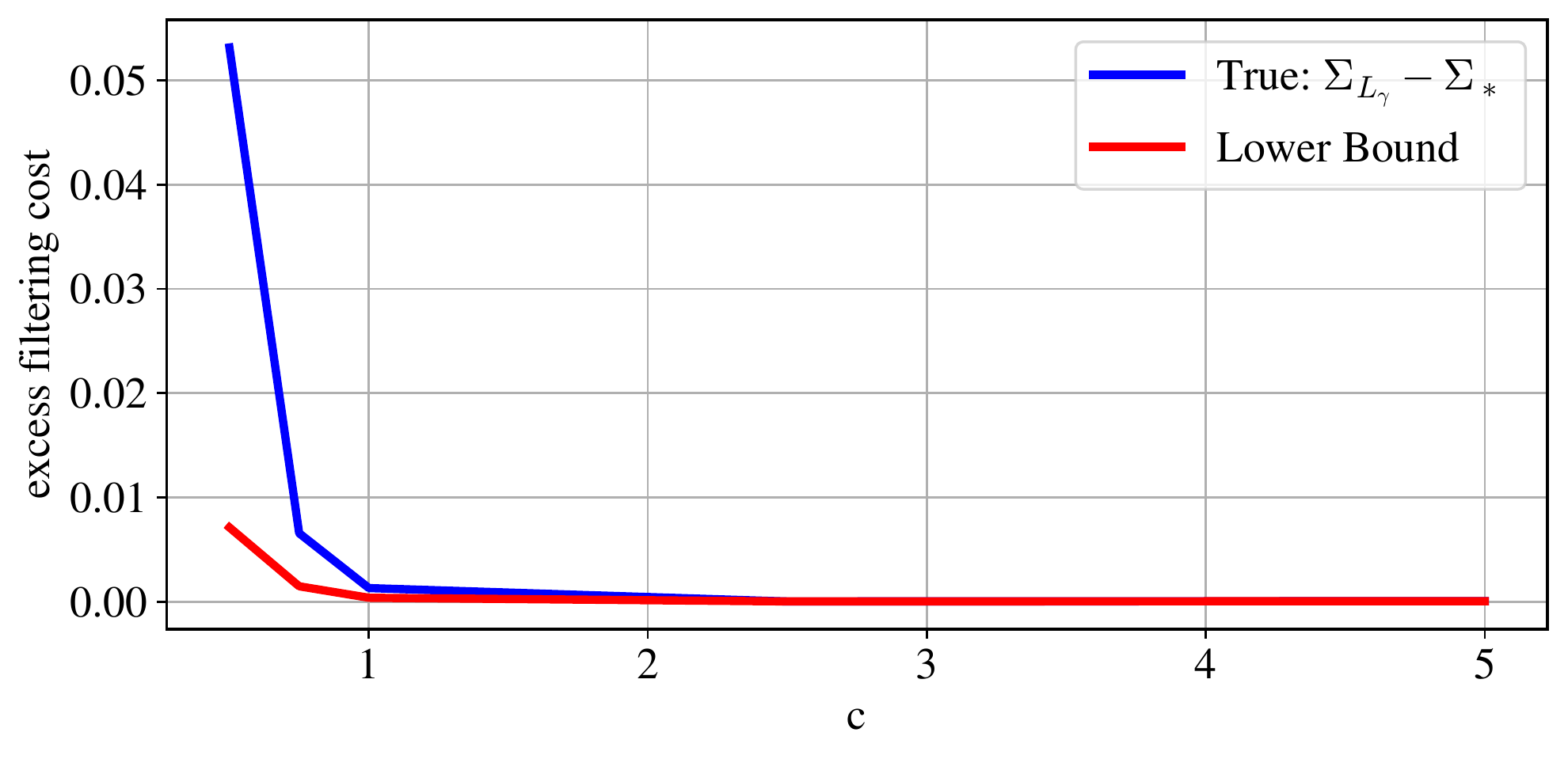}
    \caption{Lower Bound and True gap}
    \end{subfigure}
    \begin{subfigure}[b]{.45\textwidth}
    \includegraphics[width=\textwidth]{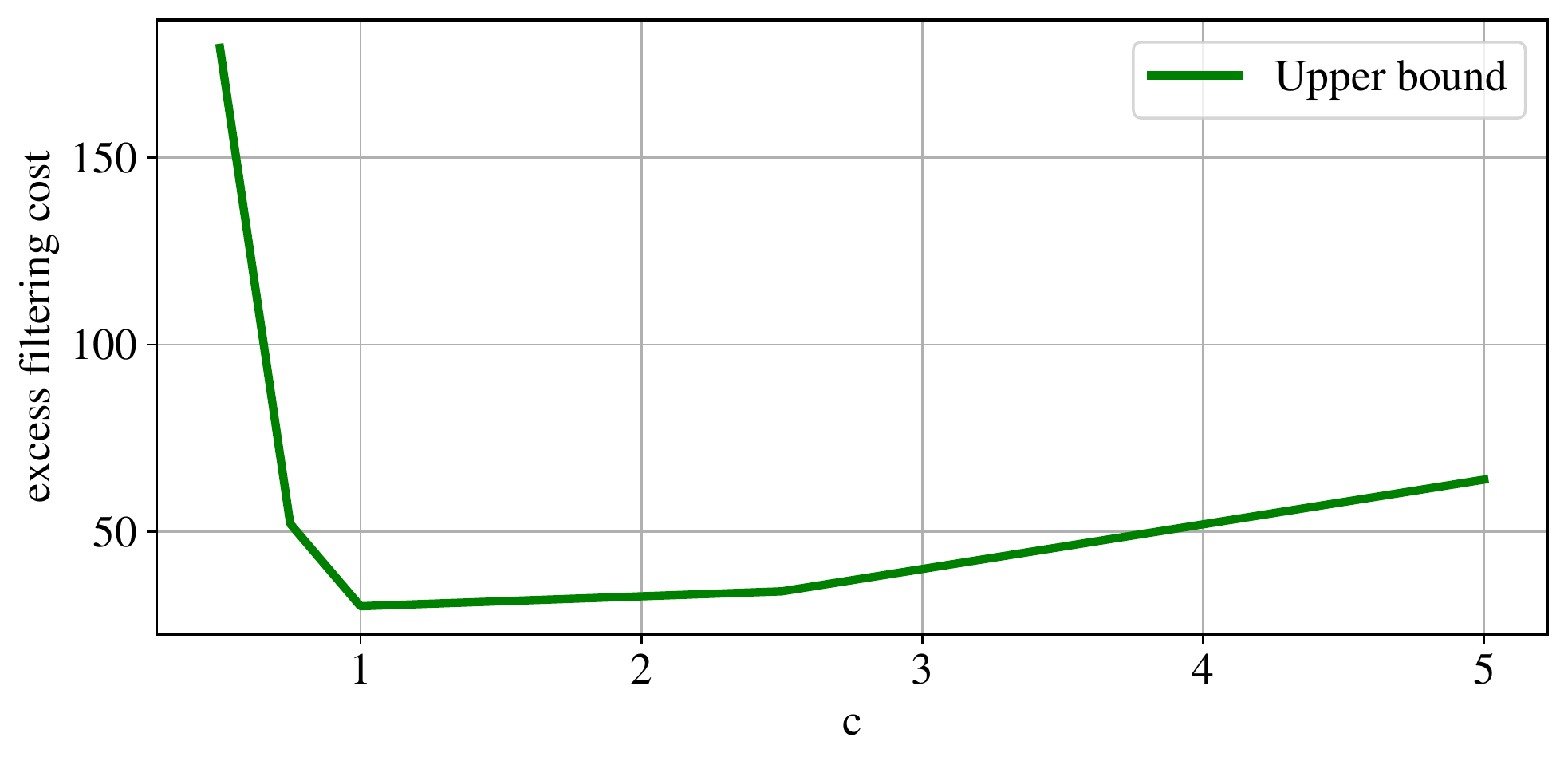}
    \caption{Upper Bound}
    \end{subfigure}
    \caption{Upper bound, lower bound, and true excess filtering cost between the robust filter designed with $\gamma=4$ and the nominal filter for a scalar system with $a=0.9$ and and varying values of $c$. As $\abs{c}$ decreases, observability becomes poor. This fact is captured in our bounds, and reflects what occurs with the true excess filtering cost.}
    \label{fig:validate scalar filtering bounds}
\end{figure}

\paragraph{Plotting the Bounds for Scalar State Estimation} In \Cref{fig:validate scalar filtering bounds}, we demonstrate the upper bound in \Cref{thm: state prediction upper bound scalar} and the lower bound in \Cref{thm: state prediction lower bound scalar} for the scalar system with $a=.9$ and varying values of $c$. We fix the robustness level using the soft penalty $\gamma = 4$. In both bounds, we substitute the true value of $\xi$, rather than the upper bound. In the upper bound,  we also use the value of $Y$ that depends on $\gamma$ rather than the uniform upper bound $\bar Y$. We see that the lower bound closely tracks the true cost error. The upper bound follows the same trends as $c$ becomes small, however, it is inflated by several orders of magnitude due to the conservative Riccati perturbation bounds applied. For the upper bound, the lower bound, and the true state prediction error gap, we see that when the observability decreases, as measured by $\abs{c}$ becoming smaller, the gap nominal cost gap between the robust filter and the nominal filter grows.

\begin{figure*}
    \centering
    \begin{subfigure}[b]{\textwidth}
    \includegraphics[width=0.3\textwidth]{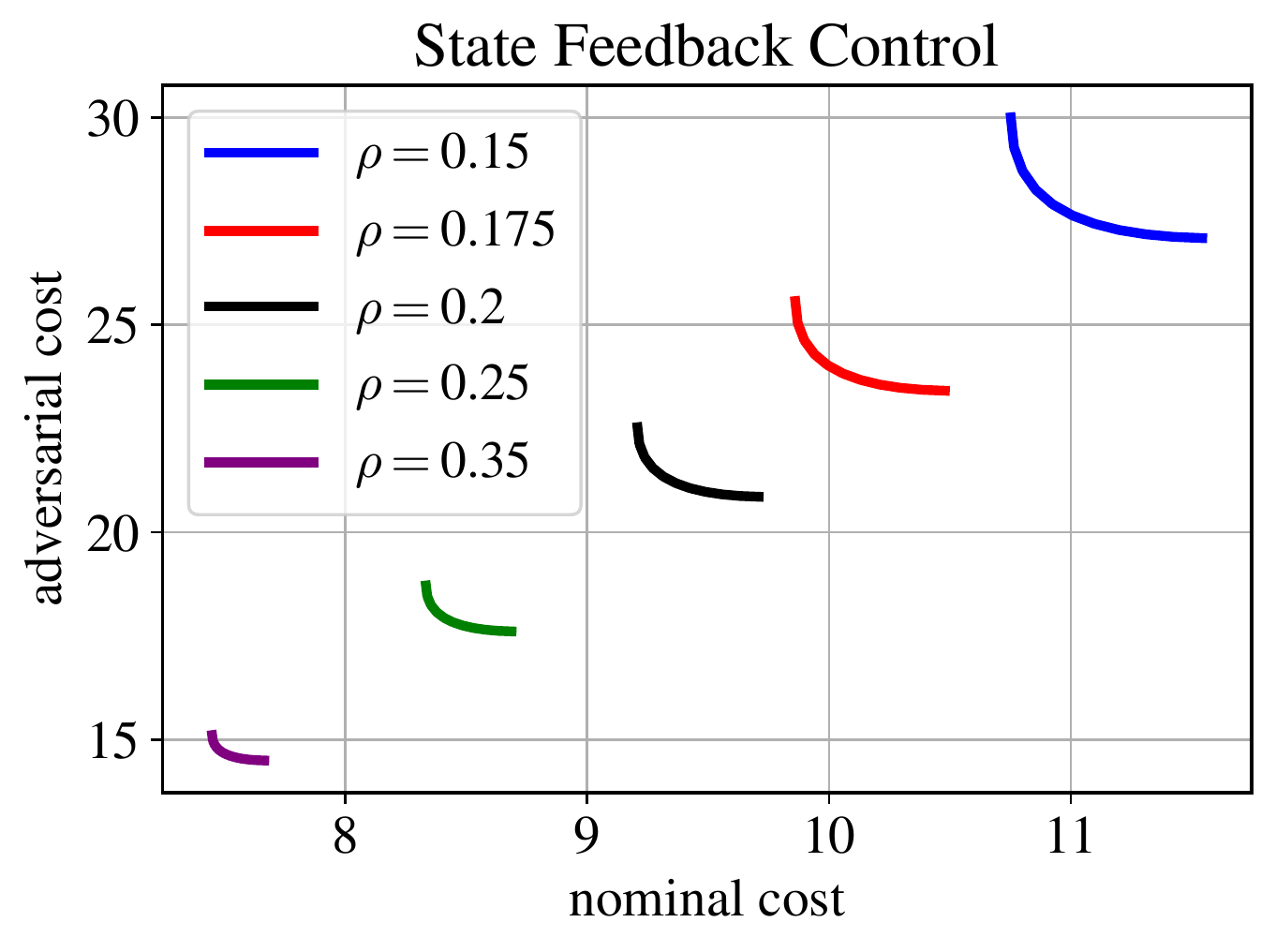}
    \includegraphics[width=0.3\textwidth]{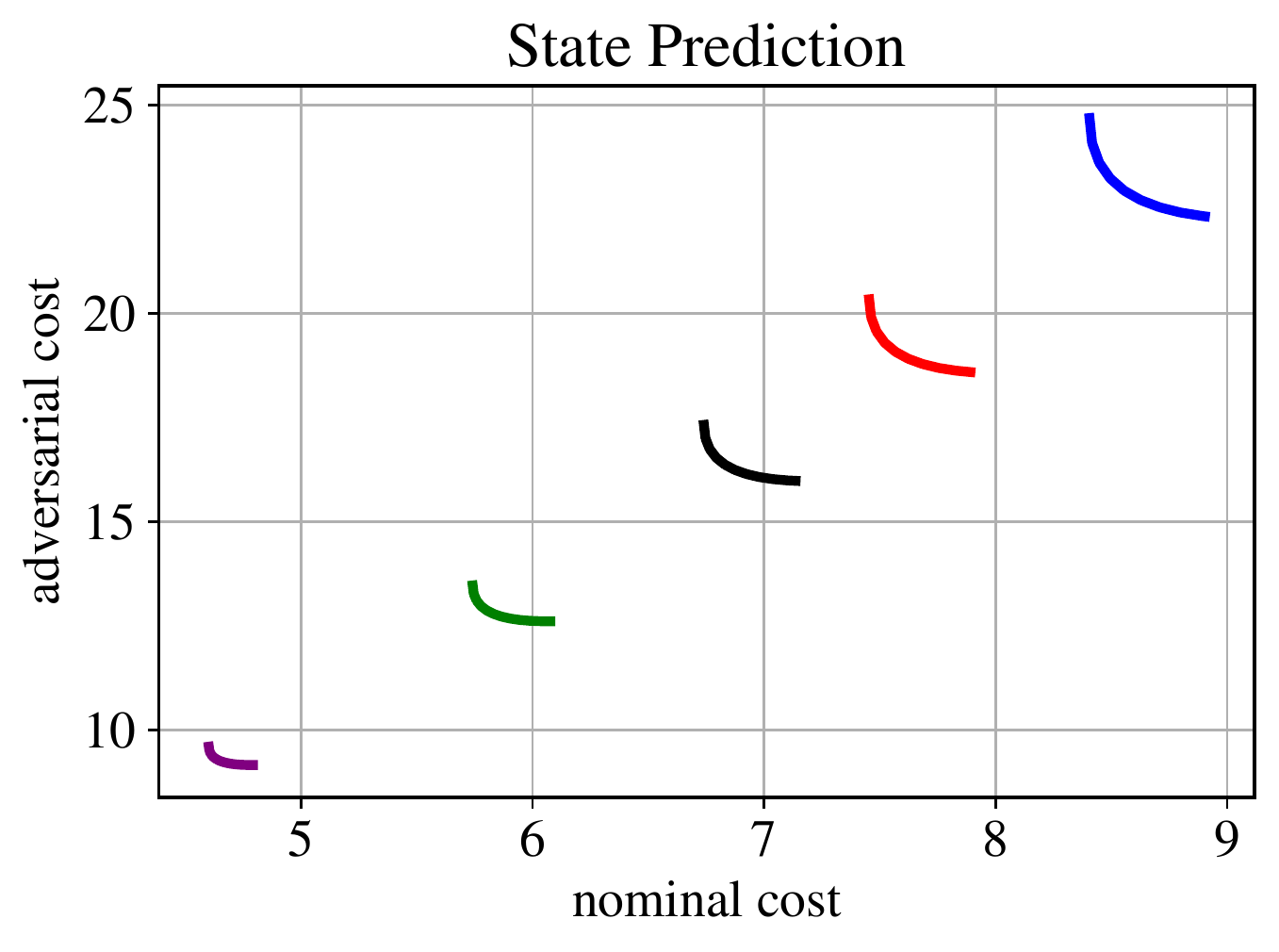}
    \includegraphics[width=0.3\textwidth]{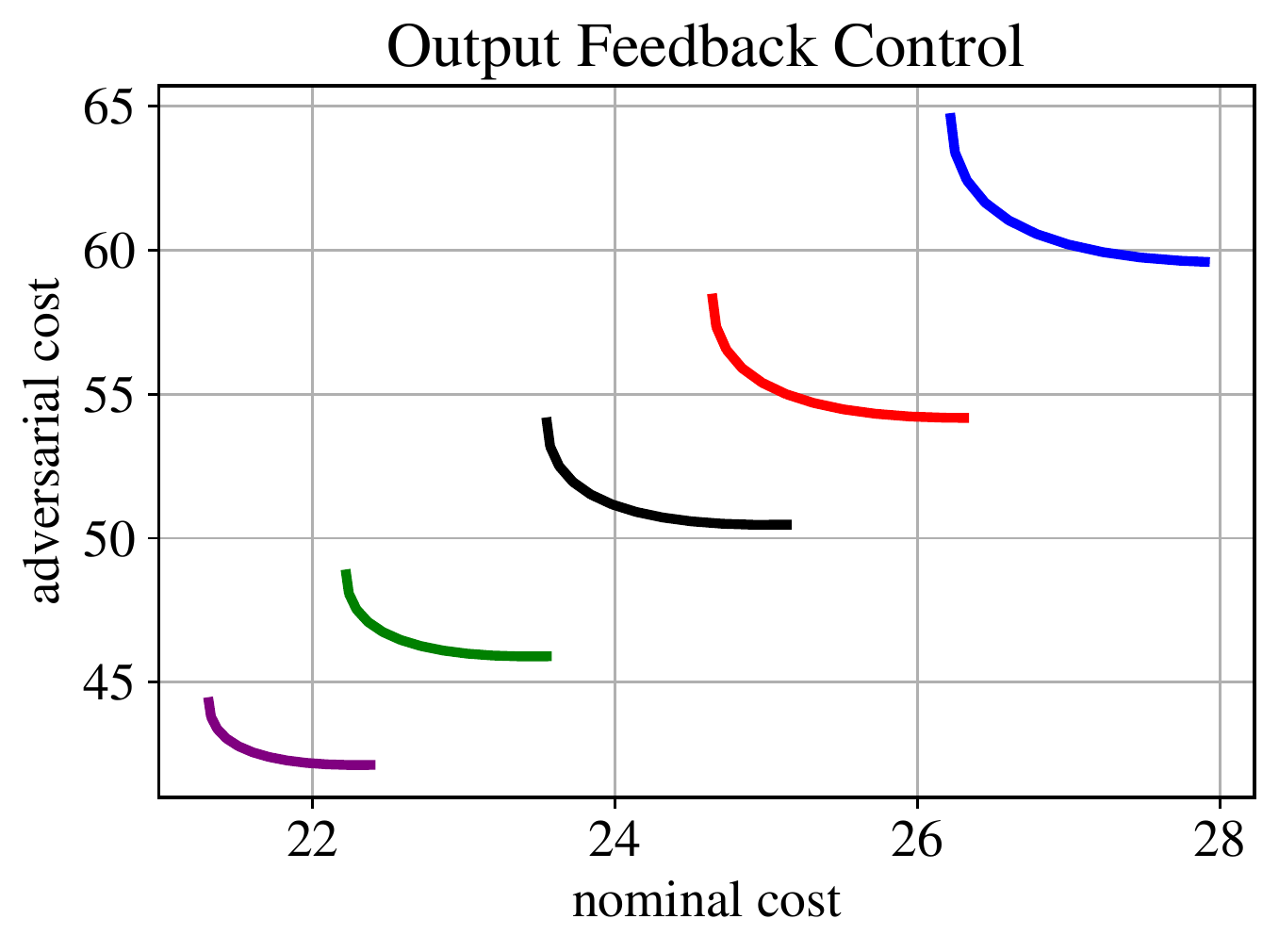}
    \caption{Tradeoff Curves}
    \label{fig: tradeoff curves}
    \end{subfigure}
    \begin{subfigure}[b]{\textwidth}
        \includegraphics[width=0.3\textwidth]{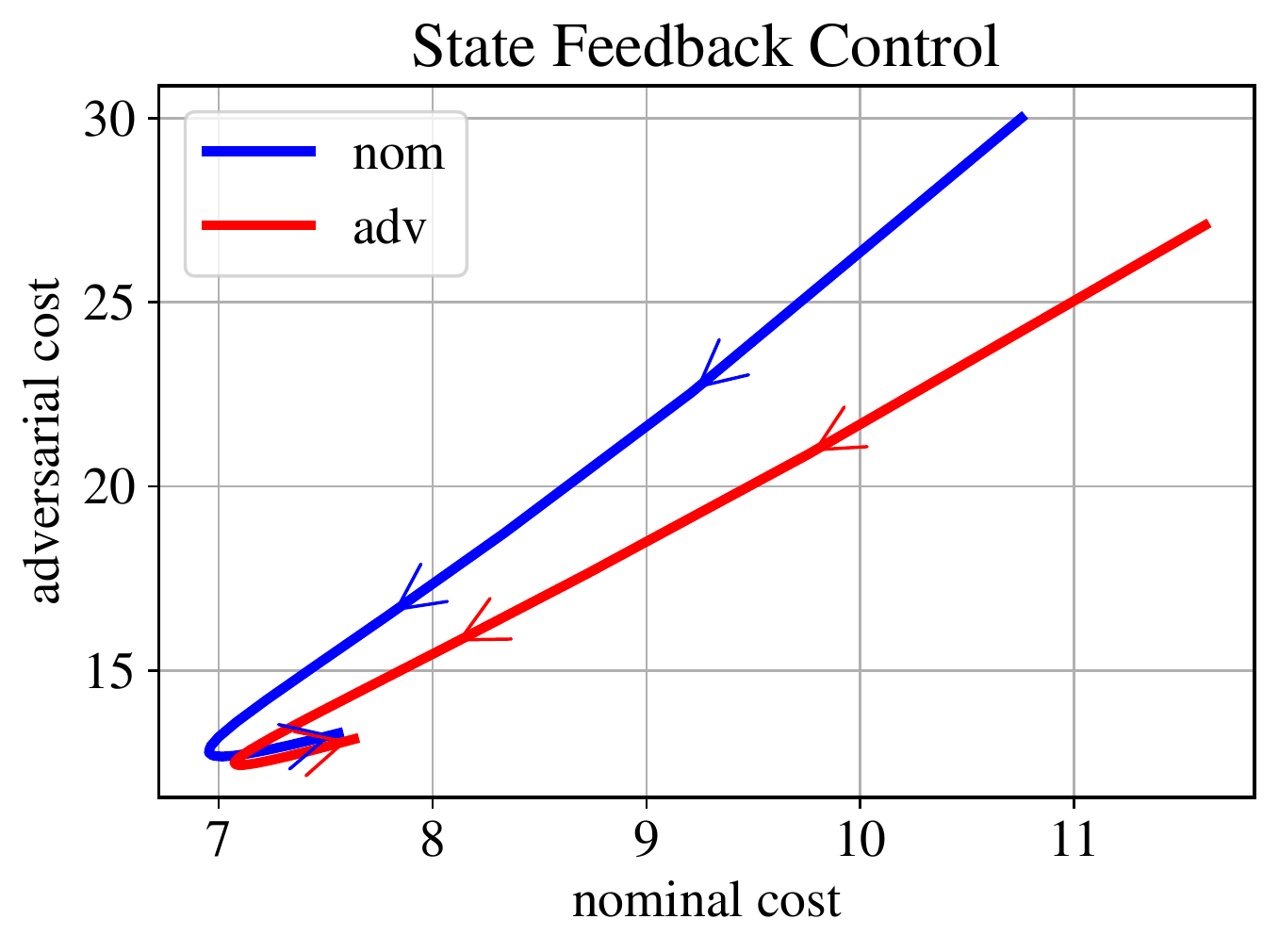}
    \includegraphics[width=0.3\textwidth]{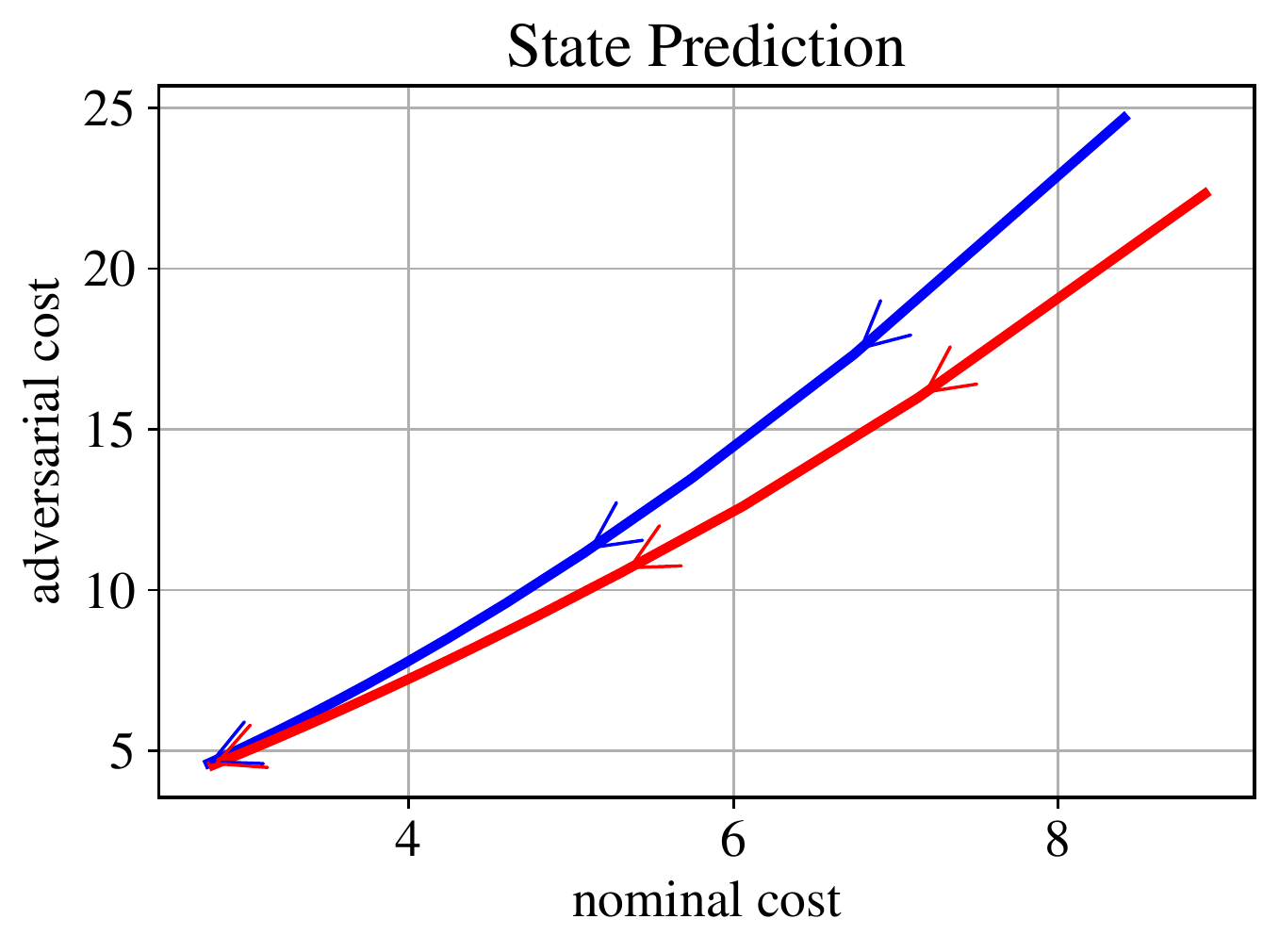}
    \includegraphics[width=0.3\textwidth]{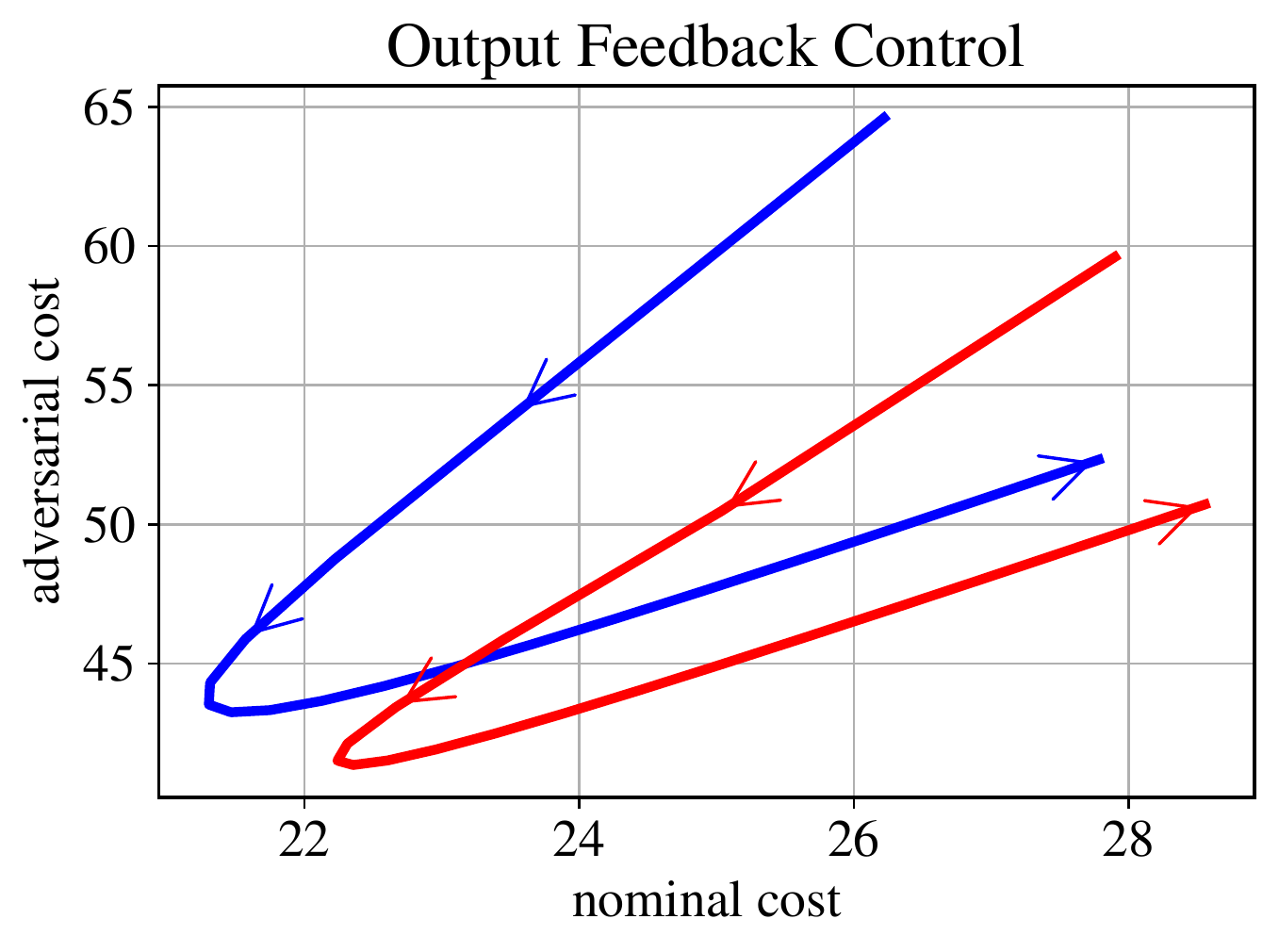}
    \caption{Tradeoff Envelopes}
    \label{fig: envelopes}
    \end{subfigure}
    \caption{We illustrate the impact of controllability and observability on the severity of the performance robustness tradeoff for state feedback control, state prediction, and output feedback control of the simple integrator system in \eqref{eq: integrator}. In \Cref{fig: tradeoff curves}, we plot the tradeoff curves for fixed values of the parameter $\rho$, which determines the controllability and/or observability of the system. The adversarial costs in this setting use $\varepsilon=0.1$, and the tradeoff curves evaluate adversarially robust controllers/observers designed with $\varepsilon \in [0,0.1]$. In \Cref{fig: envelopes}, the envelopes of the tradeoff curves are plotted by evaluating the nominal and $\varepsilon=0.1$-adversarially robust controllers/observers for $\rho \in [0.15, 1.0]$. The arrows are in the direction of increasing $\rho$.}
    \label{fig: tradeoffs and envelopes}
    \vspace{-12pt}
\end{figure*}

\paragraph{Compounding Impact of Poor Controllability and Observability}

We study the dependence of the tradeoff severity on system controllability and observability in \Cref{fig: tradeoffs and envelopes}. In this example, we consider the two dimensional integrator system defined by
\begin{align}
    \label{eq: integrator}
    x_{t+1} &= \bmat{1 & \rho \\ 0 & 1} x_t + \delta_t^x + w_t^x + \bmat{0 \\ 1} u_t, \\ y_t &= C x_t + \delta_t^y + w_t^t.
\end{align}
We consider three settings with this system: state feedback control, state prediction, and output feedback control. In all three settings, $\delta_t^x$ is an energy budget constrained adversarial input to the state, and $w_t^x$ is independent zero mean Gaussian noise with identity covariance. For state feedback control, we have $C=I$, $\delta_t^y=0$ and $w_t^y=0$. For both state prediction and output feedback control, we take $C = \bmat{1 & 0}$, with $\delta_t^y$ as an adversarial input to the measurement, and $w_t^y$ as independent zero mean Gaussian noise with covariance 1. For both the control settings (state feedback and output feedback), we set $Q=I$ and $R=1$, and the objective is to minimize the LQR cost. In state prediction, the objective is to minimize the state estimation error with $u_t = 0$, as described in \Cref{s: state prediction}. For the state feedback control problem, we study whether poor controllability increases the severity of the performance robustness tradeoffs, as predicted in \Cref{s: state feedback performance robustness tradeoffs}. The controllability of the provided system may be varied through the parameter $\rho$:  when $\rho$ is small, the system has poor controllability, and as $\rho$ increases, controllability increases. For the state prediction problem, we determine whether poor observability increases the severity of the tradeoff, as predicted in \Cref{s: state prediction}. The observability of the system may also be varied through the parameter $\rho$: when $\rho$ is small, the observability is poor, and when $\rho$ increases, observability improves. For the output feedback control setting, we study how the impacts of poor observability and controllability compound to result in a severe performance-robustness tradeoff. In \Cref{fig: tradeoff curves}, we consider the tradeoff curves traced out by fixing $\rho$ and then evaluating the nominal and adversarial ($\varepsilon=0.1$) costs of adversarially robust controllers/observers designed with budget $\varepsilon$ varying between $[0,0.1]$. We observe that as controllability and observability decrease, the tradeoff curves shift upward and also widen. This corroborates the trends described in Theorem~\ref{thm: nominal cost gap upper bound} and Theorem~\ref{thm: state prediction upper bound}, where we show the bound on the nominal cost gap between adversarially robust and nominal controllers/observers (i.e.\ width of the tradeoff curve) grows larger as controllability and obserbability decrease, respectively. These trends are further illustrated in Figure~\ref{fig: envelopes}, where we plot the nominal and adversarial ($\varepsilon = 0.1$) costs attained by the nominal and adversarially robust ($\varepsilon = 0.1$) controllers/observers as a function of $\rho \in [0.15, 1]$. We observe that for small $\rho$, the system has poor controllability and observability, hence the distance between the nominal and adversarial costs is large. As controllability and observability increase, this gap decreases monotonically. Note that the costs do not monotonically improve as $\rho$ increases after some point for the control problems, as the amplification of disturbances from the integrator eventually outstrips the benefits of better controllability or observability.

\begin{figure}
    \centering
    \begin{subfigure}[b]{0.3\textwidth}
    
    \includegraphics[width=\textwidth]{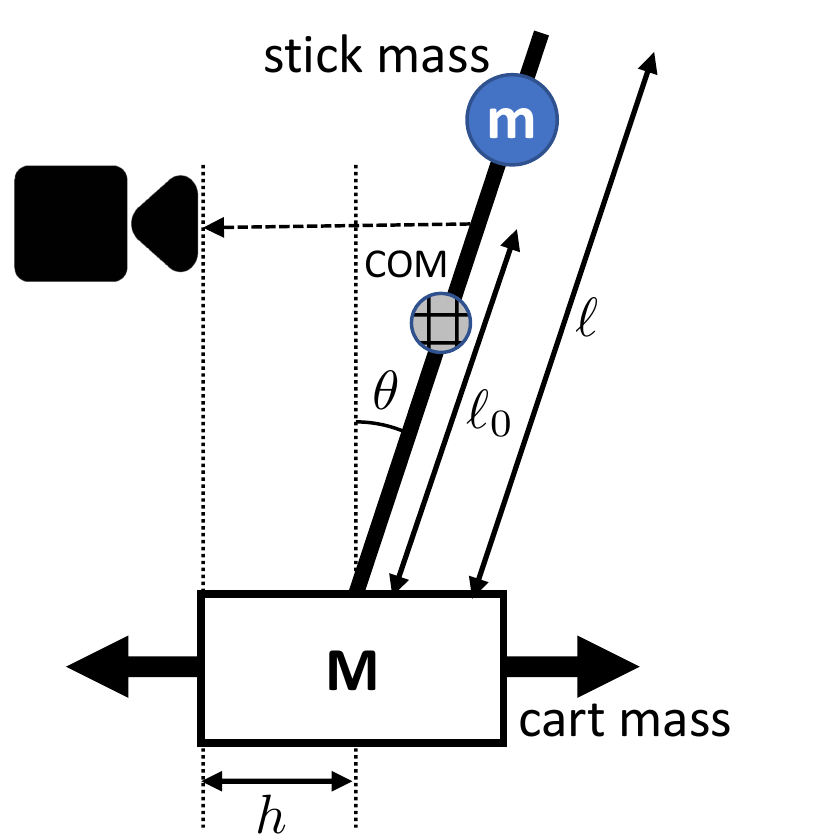}
    \caption{Cartpole setup}
    \label{fig: cartpole diagram}
    
    \end{subfigure}
    \hspace{0.1in}
    \begin{subfigure}[b]{0.4\textwidth}
    
    \includegraphics[width=\textwidth]
    {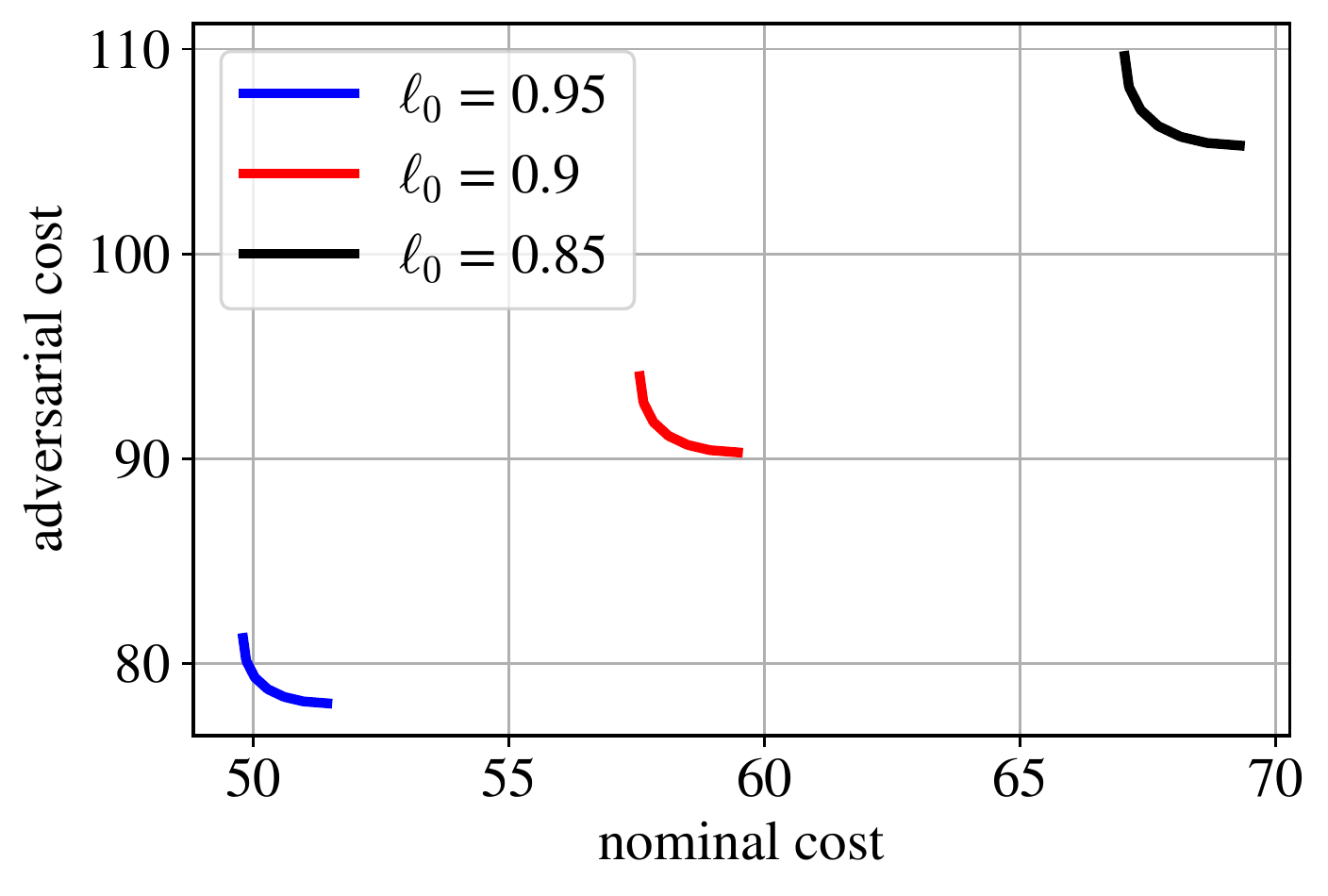}
    
    \caption{Tradeoff Curves}
    
    \label{fig: cartpole tradeoffs}
    
    \end{subfigure}
    \caption{We study the performance-robustness tradeoffs in visual feedback pole-balancing for various fixation points. As the fixation point decreases ($\ell_0$ becomes smaller), the observability of the system decreases. The adversarial cost is evaluated with $\varepsilon=0.125$, and the tradeoffs are generated by synthesizing adversarially robust controllers for $\varepsilon \in [0, 0.125]$. We see as the fixation point decreases, the tradeoffs for the linearized system become more severe.}
    \vspace{-12pt}
    \label{fig:linearized cartople}
\end{figure}

\paragraph{Tradeoffs of Linearized Pole Balancing} To further illustrate the impact of observability upon the severity of the performance-robustness tradeoff, we consider the example of pole-balancing using visual feedback from a single fixation point on the pole, illustrated in Figure~\ref{fig:linearized cartople}. The dynamics of a pole mounted on a cart may be represented as 
\ifnum\value{tac}>0{
\begin{equation}
 \begin{aligned}
    \label{eq: cartpole dynamics}
     u &+ w + \delta \\
     &= (M+m)(\ddot{h} + d_h \dot h)+ m\ell((\ddot{\theta}+d_\theta \dot \theta)\cos\theta-\dot{\theta}^2\sin\theta), \\
     0&=m((\ddot{h} + d_h \dot h)\cos\theta+\ell(\ddot{\theta}+d_\theta \dot \theta)-g\sin\theta),\\
     y &= h+\ell_0\sin\theta.
 \end{aligned}
 \end{equation}
}\else{
\begin{equation}
 \begin{aligned}
    \label{eq: cartpole dynamics}
     u + w + \delta &= (M+m)(\ddot{h} + d_h \dot h)+ m\ell((\ddot{\theta}+d_\theta \dot \theta)\cos\theta-\dot{\theta}^2\sin\theta), \\
     0&=m((\ddot{h} + d_h \dot h)\cos\theta+\ell(\ddot{\theta}+d_\theta \dot \theta)-g\sin\theta),\\
     y &= h+\ell_0\sin\theta.
 \end{aligned}
 \end{equation}}\fi
 This model was proposed in \cite{leong2016understanding} to illustrate the fundamental limitations of control. It was also studied in \cite{xu2021learned} to determine how these limitations interact with the sample complexity of learning controllers from data. 
 As depicted in \Cref{fig: cartpole diagram}, $M$ is the mass of the cart, $h$ is the position of the cart, $\theta$ is the angle of the pole, $\ell$ is the length of the pole, and $\ell_0$ is the fixation point which the camera observes. The mass of the pole is denoted by $m$ and the acceleration due to gravity by $g$. The damping coefficients are $d_h$ and $d_\theta$.  Our measurement $y$ the distance from the camera to the fixation point. The actuation noise is denoted $w$ and the adversarial input is denoted $\delta$. We consider an instance of this system with $(M,m,\ell, g, d_h, d_\theta)= (1, 0.1, 1, -10, 0.2, 0.2)$, and various fixation points $\ell_0 < \ell$. We discretize the dynamics in \eqref{eq: cartpole dynamics} using Euler discretization with stepsize $0.04$, and linearize the system about the upright equilibrium point. The stochastic noise $w$ is i.i.d. $\calN(0,1)$, and the adversarial input $\delta$ is chosen to maximize the cost while falling within the budget $\norm{\delta}^2 \leq \varepsilon$ on average, where $\varepsilon$ varies as discussed in \Cref{fig:linearized cartople}. The cost is specified by $Q=I$, $R=1$. Varying the fixation point varies the observability of the system. In particular, as $l_0$ decreases, the observability becomes poor.  We plot the tradeoff curves for this system as $l_0$ varies in \Cref{fig: cartpole tradeoffs}. As in the integrator experiments, the tradeoff becomes more severe as observability decreases. 

\begin{figure*}
    \centering
    \includegraphics[width=0.3\textwidth]{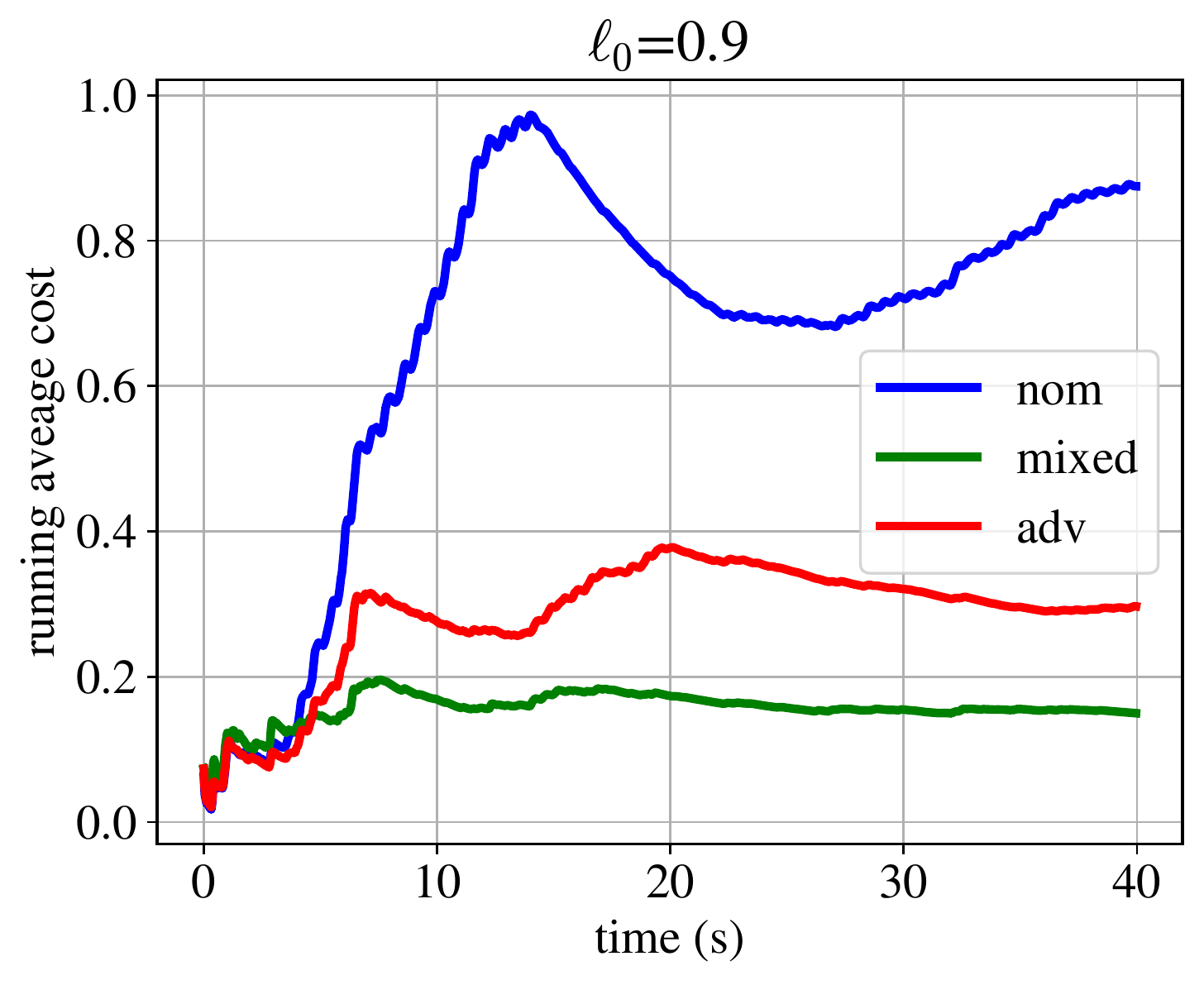}
    \includegraphics[width=0.3\textwidth]{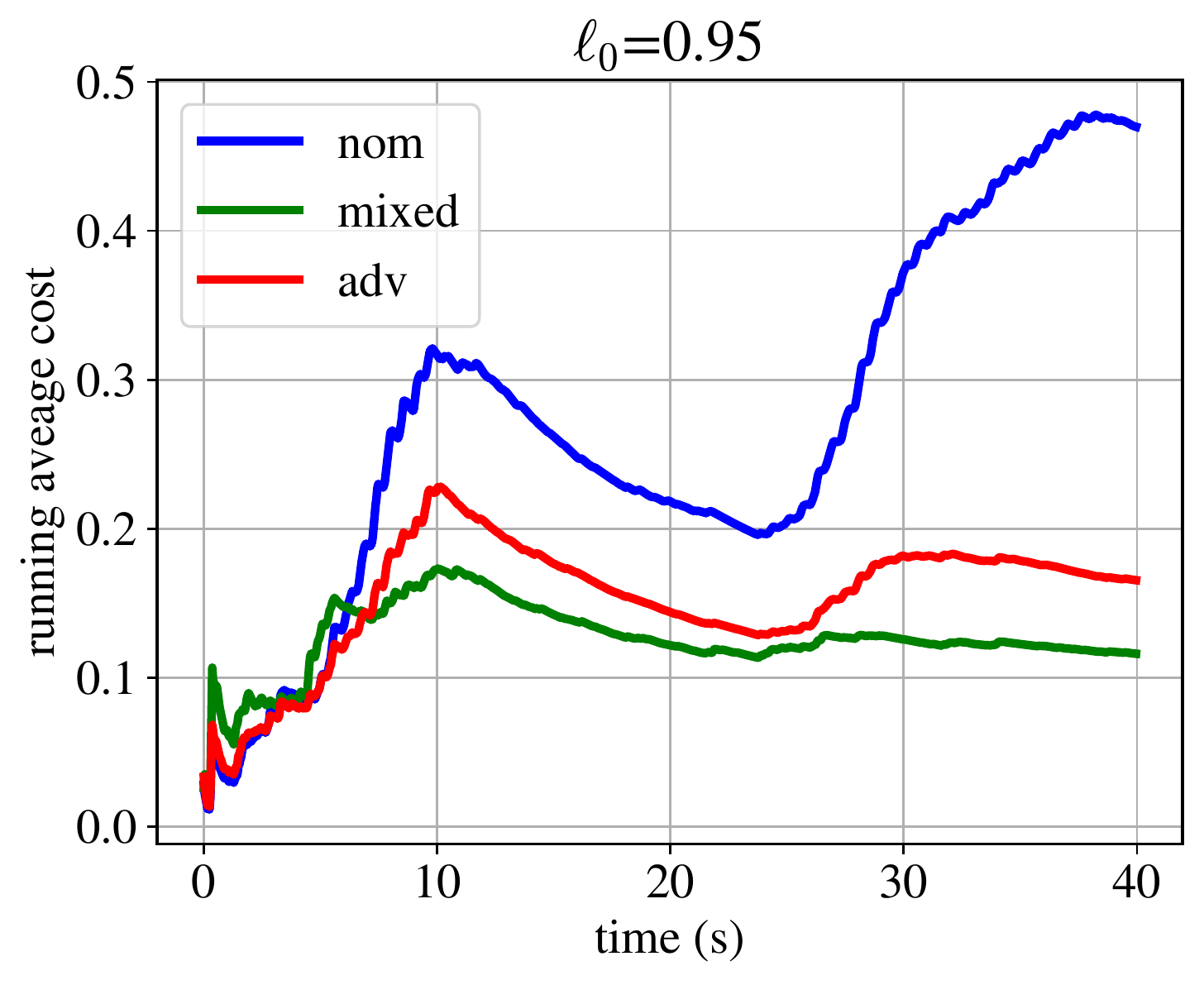}
    \includegraphics[width=0.3\textwidth]{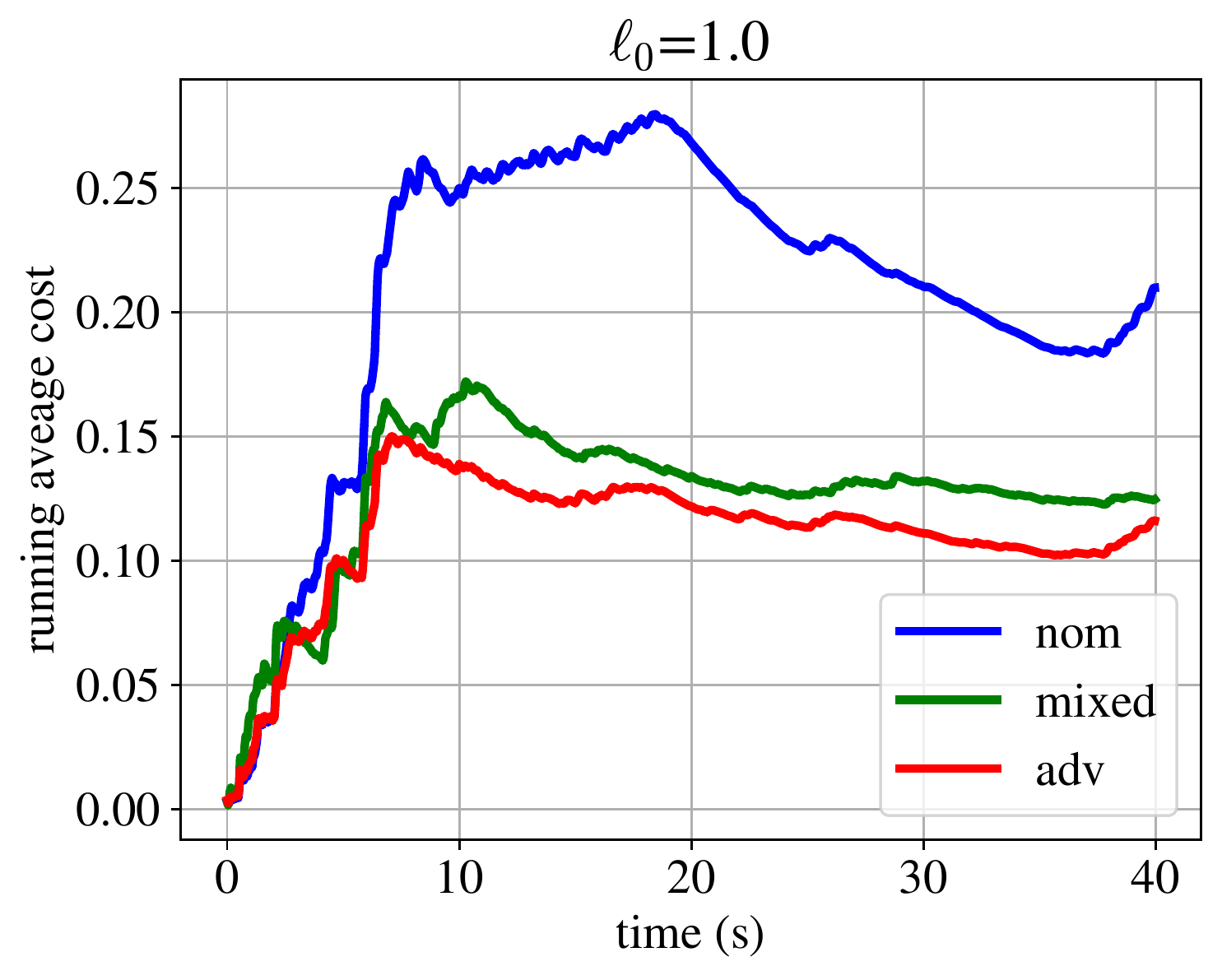}
    \caption{Simulations of pole-balancing on the nonlinear continuous time system using linear discrete time controllers. The ``nom'' controller is the LQG, while the ``adv'' controller is the adversarially robust controller synthesized with $\varepsilon=0.015$. A mixed $\calH_2/\calH_\infty$ baseline is also included. For three different fixation points $\ell_0$, the adversarially robust controller outperforms the nominal controller, and similarly to the mixed $\calH_2/\calH_\infty$ controller.}
    \label{fig:cartpole simulation}
    \vspace{-12pt}
\end{figure*}

\paragraph{Performance of Adversarially Robust Control for Pole Balancing}
We now use the cartpole system described in the previous example in order to test the performance of the adversarially robust controller in Figure~\ref{fig:cartpole simulation}. In particular, we simulate the nonlinear continuous time system in \eqref{eq: cartpole dynamics} under discrete time linear controllers synthesized using the discretized and linearized version of the dynamics. The system parameters and cost matrices are the same as in the previous example, however $\delta$ is modified to enter all states and measurements of the system through an identity mapping to better enhance robustness to the linearization error. We plot the running average cost on the $y$-axis, and the time in seconds on the $x$-axis. In the simulations, the only perturbations impacting the system are
zero mean stochastic noise with standard deviation $0.002$ entering the input. The simulations are run for three different fixation points, $\ell_0 = 0.85, 0.9$ and $0.95$. For all fixation points we see that the adversarially robust controller performs significantly better than the nominal LQG controller. It is also worth noting that as the fixation point decreases, the total cost increases, and the gap between the two controllers also increases. In particular, as observabilitiy becomes poor, choosing the appropriate level of robustness becomes more essential.  A mixed $\calH_2/\calH_\infty$ controller is included as an additional baseline.  The mixed controller is synthesized in continuous time using the matlab function $\texttt{h2hinfsyn}$ to minimize the $\calH_2$ norm of the map subject to a constraint that the $\calH_\infty$ norm is at most 20.  We see that the performance of the adversarially robust controller is similar to the mixed $\calH_2/\calH_\infty$ controller in all settings. We emphasize that our intent is not to propose a controller which uniformly outperforms mixed $\calH_2/\calH_\infty$ controllers, but rather to propose a controller for which we can effectively certify robustness while providing quantitative performance-robustness tradeoffs. The purpose of this experiment is to highlight that the synthesis procedure is effective at producing a controller that enhances general robustness, and achieves performance comparable to existing robust synthesis approaches. 
\section{Conclusion}

We proposed an adversarially robust LQ control problem, and provided sufficient conditions for the optimal solution to this problem, along with an algorithm that converges to the optimal solution when these conditions are satisfied. The solution is closely related to a central suboptimal $\calH_\infty$ controller. An interesting aspect of this solution is that unlike pure $\calH_2$ controllers, the adversarially robust controller depends upon the noise statistics. 
Experiments show that the adversarially robust controller performs similarly to mixed $\calH_2/\calH_\infty$ controllers on a simple linear system, and can beat out the LQG in the face of model error arising from linearization.

We used the adversarially robust control problem as a means to study performance-robustness tradeoffs in control. In particular, we derived quantitative upper and lower bounds on the performance gap between the nominal state feedback controller and the adversarially robust state feedback controller. The bounds show that systems with uniformly good controllability have small performance-robustness tradeoffs, while closed-loop nominal systems with a highly controllable mode in the disturbance channel will have a large performance-robustness tradeoff. We also derived bounds on the nominal state estimation error gap of the adversarially robust state estimator compared to the Kalman Filter. In this case, it was shown that systems with uniformly good observability have small tradeoffs. These trends are corroborated by experiments on a simple linear system by tracing out tradeoff curves. One direction for future work is to consider how other adversarial training techniques can be translated to robust controller synthesis and analysis.

\section*{Acknowledgements}
Bruce D.\ Lee is supported by the DoD through the National Defense Science \& Engineering Graduate Fellowship Program. The research of Hamed Hassani is supported by NSF Grants 1837253, 1943064, 1934876, AFOSR Grant FA9550-20-1-0111, and DCIST-CRA. Nikolai Matni is funded by NSF awards CPS-2038873, CAREER award ECCS-2045834, and ECCS-2231349. 

\bibliographystyle{abbrvnat}

\bibliography{refs}
\appendix

\section{Proofs from Section \ref{sec: advrobust lq control}: Controller Synthesis}

\subsection{Proof of Lemma~\ref{lem:adversarial LQR sol}: Soft-constrained State Feedback Solution}
\label{ap: soft-constrained state feedback sol}
\textit{Proof:}
First define the cost to go for a horizon $T$ at step $\tau$. 
\[
    V_{\tau, T}(x_\tau) := \min_{u_{\tau:T-1}} \Ex_{w_{\tau:T-1}} \brac{ \max_{\substack{\delta_{\tau:T-1} \\ causal}} x_T^\top Q x_T + \sum_{t=\tau}^{T-1} x_t Q x_t + u_t^\top R u_t - \gamma^2 \delta_t^\top \delta_t}.
\]
Note that for any $T$, $V_{T,T}(x) = x^\top Q x$. We will show inductively that $V_{t,T}(x) = x^\top P_{t,T} x + q_{t,T}$ for any $t$. To do so, begin with the inductive hypothesis $V_{t+1, T}(x) = x^\top P_{t+1, T} x + q_{t+1}$. Applying this hypothesis, we have
\begin{align*}
     V_{\tau, T}(x_\tau) &= x_\tau^\top Q x_\tau + \min_{u_\tau} \curly{u_\tau^\top R u_\tau + \min_{u_{\tau+1:T-1}}  \Ex_{w_{\tau:T-1}} \brac{ \max_{\substack{\delta_{\tau:T-1} \\ causal}} x_T^\top Q x_T -\delta_\tau^\top \delta_\tau + \sum_{t=\tau+1}^{T-1} x_t Q x_t + u_t^\top R u_t - \gamma^2 \delta_t^\top \delta_t}} \\
    &= x_\tau^\top Q x_\tau + \min_{u_\tau} \bigg\{u_\tau^\top R u_\tau \\
    &+ \min_{u_{\tau+1:T-1}}  \Ex_{w_\tau} \brac{\max_{\delta_\tau} -\gamma^2 \delta_\tau^\top \delta_\tau + \Ex_{w_{\tau+1:T-1}} \brac{\max_{\substack{ \delta_{\tau+1:T-1} \\ causal}} x_T^\top Q x_T  + \sum_{t=\tau+1}^{T-1} x_t Q x_t + u_t^\top R u_t - \gamma^2 \delta_t^\top \delta_t}} \bigg\}  \\
     &= x_\tau^\top Q x_\tau + \min_{u_\tau} \curly{u_\tau^\top R u_\tau + \min_{u_{\tau+1:T-1}}  \Ex_{w_\tau} \brac{\max_{\delta_\tau} -\gamma^2 \delta_\tau^\top \delta_\tau + V_{\tau+1,T}(Ax_\tau + B_0 w_\tau + B_1 \delta_\tau + B_2 u_\tau)}} \\
     &= x_\tau^\top Q x_\tau + \min_{u_\tau} \bigg\{u_\tau^\top R u_\tau \\
     &+ \Ex_{w_\tau} \brac{\max_{\delta_\tau} -\gamma^2 \delta_\tau^\top \delta_\tau + (Ax_\tau + B_0 w_\tau + B_1 \delta_\tau + B_2 u_\tau)^\top P_{\tau+1, T} (Ax_\tau + B_0 w_\tau + B_1 \delta_\tau + B_2 u_\tau) + q_{\tau+1, T}}\bigg\}
\end{align*}
Letting $\phi_t := (Ax_\tau + B_0 w_\tau + B_2 u_\tau)$, we see that the maximization with respect to $\delta_\tau$ is equivalent to
\begin{align*}
    \max_{\delta_\tau} \delta_\tau^\top \paren{B_1^\top P_{\tau+1, T} B_1 - \gamma^2 I} \delta_\tau^\top + 2 \phi_\tau^\top P_{\tau+1, T} B_1 \delta_\tau.
\end{align*}
If $B_1^\top P_{\tau+1, T} B_1 \prec \gamma^2 I$, then there is a unique optimal solution given by $\delta_\tau = -(B_1^\top P_{\tau+1, T} B_1 - \gamma^2 I)^{-1} B_1^\top P_{\tau+1, T} \phi_\tau$. Substituting this back in, we see that 
\begin{align*}
    V_{\tau, T}(x_\tau) &= q_{\tau+1, T} + x_\tau^\top Q x_\tau + \min_{u_\tau} \bigg\{u_\tau^\top R u_\tau + \Ex_{w_\tau} \bigg \lbrack \phi_\tau^\top (-\gamma^2 P_{\tau+1, T} B_1 (B_1^\top P_{\tau+1, T} B_1 - \gamma^2 I)^{-2} B_1^\top P_{\tau+1, T} \\
    &\quad \quad + (I -  P_{\tau+1, T} B_1 (B_1^\top P_{\tau+1, T} B_1- \gamma^2 I)^{-1} B_1^\top) P_{\tau+1, T} (I - B_1 (B_1^\top P_{\tau+1, T} B_1 - \gamma^2 I)^{-1} B_1 P_{\tau+1, T})) \phi_\tau \bigg \rbrack\bigg\} \\
    &= q_{\tau+1, T} + x_\tau^\top Q x_\tau + \min_{u_\tau} \bigg\{u_\tau^\top R u_\tau + \Ex_{w_\tau} \bigg \lbrack \phi_\tau^\top (P_{\tau+1, T} + P_{\tau+1, T} B_1 (\gamma^2 I - B_1^\top P_{\tau+1, T} B_1 )^{-1}  B_1^\top P_{\tau+1}) \phi_\tau \bigg \rbrack\bigg\}.
\end{align*}
Defining $M_{\tau+1, T} := P_{\tau+1, T} + P_{\tau+1, T} B_1 (\gamma^2 I - B_1^\top P_{\tau+1, T} B_1 )^{-1}  B_1^\top P_{\tau+1}$, the above expression reduces to
\begin{align*}
     q_{\tau+1, T} &+ x_\tau^\top Q x_\tau + \min_{u_\tau} \bigg\{u_\tau^\top R u_\tau + \Ex_{w_\tau} \bigg \lbrack \phi_\tau^\top M_{\tau+1} \phi_\tau \bigg \rbrack\bigg\} \\
     &=  q_{\tau+1, T} + x_\tau^\top Q x_\tau + \min_{u_\tau} \bigg\{u_\tau^\top R u_\tau + (Ax_t + B_2 u_t)^\top M_{\tau+1,T} (Ax_t + B_2 u_t)\bigg\} + \trace(M_{\tau+1,T} B_0 B_0^\top)
\end{align*}
Minimizing with respect to $u_\tau$ provides
\[
    u_\tau = -(R + B_2^\top M_{\tau+1, T} B_2)^{-1} B_2^\top M_{\tau+1, T} A x_\tau.
\]
Substituting this back in, we have
\begin{align*}
    V_{\tau, T}(x_\tau) &= q_{\tau+1, T} + \trace(M_{\tau+1, T} B_0 B_0^\top)  + x_\tau^\top \paren{ Q + A^\top M_{\tau+1, T} A - A^\top M_{\tau+1, T} B_2 (R+B_2 ^\top M_{\tau+1 ,T}B_2)^{-1} B_2^\top 
    M_{\tau+1, T} A} x_\tau  \\
    &=: q_{\tau, T} + x_\tau^\top P_{\tau, T}x_\tau
\end{align*}

Thus the cost to go is given as 
\begin{align*}
    V_{\tau, T}(x_\tau) &= q_{\tau, T} + x_\tau^\top P_{\tau, T} x_\tau,
\end{align*}
where 
\begin{align*}
    P_{\tau, T} &= Q + A^\top M_{\tau+1, T} A - A^\top M_{\tau+1, T} B_2(R+B_2^\top M_{\tau+1 ,T}B_2)^{-1} B_2^\top 
    M_{\tau+1, T} A \\
     M_{\tau, T} &= P_{\tau, T} +  P_{\tau, T} B_1 \paren{\gamma^2 I - B_1^\top P_{\tau, T} B_1}^{-1} B_1^\top P_{\tau, T} \\
    q_{\tau, T} &= q_{\tau+1, T+1} +  \trace(M_{\tau+1, T} B_0B_0^\top) \\
    q_{T,T} &= 0\\ 
    P_{T,T} &= Q.
\end{align*}
Taking the limit as $T \to \infty$, and denoting $P_\tau := P_{\tau, \infty}$ observe that if the iteration
\begin{align*}
    P_{\tau} &= Q + A^\top M_{\tau+1} A - A^\top M_{\tau+1} B_2(R+B_2^\top M_{\tau+1}B_2)^{-1} B_2^\top 
    M_{\tau+1} A \\
     M_{\tau} &= P_{\tau} +  P_{\tau} B_1 \paren{\gamma^2 I - B_1^\top P_{\tau}B_1}^{-1} B_1^\top P_{\tau} 
\end{align*}
converges to a fixed point $P$, then the fixed point must satisfy 
\begin{align*}
    P_\gamma &= Q + A^\top M_\gamma A - A^\top M_\gamma B_2(R+B_2^\top M_\gamma B_2)^{-1} B_2^\top 
    M_\gamma A \\
    M_\gamma &= P_\gamma +  P_\gamma B_1 \paren{\gamma^2 I - B_1^\top P_\gamma B_1}^{-1} B_1 P_\gamma.
\end{align*}
Finding a pair $(P_\gamma,M_\gamma)$ that satisfy the above conditions is equivalent to finding a $P_\gamma$ such that 
\begin{align}\label{eq: gamma-suboptimal DARE}
    P_\gamma = Q + A^\top P_\gamma A - A^\top P_\gamma \bmat{B_1 & B_2}\paren{\bmat{B_1 & B_2}^\top P_\gamma \bmat{B_1 & B_2} + \bmat{-\gamma^2 I & \\ & R}}^{-1}\bmat{B_1 & B_2}^\top P_\gamma A.
\end{align}
To see that this is the case, note that
\begin{align*}
    &P_\gamma - P_\gamma \bmat{B_1 & B_2} \bmat{B_1^\top P_\gamma B_1 - \gamma^2 I  & B_1^\top P_\gamma B_2 \\ B_2^\top P_\gamma B_1 & B_2^\top P_\gamma B_2  + R}^{-1} \bmat{B_1 & B_2}^\top P_\gamma   \\
    & \hspace{-.0 in}= P_\gamma - P_\gamma\bmat{B_1^\top \\ B_2^\top}^\top \bmat{ \Phi_1^{-1}  + \Phi_1^{-1} B_1^\top P_\gamma B_2 \Phi_2^{-1} B_2^\top P_\gamma B_1 \Phi_1^{-1} & \Phi_1^{-1} B_1^\top P_\gamma B_2 \Phi_2^{-1} \\ \Phi_2^{-1} B_2^\top P_\gamma B_1 \Phi_1^{-1} & \Phi_2^{-1}} \bmat{B_1^\top \\ B_2^\top} P_\gamma \\
    &= M_\gamma - M_\gamma B_2(B_2^\top M_\gamma B_2^\top +R)^{-1}  B_2^\top M_\gamma
\end{align*}
where $\Phi_1 = \gamma^2 I - B_1^\top P B_1$ and $\Phi_2 = B_2^\top M B_2  + R$. The first equality above follows from the inversion formula for a block $2\times2$ matrix. 

By our assumption that there exists a positive semidefinite solution to \eqref{eq:modified DARE}, the iterations converge to this matrix. This can be seen by verifying the monotonicty of the sequence $P_\tau$, as in Lemma 3.3 of \cite{basar1991hinf}. 

We thus have that in the limit, the cost converges to $\trace(M_\gamma B_0 B_0^\top)$,
where $M_\gamma = P_\gamma + P_\gamma B_1 (\gamma^2 I - B_1^\top P_\gamma B_1)^{-1} B_1^\top P_\gamma $ and $P_\gamma$ solves
\begin{align*}
    P_\gamma = Q + A^\top P_\gamma A - A^\top P_\gamma \bmat{B_1 & B_2}\paren{\bmat{B_1 & B_2}^\top P_\gamma \bmat{B_1 & B_2} + \bmat{-\gamma^2 I & \\ & R}}^{-1}\bmat{B_1 & B_2}^\top P_\gamma A.
\end{align*}
Thus a steady state adversarially robust LQR controller at level $\gamma$ is given by 
\begin{equation*}
\begin{aligned}
\label{eq:advrob controller}
    u_t &= F_\gamma x_t \\
    F_\gamma &= -(R + B_2^\top M B_2)^{-1} B_2^\top M A \\
    M_\gamma &= P_\gamma+P_\gamma B_1 (\gamma^2 I - B_1^\top P_\gamma B_1)^{-1} B_1^\top P_\gamma \\
    P_\gamma &\mbox{ solves \eqref{eq:modified DARE}},
\end{aligned}
\end{equation*}
and the optimal adversarial perturbation is given by
\[
    \delta_t = (\gamma^2 I - B_1^\top  P_\gamma B_1)^{-1} B_1^\top P_\gamma ((A+BKx_t) + B_0 w_t).
\]
This may be expressed compactly as 
\begin{align*}
    \bmat{\delta_t \\ u_t} &= \bmat{E_\gamma \\ F_\gamma} x_t + \bmat{\Delta_\gamma \\ 0} w_t \\
    \bmat{E_\gamma \\ F_\gamma} &= -\paren{\bmat{B_1 & B_2}^\top  P_\gamma \bmat{B_1 & B_2} + \bmat{-\gamma^2 I \\& R}}^{-1} \bmat{B_1 & B_2}^\top P_\gamma A \\
    \Delta &= (\gamma^2 I - B_1^\top P_\gamma B_1 )^{-1} B_1^\top P_\gamma B_0.
\end{align*}
To verify that the controller is stabilizing, it suffices to note that the cost is finite. Leveraging the fact that $(Q^{1/2}, A)$ is detectable thus tells us that the state must converge to zero in the absence of noise. 

For necessity of the given conditions, first note that if there is no PSD solution to the Riccati equation, then the value does not converge. If the condition $B_1^\top P_\gamma B_1 \prec \gamma^2 I$ is not satisfied, then the adversarial maximization problem is not strictly concave, and it may supply arbitrarily large perturbations. 
\hfill $\blacksquare$

\subsection{Proof of Lemma~\ref{lem:adversarial LQG sol}: Soft-Constrained Output Feedback Solution}
\label{ap: soft constrained output feedback sol}

We leverage the approach developed in \cite{doyle1989state} (see \cite{zhou1996robust} for reference), which was extended to \cite{iglesias1991state} for discrete time systems. In particular, we solve this problem via a combination of the state feedback solution, and the full control solution. To do so, we demonstrate that the optimal solution for output feedback control reduces to estimating the optimal state feedback control action. The resulting problem is known as output estimation, which may be reduced to a full control problem, and solved via dynamic programming. 

\begin{definition}
The soft constrained adversarially robust output feedback, full control, and output estimation problems all attempt to minimize the objective
\[
    \limsup_{T \to \infty} 
    \frac{1}{T} \Ex \brac{\max_{\boldsymbol{\delta} \mbox{ causal}} \sum_{t=0}^{T-1} \norm{z_t}^2 - \gamma^2 \norm{ \delta_t}^2}.
\]
The difference lies in the information available to the controller, and the channel that the control enters. 
\begin{itemize}
    \item The \textbf{adversarially robust output feedback  control problem} may consider a general plant of the form
    \begin{align*}
        \left[\begin{array}{c|ccc}
           A & B_0 & B_1 & B_2 \\ 
           \hline 
           C_1 & D_{10} & D_{11} & D_{21}  \\ 
           C_2 & D_{20} & D_{21} & 0 
        \end{array}\right].
        \end{align*}
    \item The \textbf{adverarially robust output estimation problem} considers a plant of the form
    \begin{align*}
        \left[\begin{array}{c|ccc}
           A & B_0 & B_1 & B_2 \\ 
           \hline 
           C_1 & D_{10} & D_{11} & I  \\ 
           C_1 & D_{20} & D_{21} & 0 
        \end{array}\right].
        \end{align*}
    \item The \textbf{adversarially robust full control problem} assumes complete actuation of both the state and the performance signal:
    \begin{align*}
        \left[\begin{array}{c|ccc}
           A & B_0 & B_1 & \bmat{I & 0} \\ 
           \hline 
           C_1 & D_{10} & D_{11} & \bmat{0 & I} \\ 
           C_2 & D_{20} & D_{21} & 0 
        \end{array}\right].
        \end{align*}
\end{itemize}
\end{definition}

Our first objective will be to reduce the output feedback problem to an output estimation problem. This motivates the following lemma. 
\begin{lemma}
    Suppose that $P_\gamma$ solves
    \begin{align}
        \label{eq: full info riccati}
        P_\gamma &= A^\top P_\gamma A + Q - A^\top P_\gamma \bmat{B_1 & B_2}\paren{\bmat{B_1 & B_2}^\top P_\gamma \bmat{B_1 & B_2} + \bmat{-\gamma^2 I \\ & R}}^{-1} \bmat{B_1 & B_2} P_\gamma A, 
    \end{align}
    and that there exists a nonsingular matrix $\bmat{\hat T_{11} & 0 \\ \hat T_{21} & \hat T_{22}}$ with $\hat T_{11} \in \mathbb{R}^{n_\delta \times n_\delta}$, and $\hat T_{22} \in \mathbb{R}^{n_u \times n_u}$ such that 
    \begin{align}
        \label{eq: decompose psi 2}
        \Psi := \bmat{B_1 & B_2}^\top P_\gamma \bmat{B_1 & B_2} + \bmat{-\gamma^2 I \\& R} = \bmat{\hat T_{11} & 0 \\ \hat T_{21} & \hat T_{22} }^\top \bmat{-I \\ & I} \bmat{\hat T_{11} & 0 \\ \hat T_{21} & \hat T_{22} }.
    \end{align}
    Then for the closed-loop system under any controller $K$, 
    \[
        \limsup_T \frac{1}{T} \Ex_w \brac{\sup_{\delta \textrm{causal}} \sum_{t=0}^{T-1} \norm{z_t}^2  - \gamma^2 \norm{\delta_t}^2 } =  \limsup_T \frac{1}{T} \Ex_w \brac{\sup_{r \textrm{causal}} \sum_{t=0}^{T-1} \norm{\nu_t}^2  - \gamma^2 \norm{r_t}^2 + w_t^\top J_\gamma w_t},
    \]
    where 
     \begin{align*}
        r_t &= \hat T_{11} (\delta_t - E_\gamma x_t - \Delta_\gamma w_t)  \\
        \nu_t &= \hat T_{21} \delta + \hat T_{22} u - \bmat{\hat T_{21} & \hat T_{22} } \paren{\bmat{E_\gamma \\ F_\gamma} x_t + \bmat{ \Delta_\gamma \\ 0} w_t} \\ 
        J_\gamma &= B_0^\top P_\gamma B_1 \paren{\gamma^2 I - B_1^\top P_\gamma B_1}^{-1} B_1^\top P_\gamma B_0 + B_0^\top P_\gamma B_0,
    \end{align*}
    and
    \begin{equation}
    \begin{aligned}
        \label{eq: full info gains}
        \bmat{E_\gamma \\ F_\gamma} &= \Psi^{-1} \bmat{B_1 & B_2}^\top P A, \\
    \Delta_\gamma &= (\gamma^2 I - B_1^\top P B_1 )^{-1} B_1^\top P B_0.
    \end{aligned}   
    \end{equation}
\end{lemma}

\textit{Proof:}
For any $u_t$, $\delta_t$, and $w_t$, 
\begin{align*}
    & \paren{\bmat{\delta_t \\ u_t} - \bmat{E_\gamma \\ F_\gamma} x_t - \bmat{\Delta_\gamma \\ 0} w_t}^\top \Psi \paren{\bmat{\delta_t \\ u_t} - \bmat{E_\gamma \\ F_\gamma} x_t - \bmat{\Delta_\gamma \\ 0} w_t} \\
    &=  \bmat{\delta_t \\ u_t}^\top  \paren{\bmat{0 \\ & R} + \bmat{B_1 & B_2}^\top P_\gamma \bmat{B_1 & B_2}} \bmat{\delta_t \\ u_t} - \gamma^2 \norm{\delta_t}^2 \\
    & - \paren{\bmat{E_\gamma \\ F_\gamma} x_t +\bmat{\Delta_\gamma \\ 0} w_t}^\top \Psi \bmat{\delta_t \\ u_t} - \bmat{\delta_t \\ u_t}^\top \Psi \paren{\bmat{E_\gamma \\ F_\gamma} x_t +\bmat{\Delta_\gamma \\ 0} w_t} + \paren{\bmat{E_\gamma \\ F_\gamma} x_t +\bmat{\Delta_\gamma \\ 0} w_t}^\top \Psi \paren{\bmat{E_\gamma \\ F_\gamma} x_t +\bmat{\Delta_\gamma \\ 0} w_t} \\
    & = \norm{z_t}^2 - \gamma^2 \norm{\delta_t}^2 - x_t^\top \paren{C_1^\top C_1 - \bmat{E_\gamma \\ F_\gamma}^\top \Psi \bmat{E_\gamma \\ F_\gamma}} x_t + \bmat{\delta_t \\ u_t}^\top  \bmat{B_1 & B_2}^\top P_\gamma \bmat{B_1 & B_2} \bmat{\delta_t \\ u_t}^\top \\
    & \qquad + 2\paren{\bmat{B_1 & B_2}^\top P_\gamma A x_t + \bmat{B_1 &  B_1(\gamma^2 I - B_1^\top P_\gamma B_1)^{-1} B_1^\top P_\gamma B_2}^\top P_\gamma B_0 w_t}^\top \bmat{\delta_t \\ u_t} + 2x_t^\top A^\top P_\gamma \bmat{B_1 & B_2} \bmat{\Delta_\gamma \\ 0} w_t \\
    & \qquad + w_t^\top \bmat{\Delta_\gamma \\ 0}^\top \Psi \bmat{\Delta_\gamma \\ 0} w_t \\
    & = \norm{z_t}^2 - \gamma^2 \norm{\delta_t}^2 - x_t^\top P_\gamma x_t + \paren{Ax_t + \bmat{B_1 & B_2} \bmat{\delta_t \\ u_t}}^\top P_\gamma \paren{Ax_t + \bmat{B_1 & B_2} \bmat{\delta_t \\ u_t}}  \\
    &  \qquad + 2\paren{\bmat{B_1 &  B_1(\gamma^2 I - B_1^\top P_\gamma B_1)^{-1} B_1^\top P_\gamma B_2}^\top P_\gamma B_0 w_t}^\top \bmat{\delta_t \\ u_t} + 2x_t^\top A^\top P_\gamma \bmat{B_1 & B_2} \bmat{\Delta_\gamma \\ 0} w_t + w_t^\top \bmat{\Delta_\gamma \\ 0}^\top \Psi \bmat{\Delta_\gamma \\ 0} w_t \\ 
    & = \norm{z_t}^2 - \gamma^2 \norm{\delta_t}^2 - x_t^\top P_\gamma x_t + \paren{x_{t+1}}^\top P_\gamma \paren{x_{t+1}} + 2 u_t^\top B_2^\top \paren{P_\gamma B_1 (\gamma^2 I - B_1^\top P_\gamma B_1)^{-1} B_1^\top P_\gamma - P_\gamma} B_0 w_t \\
    &  \qquad + 2x_t^\top A^\top (P_\gamma B_1 (\gamma^2 I -B_1^\top P_\gamma B_1)^{-1}  B_1^\top P_\gamma - P_\gamma) B_0 w_t - w_t^\top B_0^\top \paren{P_\gamma B_1 (\gamma^2 I - B_1^\top P_\gamma B_1)^{-1} B_1^\top P_\gamma + P_\gamma} B_0  w_t 
    ,
 \end{align*}
where we have made use of \eqref{eq: full info riccati} and \eqref{eq: full info gains}.
Summing over time, maximizing over causal $\delta$, taking the expectation with respect to $w$, and taking the limit yields
\begin{align*}
    & \limsup_T \frac{1}{T} \Ex_{w} \brac{\sup_{\delta \textrm{ causal}} \sum_{t=0}^{T-1} \paren{\bmat{\delta_t \\ u_t} - \bmat{E_\gamma \\ F_\gamma} x_t - \bmat{\Delta_\gamma \\ 0} w_t}^\top \Psi
    \paren{\bmat{\delta_t \\ u_t} - \bmat{E_\gamma \\ F_\gamma} x_t - \bmat{\Delta_\gamma \\ 0} w_t}}\\
    &= \limsup_T \frac{1}{T} \Ex_{w} \brac{ \sup_{\delta \textrm{ causal}} \norm{z}^2 - \gamma^2 \norm{\delta}^2 + w_t^\top J_\gamma w_t} 
\end{align*}
where $J_\gamma = B_0^\top\paren{ P_\gamma B_1 (\gamma^2 I - B_1^\top P_\gamma B_1)^{-1} B_1^\top P_\gamma + P_\gamma}B_0$. Note that we have used the fact that $w_t$ is independent from $x_t$ and $u_t$ due to causality. Using the decomposition of $\Psi$ in \eqref{eq: decompose psi 2}, we find
\begin{align*}
    &\paren{\bmat{\delta_t \\ u_t} - \bmat{E_\gamma \\ F_\gamma} x_t - \bmat{\Delta_\gamma \\ 0} w_t}^\top \Psi
    \paren{\bmat{\delta_t \\ u_t} - \bmat{E_\gamma \\ F_\gamma} x_t - \bmat{\Delta_\gamma \\ 0} w_t} \\
    &= \norm{\hat T_{21} \delta_t + \hat T_{22} u_t  - \bmat{\hat T_{21} & \hat T_{22}}\paren{\bmat{E_\gamma \\ F_\gamma} x_t+\bmat{\Delta_\gamma \\ 0} w_t}}^2 - \norm{\hat T_{11}(\delta_t-E_\gamma x_t - \Delta_\gamma w_t)}^2 \\
    &= \norm{\nu_t}^2 - \norm{r_t}^2.
\end{align*}
\hfill $\blacksquare$

We have thus reduced the adversarially robust partially observed control problem to the adversarially robust partially observed problem for the following system: 
\begin{align*}
    x_{t+1} &= (A+B_1 E_\gamma) x_t + (B_0 + B_1 \Delta_\gamma) w_t + B_1 \hat T_{11}^{-1} r_t + B_2 u_t \\
    \nu_t &= - \hat T_{22} F_\gamma x_t + \hat T_{21} \hat T_{11}^{-1} r_t + \hat T_{22} u_t \\
    y_t &= (C_2 + D_{21} E_\gamma)x_t + (D_{20} +D_{21} \Delta_\gamma) w_t + D_{21} \hat T_{11}^{-1} r_t.
\end{align*}
As $\hat T_{22}$ is invertible, the optimal solution to the above problem is given by $\hat T_{22}^{-1} K$, where $K$ solves the output estimation problem
\[
     \inf_K \limsup_T \frac{1}{T} \Ex_w \brac{\sup_{r \textrm{ causal}} \sum_{t=0}^{T-1} \norm{\nu_t}^2  - \norm{r_t}^2}
\]
under the system
\begin{align*}
    x_{t+1} &= (A+B_1 E_\gamma) x_t + (B_0 + B_1 \Delta_\gamma) w_t + B_1 \hat T_{11}^{-1} r_t + B_2 \hat T_{22}^{-1} u_t \\
    \nu_t &= - \hat T_{22} F_\gamma x_t + \hat T_{21} \hat T_{11}^{-1} r_t + u_t \\
    y_t &= (C_2 + D_{21} E_\gamma)x_t + (D_{20} +D_{21} \Delta_\gamma) w_t + D_{21} \hat T_{11}^{-1} r_t.
\end{align*}
We therefore recall the following change of variables:
 \begin{align*}
            \left[\begin{array}{c|ccc}
               \hat A & \hat B_0 & \hat B_1 & \hat B_2 \\ 
               \hline 
               \hat C_1 & 0 & \hat D_{11} & I  \\ 
               \hat C_2 & \hat D_{20} & \hat D_{21} & 0 
            \end{array}\right] = \left[\begin{array}{c|ccc}
               A + B_1 E_\gamma & B_0 + B_1 \Delta_\gamma & \hat  B_1 \hat T_{11}^{-1} &  B_2 \hat T_{22}^{-1} \\ 
               \hline 
               -\hat T_{22} F_\gamma & 0 & \hat T_{21} \hat T_{11}^{-1} & I  \\ 
               C_2 + D_{21} E_\gamma & D_{20} + D_{21} \Delta_\gamma & \hat D_{21} \hat T_{11}^{-1} & 0 
            \end{array}\right].
        \end{align*}
Note that by the fact that the controller gain and adversarial pertubation are derived from Lemma~\ref{lem:adversarial LQR sol}, the matrix $\hat A - \hat B_2 \hat C_1$ is stable. Then the following lemma allows us to reduce the above output estimation problem to a full control problem, i.e. a problem where the input directly enters the state and the performance signal. 

\begin{lemma}
If $\hat A-\hat B_2\hat C_1$ is stable, then the adversarially robust output estimation problem defined by the system
\begin{align*}
    G_{AOE} = 
    \left[\begin{array}{c|ccc}
       \hat A & \hat B_0 & \hat B_1 & \hat B_2 \\ 
       \hline 
       \hat C_1 & 0 & \hat D_{11} & I  \\ 
       \hat C_2 & \hat D_{20} & \hat D_{21} & 0 
    \end{array}\right]
\end{align*}
may be reduced to the adversarially robust full control problem for a system 
\begin{align*}
    G_{AFC} =
    \left[ \begin{array}{c|ccc}
    \hat A  & \hat B_0 & \hat B_1 & \bmat{I & 0} \\ \hline \hat C_1  & 0 & \hat D_{11} & \bmat{0 & I} \\ \hat C_2 & \hat D_{20} & \hat D_{21} & 0
    \end{array}\right].
\end{align*}

In particular, suppose that 
$K_{AFC}$ solves the adversarially robust full control problem. Then $K_{AOE}$ defined by $\calF_l(P_{AOE}, K_{AFC})$ with 
\begin{align*}
    P_{AOE} = 
    \left[\begin{array}{c| cc}    
    \hat A-\hat B_2 \hat C_1 & 0 & \bmat{I & -B_2} \\
    \hline \hat C_1  & 0 & \bmat{0 & I} \\
    \hat C_2 &  I & 0
    \end{array}\right]
\end{align*}
solves the adversarially robust output estimation problem. Similarly, if $K_{AOE}$ solves the adversarially robust output estimation problem, then $K_{AFC}$ defined by $\bmat{\hat B_2 \\ I} K_{AOE}$ solves the adversarially robust full control problem.  
\end{lemma}

\textit{Proof:} The result of Theorem 12.6 of \cite{zhou1998robust} tells us that if $K_{AOE}$ internally stabilizes $G_{AOE}$, then $K_{AFC}=\bmat{\hat B_2 \\ I}K_{AOE}$ internally stabilizes $G_{AOE}$, and $\calF_{l}(G_{AFC}, K_{AFC}) = \calF_{l}(G_{AOE}, K_{AOE})$. Similarly, if $K_{AFC}$ internally stabilizes $G_{AFC}$, then $K_{AOE} = \calF_l(P_{AOE}, K_{AFC})$ internally stabilzes $G_{AFC}$, and $\calF_{l}(G_{AFC}, K_{AFC}) = \calF_{l}(G_{AOE}, K_{AOE})$. As the transfer functions are equal, a solution to either the adversarially robust output estimation problem or the adversarially robust full control problem provides the solution to the other. \hfill $\blacksquare$

\begin{lemma}
 Consider the adversarially robust full control problem defined by 
    \begin{align*}
        G_{AFC} = 
        \left[\begin{array}{c|ccc} \hat A  & \hat B_0 & \hat B_1 & \bmat{I & 0} \\ \hline \hat C_1  & 0 & \hat D_{11} & \bmat{0 & I} \\ \hat C_2  & \hat D_{20} & \hat D_{21} & 0
        \end{array}\right].
    \end{align*}
    Suppose there exists real matrices $L_\gamma, N_\gamma, \Sigma_\gamma, Y_\gamma$ with $\Sigma_\gamma$ and $Y_\gamma$ symmetric positive definite that satisfy
        \begin{equation}
        \label{eq: full control conditions appendix}
        \begin{aligned}
            \tilde A &= \hat A+L_\gamma \hat C_2 \\
            \tilde B_0 &= \hat B_0 + L_\gamma \hat D_{20} \\
            \tilde B_1 &= \hat B_1 + L_\gamma\hat D_{21} \\
            \tilde C_1 &= \hat C_1 + N_\gamma \hat C_2 \\
            \tilde D_{11} &= \hat D_{11} + N_\gamma \hat D_{21} \\
            \tilde D_{10} &= N_\gamma \hat D_{20} \\
            \Phi &=  I - \tilde D_{11}^\top \tilde D_{11} - \tilde B_1^\top Y_\gamma \tilde B_1 \succ 0 \\
            Y_\gamma &= \tilde C_1^\top \tilde C_1 + \tilde A^\top Y_\gamma \tilde A + \paren{\tilde D_{11}^\top \tilde C_1 + \tilde B_1^\top Y_\gamma \tilde A}^\top \Phi^{-1} \paren{\tilde D_{11}^\top \tilde C_1 + \tilde B_1^\top Y_\gamma \tilde A} \\
            \Sigma_\gamma &= \brac{(I+\tilde B_1 \Phi^{-1} \tilde B_1^\top Y_\gamma) \tilde A + \tilde B_1 \Phi^{-1} \tilde D_{11}^\top \tilde C_1} \Sigma_\gamma \brac{(I+\tilde B_1 \Phi^{-1} \tilde B_1^\top Y_\gamma) \tilde A + \tilde B_1 \Phi^{-1} \tilde D_{11}^\top \tilde C_1}^\top \\
            &\qquad + \brac{(I+\tilde B_1 \Phi^{-1} \tilde B_1^\top Y_\gamma) \tilde B_0 + \tilde B_1^\top \Phi^{-1} \tilde D_{11}^\top \tilde D_{10} } \brac{(I+\tilde B_1 \Phi^{-1} \tilde B_1^\top Y_\gamma) \tilde B_0 + \tilde B_1^\top \Phi^{-1} \tilde D_{11}^\top \tilde D_{10} }^\top \\
            L_\gamma, N_\gamma &\in \argmin_{L, N}  \trace\paren{\tilde D_{10}^\top \tilde D_{10} +\tilde B_0^\top Y_\gamma \tilde B_0 +(\tilde D_{10}^\top \tilde D_{11} + \tilde B_0^\top Y_\gamma \tilde B_1)  \Phi^{-1} (\tilde D_{10}^\top \tilde D_{11} + \tilde B_0^\top Y_\gamma \tilde B_1)^\top } \\ 
        &\qquad\qquad + \trace\paren{\Sigma_\gamma \brac{\tilde C_1^\top \tilde C_1 + \tilde A^\top Y_\gamma \tilde A + (\tilde D_{11}^\top \tilde C_1 + \tilde B_1^\top Y_\gamma \tilde A )^\top\Phi^{-1}(\tilde D_{11}^\top \tilde C_1 + \tilde B_1^\top Y_\gamma \tilde A )  } },
        \end{aligned}
        \end{equation}
        where $Y_\gamma$ and $\Sigma_\gamma$ are treated as fixed parameters in the minimization problem in the final lines. 
    Then an optimal solution to the adversarially robust full control problem is given by the static controller with gain $\bmat{L_\gamma \\ N_\gamma}$.
\end{lemma}

\textit{Proof:}
We will first solve the finite horizon version of the problem under the assumption that the optimal controller consists of time varying static gains of the form $\bmat{L_t \\ N_t}$. The finite horizon $\gamma$-adversarially robust full control problem under this assumption may be written as 
\begin{align*}
    J_\star = \minimize_{L_0, N_0, \dots, L_{T-1}, N_{T-1}} \Ex_{x_0, w_{0:T-1}} \brac{\sup_{r \textrm{ causal}} \sum_{t=0}^{T-1} \norm{\hat C_1 x_t + \hat D_{11} r_t + N_t y_t}^2 - \norm{r_t}^2} 
\end{align*}
where $x_0 \sim \calN(0, \Sigma_0)$, and $x_t$ progresses according to dynamics
\begin{align*}
    x_{t+1} = (\hat A + L_t \hat C_2) x_t + (\hat B_0 + L_t \hat D_{20}) w_t + (\hat B_1 + L_t \hat D_{11}) r_t.
\end{align*}
We will see that under the optimal adversarial pertubation and filter gain, the state is distributed as $x_t \sim \calN(0, \Sigma_t)$ for $t=0, \dots, T$. We will define the cost to go in terms of the variance of the state. In particular, define
\begin{align*}
    V_{\tau}(\Sigma_\tau) = \minimize_{L_\tau, N_\tau \dots, L_{T-1}, N_{T-1}} \Ex_{x_\tau, w_{\tau:T-1}} \brac{\sup_{r \textrm{ causal}} \sum_{t=\tau}^{T-1} \norm{\hat C_1 x_t + \hat D_{11} r_t + N_t y_t}^2 - \norm{r_t}^2} 
\end{align*}
so that $J_\star = V_0(\Sigma_0)$. To derive the optimal gains $L_t$ and $N_t$, we will show that there exists some $Y_\tau \succeq 0$ such that  $V_\tau(\Sigma_\tau) = \trace(\Sigma_\tau Y_\tau) + q_\tau$ for all $\tau$. The base case is clear $V_T(\Sigma_T) = \trace(\Sigma_T Y_T)$ where $Y_T = 0$. Assume that this is true at time $\tau+1$, i.e. assume that $V_{\tau+1}= \trace(\Sigma_{\tau+1} Y_{\tau+1}) + q_{\tau+1}$. Then we have that 
\begin{align*}
    V_\tau(\Sigma_\tau) &= \minimize_{L_\tau, N_\tau,  \dots, L_{T-1}, N_{T-1}} \Ex_{x_\tau, w_{\tau:T-1}} \brac{\max_{r_{\tau:T-1} \textrm{causal}} \sum_{t=\tau}^{T-1} \norm{\hat C_1 x_t + \hat D_{11} r_t + N_t y_t}^2 - \norm{r_t}^2} \\
    &= \minimize_{L_\tau, N_{\tau}} \Ex_{x_\tau, w_\tau} \brac{\sup_{r_\tau} \norm{\hat C_1 x_\tau + \hat D_{11} r_\tau +N_\tau y_\tau}^2 - \norm{r_\tau}^2 + V_{\tau+1}(\Sigma_{\tau+1})} \\
    &=  \minimize_{L_\tau, N_\tau} \Ex_{x_\tau, w_\tau} \brac{\sup_{r_\tau} \norm{\hat C_1 x_\tau + \hat D_{11} r_\tau + N_t y_t}^2 -  \norm{r_\tau}^2 + x_{\tau+1} Y_{\tau+1} x_{\tau+1} + q_{\tau+1}} \\
    &=  \minimize_{L_\tau, N_\tau} \Ex_{x_\tau, w_\tau} \bigg[\sup_{r_\tau} \norm{(\hat C_1 + N_\tau \hat C_2) x_\tau + (\hat D_{11} + N_\tau \hat D_{21}) r_\tau + N_\tau \hat D_{20} w_\tau}^2 - \norm{r_\tau}^2 + ((\hat A + L_\tau \hat C_2) x_\tau  \\
    &\qquad\qquad\qquad + (\hat B_0 + L_\tau \hat D_{20}) w_\tau+ (\hat B_1 + L_\tau \hat D_{21}) r_\tau) Y_{\tau+1} (\cdot)  + q_{\tau+1} \bigg] 
\end{align*}
Maximizing over $r_\tau$ provides
\begin{align*}
    \tilde A &= \hat A+L_\tau \hat C_2 \\
    \tilde B_0 &= \hat B_0 + L_\tau \hat  D_{20} \\
    \tilde B_1 &= \hat B_1 + L_\tau \hat D_{21} \\
    \tilde C_1 &= \hat C_1 + N_\tau \hat C_2 \\
    \tilde D_{11} &= \hat  D_{11} + N_\tau \hat D_{21} \\
    \tilde D_{10} &= N_\tau D_{20} \\
    \Phi &=  I - \tilde D_{11}^\top \tilde D_{11} - \tilde B_1^\top Y_{\tau+1} \tilde B_1 \\
    \phi_\tau &= \tilde A x_\tau  + \tilde B_0 w_\tau \\
    r_\tau &= \Phi^{-1} (\tilde D_{11}^\top (\tilde C_1 x_\tau + \tilde D_{10} w_\tau)+ \tilde B_1^\top Y_{\tau+1} \phi_\tau),
\end{align*}
where we have assumed that $\Phi$ is PD for all $\tau$. 
Substituting the above quantities back into the objective, the quantity inside of the expectation simplifies to 
\begin{align*}
    ( &\tilde C_1 x_t + \tilde D_{10} w_t)^\top (\tilde C_1 x_t + \tilde D_{10} w_t) + \phi_t^\top Y_{\tau+1}\phi_t  \\
    & + (\tilde D_{1,1}^\top \tilde  C_1 x_t + \tilde D_{11}^\top \tilde D_{10} w_\tau +  \tilde B_1^\top Y_{\tau+1} \phi_\tau)^\top \Phi^{-1} (\tilde D_{1,1}^\top \tilde C_1 x_t + \tilde D_{11}^\top \tilde D_{10} +  \tilde B_1^\top Y_{\tau+1} \phi_\tau) + q_{\tau+1}.
\end{align*}
Substituting in $\phi_\tau = \tilde A x_\tau + \tilde B_0 w_\tau$, we may evaluate the expectation of the above quantity, recalling that $x_\tau$ and $w_\tau$ are independent, and $w_\tau$ is zero mean. 
\begin{align*}
    \Ex \bigg[
        ( &\tilde C_1 x_t + \tilde D_{10} w_t)^\top (\tilde C_1 x_t + \tilde D_{10} w_t) + \phi_t^\top Y_{\tau+1}\phi_t  \\
    & + (\tilde D_{1,1}^\top \tilde  C_1 x_t + \tilde D_{11}^\top \tilde D_{10} w_\tau +  \tilde B_1^\top Y_{\tau+1} \phi_\tau)^\top \Phi^{-1} (\tilde D_{1,1}^\top \tilde C_1 x_t + \tilde D_{11}^\top \tilde D_{10} +  \tilde B_1^\top Y_{\tau+1} \phi_\tau) + q_{\tau+1}
    \bigg] \\
    &= q_{\tau+1} + \trace\paren{ \Sigma_w \brac{ \tilde D_{20}^\top \tilde D_{20}  +\tilde B_0^\top Y_{\tau+1} \tilde B_0 +(\tilde D_{10}^\top \tilde D_{11} + \tilde B_0^\top Y_{\tau+1} \tilde B_1)  \Phi^{-1} (\tilde D_{10}^\top \tilde D_{11} + \tilde B_0^\top Y_{\tau+1} \tilde B_1)^\top }} \\ 
    &\qquad\qquad + \trace\paren{\Sigma_\tau \brac{\tilde C_1^\top \tilde C_1 + \tilde A^\top Y_{\tau+1} \tilde A + (\tilde D_{1,1}^\top \tilde C_1 + \tilde B_1^\top Y_{\tau+1} \tilde A )^\top\Phi^{-1}( \tilde D_{1,1}^\top \tilde C_1 + \tilde B_1^\top Y_{\tau+1} \tilde A )  } }.
\end{align*}
Thus the optimal filter gain at time $\tau$ is given by the solution to  
\begin{align*}
    \min_{L_\tau, N_\tau} & \trace\paren{ \Sigma_w \brac{\tilde D_{20}^\top \tilde D_{20} +\tilde B_0^\top Y_{\tau+1} \tilde B_0 +(\tilde D_{10}^\top \tilde D_{11} + \tilde B_0^\top Y_{\tau+1} \tilde B_1)  \Phi^{-1} (\tilde D_{10}^\top \tilde D_{11} + \tilde B_0^\top Y_{\tau+1} \tilde B_1)^\top }} \\ 
    &\qquad\qquad + \trace\paren{\Sigma_\tau \brac{\tilde C_1^\top \tilde C_1 + \tilde A^\top Y_{\tau+1} \tilde A + (\tilde D_{1,1}^\top \tilde C_1 + \tilde B_1^\top Y_{\tau+1} \tilde A )^\top\Phi^{-1}(\tilde D_{1,1}^\top \tilde C_1 + \tilde B_1^\top Y_{\tau+1} \tilde A )  } }.
\end{align*}
Note that the minimization problem is convex, as it is the result of a pointwise supremum. Therefore, it can be solved efficiently using standard solvers.

Under a gains which satisfies this, the cost function progresses according to
\begin{align*}
    Y_{\tau} &= \tilde C_1^\top  \tilde C_1 + \tilde A^\top Y_{\tau+1} \tilde A + (\tilde D_{1,1}^\top \tilde C_1 + \tilde B_1^\top Y_{\tau+1} \tilde A )^\top\Phi^{-1}(\tilde D_{1,1}^\top \tilde C_1 + \tilde B_1^\top Y_{\tau+1} \tilde A ) ).
\end{align*}
This gives us a backwards recursion to find the cost matrices. Under these adversarial perturbations and controller gains, the state dynamics progress such that $x_{\tau} \sim \calN(0, \Sigma_\tau)$, where $\Sigma_\tau$ evolves according to 
\begin{align*}
    \Sigma_{\tau+1} &= \brac{ (I + \tilde B_1 \Phi^{-1} \tilde B_1^\top Y_{t+1}) \tilde A + \tilde B_1 \Phi^{-1} \tilde D_{11}^\top \tilde C_1} \Sigma_{\tau} \brac{ (I + \tilde B_1 \Phi^{-1} \tilde B_1^\top Y_{\tau+1}) \tilde A + \tilde B_1 \Phi^{-1} \tilde D_{11}^\top \tilde C_1}^\top  \\
    &\qquad \qquad + \paren{(I+ \tilde B_1 \Phi^{-1} \tilde B_1^\top Y_{\tau+1}) \tilde B_0 + \tilde B_1 \Phi^{-1} \tilde D_{11}^\top \tilde D_{10}}\Sigma_w  \paren{(I+ \tilde B_1 \Phi^{-1} \tilde B_1^\top Y_{\tau+1}) \tilde B_0 + \tilde B_1 \Phi^{-1} \tilde D_{11}^\top \tilde D_{10}}^\top.
\end{align*}
Passing the same argument to an infinite horizon setting, we find that our gains must satisfy the conditions given in the statement.

To verify that the restriction of the parameter search space to static matrices of the form $\bmat{L_\gamma \\ N_\gamma}$ is sufficient, observe the following fact, as in \cite{doyle1989optimal}: 
\begin{align*}
   \inf_K \limsup_T & \frac{1}{T} \Ex_{w} \brac{ \sum_{t=0}^{T-1} \norm{\nu_t}^2 - \norm{r_t}^2 \vert r \textrm { given}}\\ 
   &\leq \inf_K \limsup_T \frac{1}{T} \Ex_{w} \brac{ \sup_{r \textrm{ causal}}  \sum_{t=0}^{T-1} \norm{\nu_t}^2 - \norm{r_t}^2} \\
    &\leq \limsup_T \frac{1}{T} \Ex_{w} \brac{ \sup_{r \textrm{ causal}}  \sum_{t=0}^{T-1} \norm{\nu_t}^2 - \norm{r_t}^2 \vert K \textrm{ given}}.
\end{align*}
With $K$ given by $\bmat{L_\gamma \\ N_\gamma}$, where $L_\gamma$ and $N_\gamma$ satsify \eqref{eq: full control conditions appendix}, the right hand side is equal to the left hand side, where $r$ is given as $r_t = \Phi^{-1} (\tilde D_{11}^\top \tilde C_1 x_t + \tilde B_1^\top Y_\gamma (\tilde A x_t + \tilde B_0  w_t))$. To see that this is the case, note that under this adversarial perturbation, the problem on the left reduces to a Kalman Filtering problem, where the optimal gain is static. \hfill $\blacksquare$ 

The proof of \Cref{lem:adversarial LQG sol} hinges on sequential integration of the results presented in this section. 

\subsection{Proof of Theorem~\ref{thm:adversarial controller solution}: Hard Constrained Output Feedback Solution}
\label{appendix: advlqr thm proof}

The following lemma will be useful. 

\begin{lemma}
    \label{lem: linsys ergodic}
    Consider a stable system driven by noise 
    \[
        x_{t+1} = Ax_t + B w_t,
    \]  
    where $x_0 \sim \calN(0, I)$ and $w_t \overset{(iid)}{\sim} \calN(0,\Sigma_w)$. Then for any quadratic function $f(x_t) = x_t^\top V x_t + x_t^\top b + c$, 
    \[
        \limsup_T \frac{1}{T} \sum_{t=0}^{T-1} f(x_t) = \limsup_t \mathbb{E}\brac{f(x_t)}.
    \]
\end{lemma}
\textit{Proof:}
Let $\tilde x_0 \sim \calN(0,\Sigma_\infty)$,
where 
\[
    \Sigma_\infty = \limsup_t \Ex\brac{x_t x_t^\top} = \dlyap(A, BB^\top). 
\]
Also let $\tilde x_t$ follow the same dynamics of $x_t$ rolled out under the same noise realization. We have that $\tilde x_t$ is ergodic. Then we may express
\begin{align*}
    \limsup_T&\frac{1}{T} \sum_{t=0}^{T-1} f(x_t) =  \limsup_T \frac{1}{T} \sum_{t=0}^{T-1} f(\tilde x_t + (x_t - \tilde x_t))\\
    &= \limsup_T \frac{1}{T} \sum_{t=0}^{T-1} f(\tilde x_t + (x_t - \tilde x_t)) \\
    &= \limsup_T \frac{1}{T} \sum_{t=0}^{T-1} f(\tilde x_t ) + (x_t - \tilde x_t)^\top V (x_t - \tilde x_t) + 2 (x_t - \tilde x_t)^\top V \tilde x_t + (x_t - \tilde x_t)^\top b \\
     &= \limsup_t \Ex\brac{f(\tilde x_t)} + \limsup_T \sum_{t=0}^{T-1} \paren{A^t(x_0 - \tilde x_0)}^\top V A^t (x_0 - \tilde x_0) + 2 \paren{A^t (x_ 0- \tilde x_0)}^\top V \tilde x_t + \paren{A^t (x_0 - \tilde x_0)}^\top b \\
      &= \limsup_t \Ex\brac{f( x_t)},
\end{align*}
as $\rho(A) < 1$ implies that the the quantity in the sum may be bounded by a constant. 
\hfill $\blacksquare$

\begin{lemma} (Ergodicity)
\label{lem:ergodicity}
Let $K$  be any LTI controller that internally stabilizes the system in the face of the optimal $\gamma$-soft-constrained adversarial perturbations. Then 
\begin{align*}
    \limsup_{T \to \infty} \frac{1}{T} \Ex \brac{\max_{\delta \mathrm{ causal}} \sum_{t=0}^{T-1} \norm{z_{t}}^2 - \gamma^2 \norm{\delta_t}^2} =  \limsup_{T \to \infty} \frac{1}{T} \max_{\delta \mathrm{ causal}} \sum_{t=0}^{T-1} \norm{z_{t}}^2 - \gamma^2 \norm{\delta_t}^2.
\end{align*}

\end{lemma}
\textit{Proof:}
To see that this is so, observe that the optimal adversarial perturbation may be expressed
\begin{align*}
    \delta_t = \Delta_x \bmat{x_t \\ x_t^{(K)}} + \Delta_w w_t,
\end{align*}
for some $\Delta_x$ and $\Delta_w$ defined in terms of the closed loop system and $\gamma$. Here $x_t^{(K)}$ denotes the state of the controller. Under this perturbation, the dynamics of the closed loop system becomes
\begin{align*}
    \bmat{x_{t+1} \\ x_{t+1}^{(K)}} &= A_{cl} \bmat{ x_t  \\ x_t^{(K)}} + B_{cl} w_t, \\
    z_t &= C_{cl} \bmat{ x_t  \\ x_t^{(K)}} + D_{cl} w_t,
\end{align*}
where $A_{cl}$ has eigenvalues in the open unit disc, by the fact that $K$ is internally stabilizing.  
 Then by \Cref{lem: linsys ergodic},
\begin{align*}
    &\limsup_{T \to \infty}  \max_{\delta \mathrm{ causal}} \sum_{t=0}^{T-1} \norm{z_{t}}^2 - \gamma^2 \norm{\delta_t}^2 \\
    =&\limsup_{T \to \infty}
    \frac{1}{T}  \sum_{t=0}^{T-1} \norm{C_{cl} \bmat{ x_t \\ x_t^{(K)}} + D_{cl} w_t }^2 - \gamma^2 \norm{\Delta_x \bmat{x_t \\ x_t^{(K)}} + \Delta_w w_t}^2 \\ 
    = &\limsup_{t \to \infty} \Ex\brac{\norm{C_{cl} \bmat{ x_t \\ x_t^{(K)}} + D_{cl} w_t }^2 - \gamma^2 \norm{\Delta_x \bmat{x_t \\ x_t^{(K)}} + \Delta_w w_t}^2 } \\
    = &\limsup_{t \to \infty} \frac{1}{T} \Ex\brac{ \sum_{t=0}^{T-1} \norm{C_{cl} \bmat{ x_t \\ x_t^{(K)}} + D_{cl} w_t }^2 - \gamma^2 \norm{\Delta_x \bmat{x_t \\ x_t^{(K)}} + \Delta_w w_t}^2 } \\
    &=\limsup_{T \to \infty} \frac{1}{T} \Ex \brac{\max_{\delta \mathrm{ causal}} \sum_{t=0}^{T-1} \norm{z_{t}}^2 - \gamma^2 \norm{\delta_t}^2}.
\end{align*}
\hfill $\blacksquare$


Now we are ready to prove Theorem~\ref{thm:adversarial controller solution}. 

\textit{Proof:}
First note that optimization problem
\[
     \max_{\substack{\mathbf{\delta} \mbox{ causal} \\ \norm{\mathbf{\delta}}_{\ell_2}^2 \leq T \varepsilon}}  \sum_{t=0}^{T-1} \norm{z_t}^2
\]
is the maximization of a quadratic function subject to a quadratic constraint, and therefore satisfies strong duality (see \cite{boyd2004convex}, Appendix B.1). Therefore, maximizing subject to the quadratic constraint is equivalent to maximizing the Lagrangian, then minimizing over the dual variable. 
\begin{align*}
    \max_{\substack{\mathbf{\delta} \mbox{ causal} \\ \norm{\mathbf{\delta}}_{\ell_2}^2 \leq T \varepsilon}}  \sum_{t=0}^{T-1} \norm{z_t}^2 = \min_{\gamma \geq \gamma_{LB}}
    \gamma^2 \varepsilon T + \max_{\mathbf{\delta} \mbox{ causal}} \sum_{t=0}^{T-1} \norm{z_t}^2 - \gamma^2 \norm{\delta_t}^2.
\end{align*}
Then the minimization of objective \eqref{eq:hardadversarialLQRobjective} is equivalent to 
\begin{align*}
    \minimize_{u} \limsup_T \frac{1}{T} \Ex \brac{\min_{\gamma \geq \gamma_{LB}}  \gamma^2 \varepsilon T + \max_{\mathbf{\delta} \mbox{ causal}} \sum_{t=0}^{T-1} \norm{z_t}^2 - \gamma^2 \norm{\delta_t}^2}.
\end{align*}
Our next step is to show that we may pull the minimization over $\gamma$ outside of the expectation. To see that this is the case, we will make use of ergodicity. 

By Assumption~\ref{as: system assumptions}, we have that 
 \begin{align*}
    \infty &> \min_{\gamma \geq \gamma_{LB}} \gamma^2 \varepsilon T + \limsup_{T \to \infty} \min_{u_0, \dots, u_{T-1}}
    \frac{1}{T} \Ex \brac{\max_{\delta \mbox{ causal}} \sum_{t=0}^{T-1} \norm{z_t}^2 - \gamma^2 \norm{\delta_t}^2}. 
\end{align*}
Now, for each $\gamma \geq \gamma_{LB}$ let $K_\gamma$ be the controller defined in Lemma~\ref{lem:adversarial LQG sol} and let $u_t = K_\gamma (y_{-\infty:t})$. Then
\begin{align*}
    & \quad \quad \, \, \,\min_{\gamma \geq \gamma_{LB}} \gamma^2 \varepsilon T + \limsup_{T \to \infty}  
    \frac{1}{T} \Ex \brac{\max_{\delta \mbox{ causal}} \sum_{t=0}^{T-1} \norm{z_t}^2 - \gamma^2 \norm{\delta_t}^2} \\
    &=\Ex \brac{\min_{\gamma \geq \gamma_{LB}} \gamma^2 \varepsilon T + \limsup_{T \to \infty}
    \frac{1}{T} \Ex \brac{\max_{\delta \mbox{ causal}} \sum_{t=0}^{T-1} \norm{z_t}^2 - \gamma^2 \norm{\delta_t}^2}} \\
    &= \Ex \brac{ \min_{\gamma \geq \gamma_{LB}} \gamma^2 \varepsilon T + \limsup_{T \to \infty} 
    \frac{1}{T} \max_{\delta \mbox{ causal}} \sum_{t=0}^{T-1} \norm{z_t}^2 - \gamma^2 \norm{\delta_t}^2} \\
    &\geq \limsup_{T \to \infty} \frac{1}{T} \Ex \brac{ 
     \min_{\gamma \geq \gamma_{LB}} \gamma^2 \varepsilon T + \max_{\delta \mbox{ causal}} \sum_{t=0}^{T-1} \norm{z_t}^2 - \gamma^2 \norm{\delta_t}^2}.
\end{align*}
Where the first equality is permitted by the fact that first line is not random. The second line follows from ergodicity (Lemma \ref{lem:ergodicity}). The inequality is a result of Fatou's Lemma and the $\min$-$\max$ inequality. We can similarly obtain an upper bound:
\begin{align*}
    & \quad \quad \, \, \,\min_{\gamma \geq \gamma_{LB}} \gamma^2 \varepsilon T + \liminf_{T \to \infty}  
    \frac{1}{T} \Ex \brac{\max_{\delta \mbox{ causal}} \sum_{t=0}^{T-1} \norm{z_t}^2 - \gamma^2 \norm{\delta_t}^2} \\
    &=\Ex \brac{\min_{\gamma \geq \gamma_{LB}} \gamma^2 \varepsilon T + \liminf_{T \to \infty}
    \frac{1}{T} \Ex \brac{\max_{\delta \mbox{ causal}} \sum_{t=0}^{T-1} \norm{z_t}^2 - \gamma^2 \norm{\delta_t}^2}} \\
    &= \Ex \brac{ \min_{\gamma \geq \gamma_{LB}} \gamma^2 \varepsilon T + \liminf_{T \to \infty} 
    \frac{1}{T} \max_{\delta \mbox{ causal}} \sum_{t=0}^{T-1} \norm{z_t}^2 - \gamma^2 \norm{\delta_t}^2} \\
    &\leq \liminf_{T \to \infty} \frac{1}{T} \Ex \brac{ 
     \min_{\gamma \geq \gamma_{LB}} \gamma^2 \varepsilon T + \max_{\delta \mbox{ causal}} \sum_{t=0}^{T-1} \norm{z_t}^2 - \gamma^2 \norm{\delta_t}^2}.
\end{align*}
Combining these results, we see that
\begin{align*}
    \limsup_{T \to \infty} \frac{1}{T} \Ex \brac{ 
     \min_{\gamma \geq \gamma_{LB}} \gamma^2 \varepsilon T + \max_{\delta \mbox{ causal}} \sum_{t=0}^{T-1} \norm{z_t}^2 - \gamma^2 \delta_t^\top \delta_t} \\
     = \min_{\gamma \geq \gamma_{LB}} \gamma^2 \varepsilon + \limsup_{T \to \infty}  
    \frac{1}{T} \Ex \brac{\max_{\delta \mbox{ causal}}  \sum_{t=0}^{T-1} \norm{z_t}^2 - \gamma^2 \delta_t^\top \delta_t}.
\end{align*}
Therefore, we see that the optimal solution is given by Lemma~\ref{lem:adversarial LQG sol} for some properly chosen $\gamma$. Now, observe that i) the power of the optimal adversary for the problem at some level $\gamma$, $\lim_t \Ex\brac{\paren{\delta_t^{\gamma}}^\top \delta_t^\gamma}$ is decreasing as $\gamma$ increases, and ii) the power of the optimal adversary for the hard constrained problem \eqref{eq:adversarialLQRobjective} will equal $\varepsilon$. Then bisecting such that the adversarial power of the solution to Lemma~\ref{lem:adversarial LQG sol} is correct provides the proper level $\gamma$. 

\Cref{alg:advControl} relies on the adversarial power computed in \Cref{alg:advPower}. To verify correctness of this algorithm we observe that for a closed-loop system $G_{cl} = \left[\begin{array}{c|cc}
       A_{cl} & B_{0, cl} & B_{1,cl}  \\ 
       \hline 
       C_{cl} & D_{0, cl} & D_{1, cl} 
    \end{array}\right]$, the optimal adversary found using dynamic programming arguments as in the proofs of \Cref{lem:adversarial LQR sol} and \Cref{lem:adversarial LQG sol}. In particular, we have $\delta_t^\gamma = \Phi^{-1} \Gamma_x x_t + \Phi^{-1} \Gamma_w w_t$, where $\Phi$, $\Gamma_x$ and $\Gamma_w$ are as in \Cref{alg:advPower}. Then
    \begin{align*}
        \lim_t \E \brac{\delta_t^\gamma \paren{\delta_t^\gamma}^\top} = \trace(\Gamma_w^\top \Phi^{-2} \Gamma_w) + \trace(\Gamma_x^\top \Phi^{-2} \Gamma_x \Sigma_\gamma),
    \end{align*}
    where $\Sigma_\gamma$ is the covariance of the state under the optimal adversarial perturbation and may be computed as the solution to the lyapunov equation appearing in \Cref{alg:advPower}.

\hfill $\blacksquare$ 

\section{Proofs for Section~\ref{s: state feedback performance robustness tradeoffs}}

We introduce the following recurring fact: given a solution $P_\gamma$ to the DARE in Lemma~\ref{lem:adversarial LQR sol}, we have
\begin{align} \label{eq: x0'Px0}
    x_0^\top P_\gamma x_0 &= \limsup_{T \to \infty}  
    \frac{1}{T} \paren{\max_{\delta \mbox{ causal}} x_T^\top Q x_T + \sum_{t=0}^{T-1} x_t \paren{Q + F_{\gamma}^\top R F_{\gamma}} x_t - \gamma^2 \delta_t^\top \delta_t }.
\end{align}
In other words, $x_0^\top P_\gamma x_0$ is the soft-constrained cost of the optimal adversarially robust controller against the optimal adversary at level $\gamma$ in the noiseless setting, with initial condition $x_0$.

\subsection{Proof of Lemma~\ref{lem: controller gap to DARE gap}}
The following lemma from \cite{mania2019certainty} will be helpful. 
\begin{lemma}(Lemma 1 in \cite{mania2019certainty})
    \label{lem: upper bounding gap between optimizers}
    Let $f_1$ and $f_2$ be $\mu$-strongly convex twice differentiable functions functions. Let $x_1 = \argmin_x f_1(x)$ and $x_2 = \argmin_x f_2(x)$. Suppose $\norm{\nabla f_1(x_2)} \leq \varepsilon$. Then $\norm{x_1 - x_2} \leq \frac{\varepsilon}{\mu}$. 
\end{lemma}

We also consider an analogous lemma which provides a lower bound.
\begin{lemma}
    \label{lem: lower bounding gap between optimizers}
    Suppose $f_1$ and $f_2$ are $L$-Lipschitz function, and let $x_1$ minimize $f_1$ and $x_2$ minimize $f_2$. If $\norm{\nabla f_1(x_2)} \geq \varepsilon$, then $\norm{x_1 - x_2} \geq \frac{\varepsilon}{L}$.
\end{lemma}

\textit{Proof:}
By a Taylor expansion of $\nabla f_1$ about $x_1$, we see that
\begin{align*}
    \nabla f_1(x_2) = \nabla f_1 (x_1) + \nabla^2 f_1(\tilde x) (x_2 - x_1)
\end{align*}
where $\tilde x = t x_1 + (1-t) x_2$ for some $t \in [0,1]$. As $x_1$ minimizes $f_1$, we must have $\nabla f_1(x_1) = 0$. Futhermore, by the fact that $f_1$ is $L$ Lipschitz, we have that
\begin{align*}
    \norm{\nabla f_1(x_2)} = \norm{\nabla f_1 (x_1) + \nabla^2 f_1(\tilde x) (x_2 - x_1)} = \norm{\nabla^2 f_1(\tilde x) (x_2 - x_1)} \leq L \norm{x_2 - x_1}
\end{align*}
Dividing both sides of the above inequality by $L$ and leveraging the assumption that $\norm{\nabla f_1(x_2)}\leq \varepsilon$ provides the desired result. \hfill $\blacksquare$

Armed with these lemmas, we may prove Lemma~\ref{lem: controller gap to DARE gap}.

\textit{Proof:}
First, note that both functions are strongly convex with parameter $\sigma_{\min}(R + B^\top P B)$, and Lipschitz continuous with parameter $\norm{R+B^\top M B}$. 

For any $x$, by the fact that $\nabla f_2(u_2, x) = 0$, we have \begin{align*}
    \nabla f_1(u_2, x) &= \nabla f_1(u_2, x) - \nabla f_2(u_2, x) \\
    &= (R + B^\top M B) u_2 + B^\top M Ax - (R+B^\top P B) u_2 - B^\top P A x \\
    &= B^\top (M-P) (A+B F_2) x,
\end{align*}
and thus
\begin{align*}
    \norm{\nabla f_1(u_2, x)} = \norm{B^\top (M-P) (A+B F_2) x}.
\end{align*}
Then by application of Lemma~\ref{lem: upper bounding gap between optimizers} and Lemma~\ref{lem: lower bounding gap between optimizers}, we have that for any $x$, $\norm{u_1 - u_2}=\norm{(F_1- F_2)x}$ satisfies
\begin{align*}
    \frac{
    \norm{B^\top (M-P) (A+B F_2) x}}{\norm{R+B^\top M B}} \leq \norm{(F_1-F_2)x} \leq \frac{\norm{B^\top (M-P) (A+B F_2) x}}{\sigma_{\min}(R + B^\top P B)}.
\end{align*}
Taking the supremum over $\norm{x}=1$, we find
\begin{align*}
    \frac{
    \norm{B^\top (M-P) (A+B F_2)}}{\norm{R+B^\top M B}} \leq \norm{F_1-F_2} \leq \frac{\norm{B^\top (M-P) (A+B F_2) }}{\sigma_{\min}(R + B^\top P B)}.
\end{align*}

\hfill $\blacksquare$

\subsection{Proofs for Section~\ref{sec: upper bound}}\label{appendix: upper bound}

From the upper bound in Equation~\eqref{eq:cost gap upper bound}, we need to bound $\norm{\Sigma(F_\gamma)}$ and $\norm{F_\gamma - F_\star}$. For $\norm{\Sigma(F_\gamma)}$, we have by definition:
\begin{align}
    \norm{\Sigma(F_\gamma)} &:= \norm{\sum_{t=0}^\infty (A+BF_\gamma)^t \Sigma_w {(A+BF_\gamma)^\top}^t} \nonumber\\
    &= \sigma_w^2 \norm{\sum_{t=0}^\infty (A+BF_\gamma)^t {(A+BF_\gamma)^\top}^t} \nonumber\\
    &= \sigma_w^2 \norm{W_\infty(A+BF_\gamma, I)}, \label{eq: upper bound on Sigma(K_gamma)}
\end{align}
where we recall $W_\infty(A,B)$ is the controllability gramian, and we are guaranteed from Lemma~\ref{lem:adversarial LQR sol} that $\rho(A+BF_\gamma) < 1$, since $\gamma \geq \gamma_\infty$. We note that there exists a uniform constant upper bound on $\norm{W_\infty(A+BF_\gamma, I)}$ independent of $\gamma$, since $W_\infty(A+BF_{\gamma_\infty}, I)$ and $W_\infty(A+BF_\star, I)$ are both well-defined. From the intuition that $W_\infty(A+BF_\gamma, I)$ represents the ``disturbability'' of the closed-loop system $A+BF_\gamma$, i.e.\ the controllability of $A+BF_\gamma$ treating disturbances as inputs, one may conjecture some monotone behavior of $\norm{W_\infty(A + BF_\gamma, I)}$ between $\gamma = \gamma_\infty$ and $\gamma = \infty$ (i.e.\ going from the $\calH_\infty$ robust closed-loop system to the nominal closed-loop system), however proving such behavior of the curve drawn by $\gamma$ is outside the scope of this work. Alternatively, using norm bounds on the difference of block-Toeplitz matrices (such as those in \cite{mania2019certainty}), one can show that for sufficiently large $\gamma$, an upper bound on $\norm{W_\infty(A+BF_\gamma, I)}$ using only nominal system parameters can be derived, but these bounds tend to be overly conservative, as they throw out the structure provided by the adversarial LQ formulation. For illustrative purposes, we leave this characterization of $\norm{\Sigma(F_\gamma)}$ as is.

We may use Lemma~\ref{lem: controller gap to DARE gap} to get an upper bound on $\norm{F_\gamma - F_\star}$:
\begin{align}
    \norm{F_\gamma - F_\star} &\leq \frac{\norm{B^\top(M_\gamma - P_\star) (A+BF_\star)}}{\sigma_{\min}(R + B^\top P_\star B)}.
\end{align}
This reduces our problem to upper bounding $\norm{M_\gamma - P_\star}$ As discussed in Section~\ref{sec: upper bound}, this can be further upper bounded by
\begin{align*}
    \norm{M_\gamma - P_\star} &\leq \norm{P_\gamma - P_\star} + \frac{\norm{P_\gamma}^2}{\gamma^2 - \norm{P_\gamma}}.
\end{align*}
Therefore, our problem reduces to upper bounding $\norm{P_\gamma}$ and $\norm{P_\gamma - P_\star}$. For upper bounding $\norm{P_\gamma}$, we use the following lemma:
\begin{lemma}\label{lem: gamma ordering}
    Given $(A,B,Q^{1/2})$ stabilizable and detectable, for $\gamma_1 > \gamma_2 \geq \gamma_\infty$, we have
    \begin{align*}
        0 \prec P_{\gamma_1} \prec P_{\gamma_2} \preceq P_{\gamma_\infty} \preceq \gamma_\infty^2,
    \end{align*}
    where $P_{\gamma_1}$ and $P_{\gamma_2}$ are the solutions to the DARE in Lemma~\ref{lem:adversarial LQR sol} Equation~\eqref{eq:modified DARE} at levels $\gamma_1,\gamma_2$ respectively.
\end{lemma}

\textit{Proof:} This lemma follows by observing that since a smaller $\gamma$ corresponds to a smaller penalty on the adversary, for any initial condition, the corresponding adversarial LQ cost~\eqref{eq:adversarialLQRobjective} will be larger. To see this, let $\curly{\delta^{\gamma_1}_t}$ be optimal adversarial perturbations corresponding to level $\gamma_1$. Then $\curly{\delta^{\gamma_1}_t}$ are clearly feasible adversarial perturbations to the adversarial LQ instance at level $\gamma_2$, and since $\gamma_1 > \gamma_2$, the corresponding adversarial cost at level $\gamma_2$ must be larger. The maximal adversarial perturbations at level $\gamma_2$, $\curly{\delta^{\gamma_2}_t}$, can only incur higher cost than that, and thus we can show
\begin{align*}
    x_0^\top P_{\gamma_1}x_0 &= \limsup_{T \to \infty}  
    \frac{1}{T} \paren{\max_{\delta \mbox{ causal}} x_T^\top Q x_T + \sum_{t=0}^{T-1} x_t \paren{Q + F_{\gamma_1}^\top R F_{\gamma_1}} x_t - \gamma_1^2 \delta_t^\top \delta_t } \\
    &=: \limsup_{T \to \infty}  
    \frac{1}{T} \paren{x_T^\top Q x_T + \sum_{t=0}^{T-1} x_t \paren{Q + F_{\gamma_1}^\top R F_{\gamma_1}} x_t - \gamma_1^2 {\delta^{\gamma_1}_t}^\top \delta^{\gamma_1}_t } \\
    &< \limsup_{T \to \infty}  
    \frac{1}{T} \paren{x_T^\top Q x_T + \sum_{t=0}^{T-1} x_t \paren{Q + F_{\gamma_2}^\top R F_{\gamma_2}} x_t - \gamma_2^2 {\delta^{\gamma_1}_t}^\top \delta^{\gamma_1}_t } \\
    &\leq \limsup_{T \to \infty}  
    \frac{1}{T} \paren{\max_{\delta \mbox{ causal}} x_T^\top Q x_T + \sum_{t=0}^{T-1} x_t \paren{Q + F_{\gamma_2}^\top R F_{\gamma_2}} x_t - \gamma_2^2 \delta_t^\top \delta_t } \\
    &= x_0^\top P_{\gamma_2} x_0.
\end{align*}
Thus, we have shown $x_0^\top P_{\gamma_1} x_0 < x_0^\top P_{\gamma_2} x_0$ for any $x_0$, and thus $P_{\gamma_1} \prec P_{\gamma_2}$. From Lemma~\ref{lem:adversarial LQR sol}, this further implies
$P_{\gamma_2} \prec P_{\gamma_\infty} \prec \gamma_\infty^2 I$. \hfill$\blacksquare$

Therefore, Lemma~\ref{lem: gamma ordering} immediately implies $\norm{P_\gamma} \prec \gamma_\infty^2$, and thus:
\begin{align}\label{eq: upper bound second term}
    \frac{\norm{P_\gamma}^2}{\gamma^2 - \norm{P_\gamma}} \leq \frac{\gamma_\infty^4}{\gamma^2 - \gamma_\infty^2}.
\end{align}
Upper bounding $\norm{P_\gamma - P_\star}$ is a bit more involved. As previewed in Section~\ref{sec: upper bound}, we show that defining 
\begin{align}
    \label{eq:perturbed dynamics}
    \tilde{A} &= \paren{I + \paren{\gamma^2 I - P_\gamma}^{-1} P_\gamma} A, \quad
    \tilde{B} = \paren{I  + \paren{ \gamma^2 I - P_\gamma}^{-1} P_\gamma} B
\end{align}
we can re-write the state update under the adversary as
\begin{align*}
    x_{t+1} &= Ax_t + Bu_t + \delta_t = \tilde{A}x_t + \tilde{B}u_t.
\end{align*}
We then show that the gaps $\norm{\tilde{A} - A},\;\norm{\tilde{B} - B}$ can be upper bounded in terms of $\gamma$ in the following lemma, after which we may adapt some techniques introduced in Proposition~3 from \cite{mania2019certainty} to get an upper bound on $\norm{P_\gamma - P_\star}$.

\begin{lemma}\label{lem: eps to gamma conversion}
Given an arbitrary control sequence $\curly{u_t}_{t\geq 0}$, and an adversarial noiseless LQR instance at level $\gamma > 0$, we recall the adversary $\delta_t$ can be expressed as:
\begin{align*}
    \delta_t &= \paren{\gamma^2 I - P_\gamma }^{-1} P_\gamma \paren{Ax_t + Bu_t}.
\end{align*}
Defining $\tilde{A}, \tilde{B}$ as in Equation~\eqref{eq:perturbed dynamics}, we have the following perturbation bounds 
\begin{align*}
    \norm{\tilde{A} - A} &\leq \frac{\gamma_\infty^2}{\gamma^2 - \gamma_\infty^2 } \norm{A} \\
    \norm{\tilde{B} - B} &\leq \frac{\gamma_\infty^2}{\gamma^2 - \gamma_\infty^2 } \norm{B}.
\end{align*}
Therefore, given a desired level $\zeta > 0$, choosing the adversarial level
\begin{align*}
    \gamma^2 \geq \gamma_\infty^2 + \frac{\gamma_\infty^2}{\zeta}\max\curly{\norm{A}, \norm{B}},
\end{align*}
guarantees $||\tilde{A} - A||, ||\tilde{B} - B|| \leq \zeta$.

\end{lemma}

\textit{Proof:} We can apply rudimentary matrix bounds combined with Lemma~\ref{lem: gamma ordering} to get
\begin{align*}
    \norm{\tilde{A} - A} &= \norm{\paren{P_{\gamma} - \gamma^2 I}^{-1} P_\gamma A} \\
    &\leq \norm{\paren{P_{\gamma} - \gamma^2 I}^{-1}} \norm{P_\gamma} \norm{A} \\
    &\leq \frac{\norm{P_\gamma}}{\gamma^2  - \norm{P_{\gamma}}} \norm{A} \\
    &\leq \frac{\gamma_\infty^2}{\gamma^2 - \gamma_\infty^2} \norm{A},
\end{align*}
which we may repeat for $\norm{\tilde{B} - B}$. Inverting
\[
\frac{\gamma_\infty^2}{\gamma^2 - \gamma_\infty^2} \max\curly{\norm{A},\norm{B}} \leq \zeta,
\]
for $\gamma^2$, we get the desired result. \hfill $\blacksquare$






\begin{proposition}[Upper Bound on $\gamma$-Adversarial versus Nominal DARE Solution]\label{prop: advrobust DARE upper bound ctrlbility}
Let $P_\star$ be the solution to $\dare\paren{A, B, Q, R}$, and $P_\gamma$ be the solution to $\dare\paren{A, \bmat{B & I}, Q,\bmat{R & 0 \\ 0 & -\gamma^2 I}}$, and let $(A,B, Q^{1/2})$ be controllable and detectable. Define $\kappa(Q, R) = \frac{\max\curly{\sigma_{\max}(Q), \sigma_{\max}(R)}}{\min\curly{\sigma_{\min}(Q), \sigma_{\min}(R)}}$,
$\tau(A, \rho) := \sup\curly{\norm{A^k}\rho^{-k}: k \geq 0}$, 
and $\beta = \max\curly{1, \frac{\gamma_\infty^2}{\gamma^2 - \gamma_\infty^2} \max\curly{\norm{A}, \norm{B}} \tau\paren{A, \rho} + \rho}$, where $\rho > \rho(A)$. Then for
\begin{align*}
    \gamma^2 \geq \gamma_\infty^2 + \frac{3}{2} \ell^{3/2} \beta^{\ell - 1} \sigma_{\min}(W_\ell(A,B))^{-1/2} \tau(A, \rho)^2 \paren{\norm{B} + 1} \max\curly{\norm{A}, \norm{B}}  \gamma_\infty^2,
\end{align*}
the following bound holds:
\begin{align}
    \norm{P_\gamma - P_\star} &\leq 16 \frac{\gamma_\infty^2 \max\curly{\norm{A}, \norm{B}}}{\gamma^2 - \gamma_\infty^2} \ell^{5/2} \beta^{2(\ell - 1)}  \paren{1 + \sigma_{\min}(W_\ell(A,B))^{-1/2}} \tau(A, \rho)^3  \paren{\norm{B}+1}^2 \kappa(Q,R)  \norm{P_\star}.
\end{align}

\end{proposition}

The proof of Proposition \ref{prop: advrobust DARE upper bound ctrlbility} is mostly similar to Proposition~3 in \cite{mania2019certainty}. However, we note that our requirement on the size of $\gamma$ is much more lenient than the requirement on $\norm{\tilde{A} -A},\norm{\tilde{B}-B}$ in Proposition 3 of \cite{mania2019certainty}. In short, this is because in our setting we are guaranteed $P_{\gamma} \succ P_\star$, and thus $\norm{P_\gamma - P_\star} = \lambda_{\max}(P_{\gamma} - P_\star)$. Therefore, upper bounding $\lambda_{\max}(P_\gamma - P_\star)$ suffices. Unlike the general case of perturbed DARE solutions, we do not need the gap to be small enough for the upper bound on $\lambda_{\max}(P_\gamma - P)$ to be less than $1/2$. Our only size requirement on $\gamma$ comes from guaranteeing the minimum singular value of the $\ell$-controllability gramian of $(\tilde{A}, \tilde{B})$ to be at least $1/2$ that of the nominal system $(A,B)$. 

Combining Lemma~\ref{lem: controller gap to DARE gap}, Equations~\eqref{eq: upper bound on Sigma(K_gamma)} and~\eqref{eq: upper bound second term}, and Proposition~\ref{prop: advrobust DARE upper bound ctrlbility} together yields the final upper bound on $\NC(K_\gamma) - \NC(K)$.





\subsection{Proof of Theorem~\ref{thm:lower bound}}

We start by stating and proving the spectral upper and lower bounds mentioned in Section~\ref{s: p-r tradeoffs lower bounds}.
\begin{lemma}     \label{lem: spectral UB+LB}
Under the assumption $(A,B,Q^{1/2})$ is stabilizable and detectable, and for $\gamma \geq \tilde{\gamma}_\infty$, we have the following bounds:
    \begin{align*}
        \tilde P_\gamma - P_\gamma 
        &\preceq \norm{F_\gamma-F_\star}^2 \norm{R+B^\top M_\gamma B} W_\infty(\tilde A + \tilde B F_\star, I) \\
        P_\gamma(\gamma^2 I - P_\gamma)^{-1} P_\gamma +  \tilde P_\gamma - P_\star &\succeq \frac{\sigma_{\min}(P_\star)^2}{\gamma^2 - \sigma_{\min}\paren{P_\star}} \paren{W_\infty(A+BF_\star, I)},
    \end{align*}
    where $\tilde A = \paren{I+(\gamma^2 I-\tilde P_\gamma)^{-1}\tilde P_\gamma} A$ and $\tilde B = \paren{I+(\gamma^2 I-\tilde P_\gamma)^{-1}\tilde P_\gamma} B$.
\end{lemma}

\textit{Proof:} We start with the upper bound. Begin by recalling that for any $x$, $x^\top \tilde P_\gamma x$ is the cost of applying controller $F_\star$ in the noiseless adversarial setting at level $\gamma$ from initial state $x$, while $P_\gamma$ is the cost of applying controller $F_\gamma$ in the noiseless adversarial setting at level $\gamma$ from initial state $x$. If we denote the optimal map from the state to the adversary perturbation in the cases where controller $F_\star$ and $F_\gamma$ are deployed as $\tilde E_\gamma$ and $E_\gamma$ respectively, then we may consider this as the cost gap of applying the controller
\[
    \bmat{u_t \\ \delta_t} = \bmat{F_\star \\ \tilde E_\gamma} x_t
\]
to the system 
\[
    x_{t+1}=Ax_t + \bmat{B & I} \bmat{u_t \\ \delta_t}
\]
with cost $x_t^\top Q x_t + \bmat{u_t \\ \delta_t}^\top \bmat{R & 0\\0 & -\gamma^2 I} \bmat{u_t \\ \delta_t}$. Similarly, we can consider applying the controller
\[
    \bmat{u_t \\ \delta_t} = \bmat{F_\gamma \\ E_\gamma} x_t,
\]
which is optimal in this setting. 

Then by Lemma 10 in \cite{fazel2018global}, we have that
\begin{align*}
    x^\top \tilde P_\gamma x - x^\top P_\gamma x = \sum_{t=0} \paren{x_t'}^\top \bmat{F_\star - F_\gamma \\ \tilde E_\gamma - E_\gamma}^\top\paren{\bmat{R & 0\\0 & -\gamma^2 I} + \bmat{B & I}^\top P_\gamma \bmat{B & I}}\bmat{F_\star - F_\gamma \\ \tilde E_\gamma - E_\gamma} x_t'
\end{align*}
where $x_0' = x$, and $x_{t+1}' = (A+BF_\star + \tilde E_\gamma) x_t' $. Next, observe that 
\begin{align*}
    \bmat{R & 0\\0 & -\gamma^2 I} + \bmat{B & I}^\top P_\gamma \bmat{B & I} \preceq \bmat{R+B^\top M_\gamma B & 0 \\ 0 & 0}.
\end{align*}
To see that this is the case, we will show that
\begin{align*}
   \bmat{B^\top (P_\gamma - M_\gamma) B & B^\top P_\gamma \\ P_\gamma B & P_\gamma - \gamma^2 I} \preceq 0.
\end{align*}
Expanding $M_\gamma = P_\gamma + P_\gamma (\gamma^2 I- P_\gamma)^{-1} P_\gamma$, and multiplying the expression by $-1$, we can see that this is equivalent to showing that
\begin{align*}
   \bmat{B^\top P_\gamma(\gamma^2I - P_\gamma)^{-1}P_\gamma B & -B^\top P_\gamma \\ -P_\gamma B & \gamma^2I - P_\gamma} \succeq 0.
\end{align*}
This follows from a Scur complement check for positive definiteness by the fact that $\gamma^2 I - P_\gamma \succ 0$. Then we have that
\begin{align*}
    x^\top \tilde P_\gamma x - x^\top P_\gamma x \leq  \sum_{t=0} \paren{x_t'}^\top (F_\star - F_\gamma)^\top\paren{R+B^\top M_\gamma B}(F_\star - F_\gamma) x_t',
\end{align*}
or, by the fact that $x_{t+1}' = A+BF_\star + \tilde E_\gamma x_t' = \paren{I+(\gamma^2 I-\tilde P_\gamma)^{-1}\tilde P_\gamma}(A+BF_\star) x_t'$,
\begin{align*}
    \tilde P_\gamma - P_\gamma &\preceq \sum_{t=0} \paren{\paren{\paren{I + (\gamma^2 I - \tilde P_\gamma)^{-1} \tilde P_\gamma} (A+BF_\star)}^t}^\top  (F_\star - F_\gamma)^\top \\
    &\quad\quad\quad \paren{R+B^\top M_\gamma B}(F_\star - F_\gamma) \paren{\paren{I + (\gamma^2 I - \tilde P_\gamma)^{-1} \tilde P_\gamma} (A+BF_\star)}^t 
\end{align*}
as we wanted to show.

Moving on to the lower bound, note that we may express 
\begin{align*}
    \tilde P_\gamma &= \dlyap\paren{(I + H) (A+BF_\star), Q + F_\star^\top R F_\star -\gamma^2 (A+BF_\star)^\top H^\top H (A+BF_\star)} \\
    P_\star &= \dlyap \paren{A + BF_\star, Q + F_\star^\top RF_\star} \\
    H &= (\gamma^2 I - \tilde P_\gamma)^{-1} \tilde P_\gamma
\end{align*}
Then 
\begin{align*}
    \tilde P_\gamma - P_\star &= (A+BF_\star)^\top (I+H)^\top \tilde P_\gamma (I+H) (A+BF_\star) - \gamma^2 (A+BF_\star)^\top H^\top H (A+BF_\star)  \\
    & \quad \quad - (A+BF_\star)^\top  P_\star (A+BF_\star) \\
    &= (A+BF_\star)^\top (\tilde P_\gamma-  P_\star) (A + BF_\star)  + (A+BF_\star)^\top \paren{H^\top \tilde P_\gamma + \tilde P_\gamma H + H^\top \tilde P_\gamma H - \gamma^2 H^\top H} (A+BF_\star)
\end{align*}
However, we have that 
\begin{align*}
    H^\top \tilde P_\gamma = \tilde P_\gamma H = \tilde P_\gamma (\gamma^2 I - \tilde P_\gamma) \tilde P_\gamma 
\end{align*}
and 
\begin{align*}
    H^\top \tilde P_\gamma H - \gamma^2 H^\top H =  \paren{\tilde P_\gamma - \gamma^2 I}  H = \tilde P_\gamma^\top (\gamma^2 I - \tilde P_\gamma) \tilde P_\gamma.
\end{align*}
Then 
\[
    \tilde P_\gamma -  P_\star = \dlyap(A+BF_\star, (A+BF_\star)^\top \tilde P_\gamma(\gamma^2 I - \tilde P_\gamma)^{-1} \tilde P_\gamma (A+BF_\star)).
\]

We can expand the Lyapunov solution as an infinite sum:
\[
    \tilde P_\gamma -  P_\star = \sum_{t=1}^\infty \paren{(A+BF_\star)^t}^\top \tilde P_\gamma(\gamma^2 I - \tilde P_\gamma)^{-1} P_\gamma (A+BF_\star)^t.
\] 
Next, note that
\[
    \tilde P_\gamma(\gamma^2 I - \tilde P_\gamma)^{-1} \tilde P_\gamma \succeq \frac{\sigma_{\min}(\tilde P_\gamma)^2}{\gamma^2- \sigma_{\min}\paren{\tilde P_\gamma}} I \succeq \frac{\sigma_{\min}(P_\star)}{\gamma^2 - \sigma_{\min}\paren{P_\star}} I
\]  
and thus 
\[
    \tilde P_\gamma -  P \succeq \frac{\sigma_{\min}(P_\star)^2}{\gamma^2 - \sigma_{\min}\paren{P_\star}}\sum_{t=1}^\infty \paren{(A+BF_\star)^t}^\top  (A+BF_\star)^t 
\]
Adding in the term 
\[
    P_\gamma(\gamma^2 I - P_\gamma)^{-1} P_\gamma \succeq \frac{\sigma_{\min}( P_\gamma)^2}{\gamma^2- \sigma_{\min}\paren{ P_\gamma}} I \succeq \frac{\sigma_{\min}(P_\star)}{\gamma^2 - \sigma_{\min}\paren{P_\star}} I
\]
allows us to start the sum from $0$, providing the desired result. \hfill $\blacksquare$

\begin{lemma}
    \label{lem:controller gap LB}
    As long as $\gamma$ satisfies
    \begin{align*}
        \gamma^2 &\geq \sigma_{\min}(P_\star) + \sigma_{\min}(P_\star)^2 \sigma_{\min}(A+BF_\star)^2 \frac{1}{2 \norm{R+B^\top M_\gamma B}} \\
        &\quad \quad \quad \norm{B^\top W_\infty\paren{\paren{I + (\gamma^2 I - \tilde P_\gamma)^{-1} \tilde P_\gamma}(A+BF_\star), I}}
        \norm{B^\top W_\infty(A+BF_\star, I) },
    \end{align*}
    we have the following bound
    \begin{align*}
        \norm{F_\gamma - F_\star} \geq \frac{\sigma_{\min}(P_\star)^2 \norm{B^\top W_\infty(A+BF_\star,I)} \sigma_{\min}(A+BF_\star) }{2\norm{R + B^\top  M_\gamma B} (\gamma^2 - \sigma_{\min}\paren{P_\star})}
    \end{align*}
\end{lemma}

\textit{Proof:}
Beginning with the lower bound in \eqref{eq:lb on controller gap} and applying Lemma~\ref{lem: spectral UB+LB}, we find
\begin{align*}
    \norm{F_\gamma-F_\star} &\geq  \frac{\sigma_{\min}(A+BF_\star)}{\norm{R + B M_\gamma B}} \bigg(\frac{\sigma_{\min}(P_\star)}{\gamma^2 - \sigma_{\min}(P_\star)}\norm{B^\top W_\infty(A+BF_\star, I)  } \\
    &\quad\quad\quad -\norm{B^\top W_\infty\paren{\paren{I + (\gamma^2 I - \tilde P_\gamma)^{-1} \tilde P_\gamma} (A+BF_\star), I} } \norm{F_\gamma-F_\star}^2 \norm{R+B^\top M_\gamma B}\bigg).
\end{align*}

In particular, by defining 
\begin{align*}
    a &= \sigma_{\min}(A+BF_\star) \norm{B^\top W_\infty\paren{\paren{I + (\gamma^2 I - \tilde P_\gamma)^{-1} \tilde P_\gamma} (A+BF_\star), I}}  \\
    c &= \frac{\sigma_{\min}(P_\star)^2}{\gamma^2 - \sigma_{\min}(P_\star)}\frac{\norm{B^\top W_\infty(A+BF_\star,I)}}{\norm{R + B^\top M_\gamma B}} \sigma_{\min}(A+BF_\star),
\end{align*}
the above bound can be written
\begin{align*}
    a\norm{F_\gamma-F_\star}^2 + \norm{F_\gamma-F_\star} -c \geq 0
\end{align*}
which, considering that $\norm{F_\gamma-F_\star} \geq 0$, implies that
\begin{align*}
    a\norm{F_\gamma-F_\star} \geq \sqrt{\frac{1}{4}  + ac} - \frac{1}{2}.
\end{align*}
Under the assumption that $ac \leq 2$, this may be simplified to $\norm{F_\gamma-F_\star} \geq \frac{1}{2} c$. $\blacksquare$

Theorem~\ref{thm:lower bound} then follows from the general lower bound in equation \eqref{eq:cost gap lower bound} and the bound from Lemma~\ref{lem:controller gap LB}.

\section{Proofs for \Cref{s: state prediction}}

\subsection{Proofs for \Cref{s: adv state only}}
\label{s: proofs for adv state upper}
\begin{proposition}\label{prop: advrobust covariance DARE upper bound obsility}
Let $\Sigma_\star$ be the solution to 
\[
\dare\paren{A^\top, C_2^\top, B_0B_0^\top, D_{20}D_{20}^\top},
\]
and $\Sigma_\gamma$ be the solution to 
\[  
    \dare\paren{A^\top (I + \Lambda_\gamma)^\top, C_2^\top (I+\Lambda_\gamma)^\top, B_0 B_0^\top, D_{20} D_{20}^\top},
\] 
where 
\[
    \Lambda_\gamma = B_1(\gamma^2 I - B_1^\top Y_\gamma B_1)^{-1} B_1^\top Y_\gamma.
\] Define $\bar Y$ such that $\bar Y \geq \norm{Y_\gamma}$ for all $\gamma$ such that the conditions of \eqref{eq: adv rob observer gain} are satisifed. Suppose that $(A, C_2)$ is observable. Define $\kappa(B_0 B_0^\top, D_{20} D_{20}^\top ) = \frac{\max\curly{\sigma_{\max}(B_0B_0^\top), \sigma_{\max}(D_{20} D_{20}^\top)}}{\min\curly{\sigma_{\min}(B_0 B_0^\top), \sigma_{\min}(D_{20} D_{20}^\top)}}$,
$\tau(A, \rho) := \sup\curly{\norm{A^k}\rho^{-k}: k \geq 0}$, 
and $\beta = \max\curly{1, \frac{\norm{B_1}^2 \bar Y}{\gamma^2 - \norm{B_1}^2 \bar Y} \max\curly{\norm{A}, \norm{C_2}} \tau\paren{A, \rho} + \rho}$, where $\rho > \rho(A)$. Then for
\begin{align*}
    \gamma^2 \geq  \norm{B_1}^2 \bar Y + \frac{3}{2} \ell^{3/2} \beta^{\ell - 1} \sigma_{\min}(W_\ell(A^\top,C^\top))^{-1/2} \tau(A, \rho)^2 \paren{\norm{C} + 1} \max\curly{\norm{A}, \norm{C_2}}  \norm{B_1}^2 \bar Y,
\end{align*}
the following bound holds:
\begin{align*}
    \norm{\Sigma_\gamma - \Sigma_\star} &\leq 16 \frac{\norm{B_1}^2 \bar Y}{\gamma^2 - \norm{B_1}^2 \bar Y} \max\curly{\norm{A}, \norm{C_2}} \ell^{5/2} \beta^{2(\ell - 1)} \\
    &\quad \cdot \paren{1 + \sigma_{\min}(W_\ell(A^\top,C_2^\top))^{-1/2}} \tau(A, \rho)^3  \paren{\norm{C_2}+1}^2 \kappa(B_0B_0^\top,D_{20}D_{20}^\top)  \norm{\Sigma_\star}.
\end{align*}

\end{proposition}

As with \Cref{prop: advrobust DARE upper bound ctrlbility}, the proof of \Cref{prop: advrobust covariance DARE upper bound obsility} follows the proof of Proposition~3 in \cite{mania2019certainty}. The modifications to reach \Cref{prop: advrobust DARE upper bound ctrlbility} are the same as those to reach \Cref{prop: advrobust covariance DARE upper bound obsility}. 

\scalarLBlemma*
\textit{Proof:}
Begin by observing that
\begin{align*}
    \Sigma_\gamma -  \tilde \Sigma_\gamma &= (I+\Lambda_\gamma)\tilde \Sigma_\gamma (I+\Lambda_\gamma) - \tilde \Sigma_\gamma \\
    &= (I + B_1(\gamma^2 I - B_1^\top Y_\gamma B_1)^{-1} B_1^\top Y_\gamma) \tilde \Sigma_\gamma (I + B_1(\gamma^2 I - B_1^\top Y_\gamma B_1)^{-1} B_1^\top Y_\gamma)^\top -\tilde \Sigma_\gamma.
\end{align*}
Letting $B_1^\top Y_\gamma^{1/2} = U \Sigma V^\top$, we find that 
\begin{align*}
    \Sigma_\gamma -  \tilde \Sigma_\gamma &= (I + Y_\gamma^{-1/2} V \Sigma^2(\gamma^2 I - \Sigma^2)^{-1} V^\top Y_\gamma^{1/2})^\top \tilde \Sigma_\gamma (I + Y_\gamma^{-1/2} V \Sigma^2(\gamma^2 I - \Sigma^2)^{-1} V^\top Y_\gamma^{1/2}) -\tilde \Sigma_\gamma \\
    &=  Y_\gamma^{1/2} V \gamma^2 I(\gamma^2 I - \Sigma^2)^{-1} V^\top Y_\gamma^{-1/2} \tilde \Sigma_\gamma Y_\gamma^{-1/2} V \gamma^2 I(\gamma^2 I - \Sigma^2)^{-1} V^\top Y_\gamma^{1/2} -\tilde \Sigma_\gamma \\
    &= Y_\gamma^{1/2} V \brac{\gamma^2 I(\gamma^2 I - \Sigma^2)^{-1} V^\top Y_\gamma^{-1/2} \tilde \Sigma_\gamma Y_\gamma^{-1/2} V \gamma^2 I(\gamma^2 I - \Sigma^2)^{-1} - V^\top Y_\gamma^{-1/2} \tilde \Sigma_\gamma Y_\gamma^{-1/2} V }V^\top Y_\gamma^{1/2}.
\end{align*}
Then 
\begin{align*}
    \sigma_{\min}(\Sigma_\gamma -  \tilde \Sigma_\gamma) &\geq \sigma_{\min}(Y_\gamma) \paren{\sigma_{\min}(\gamma^2 I(\gamma^2 I - \Sigma^2) V^\top Y_\gamma^{-1/2} \tilde \Sigma_\gamma Y_\gamma^{-1/2} V \gamma^2 I(\gamma^2 I - \Sigma^2)) - \norm{Y_\gamma^{-1/2} \tilde \Sigma_\gamma Y_\gamma^{-1/2}}} \\
    &\geq \sigma_{\min}(Y_\gamma) \paren{\paren{\frac{\gamma^2}{\gamma^2 - \norm{B_1^\top Y_\gamma^{1/2}}}}^2 \frac{1}{\norm{Y_\gamma}} \sigma_{\min}(\tilde \Sigma_\gamma) - \norm{\tilde \Sigma_\gamma}/\sigma_{\min}(Y_\gamma)}.
\end{align*}
\hfill $\blacksquare$

\subsection{Proofs for \Cref{s: scalar state and measurement}}

\taylorFilterGain*
\textit{Proof: }
The problem may be expressed as finding a solution to
\[
    ac \frac{\Sigma_\gamma Y_\gamma}{\gamma^2} L^2 + \paren{(a^2 - c^2)\frac{\Sigma_\gamma Y_\gamma}{\gamma^2}  +(1+c^2 \Sigma_\gamma)}L +ac\Sigma_\gamma - ac \frac{\Sigma_\gamma Y_\gamma}{\gamma^2}.
\]

Viewing this as a general quadratic formula $c_1 L^2 + c_2 L + c_3 = 0$, the solution of interest is 
\begin{align*}
    L = \frac{-c_2 + \sqrt{c_2^2 - 4 c_1 c_3}}{2c_1} &= \frac{c_2}{2c_1} \paren{-1 + \sqrt{1 - 4 \frac{c_1 c_3}{c_2^2}}}
\end{align*}
Taylor's formula tells us that for some $c \in [0,x]$, 
\[
    \sqrt{1+x} = 1 + \frac{x}{2} -\frac{x^2}{8(1+c)^{3/2}}
\]
Therefore, if $\abs{x} < \frac{1}{2}$, 
\[
    \sqrt{1+x} = 1 + \frac{x}{2} + \alpha x^2,
\]
where  $\abs{\alpha}<\frac{1}{2}$. Substituting this in, we find that as long as $4\frac{\abs{c_1c_3}}{c_2^2} \leq \frac{1}{2}$, the solution of interest for the quadratic formula is
\begin{align*}
    L &= \frac{c_2}{2c_1} \paren{-2\frac{c_1c_3}{c_2^2} +\alpha 16 \frac{(c_1c_3)^2}{c_2^4}} \\
    &= -\frac{c_3}{c_2} \paren{1 + 8\alpha \frac{c_1c_3}{c_2^2}}
\end{align*}
where $\abs{\alpha}<\frac{1}{2}$. Substituting in values of $c_1$, $c_2$ and $c_3$, we may therefore show that as long as \eqref{eq: gamma bounds taylor filter gain} is satisfied, then 
\[
    L = \frac{-ac \Sigma_\gamma}{1 + c^2 \Sigma_\gamma} +  \xi,
\]
where 
\begin{align*}
    \xi &=  \paren{-\frac{c_3}{c_2} - \frac{-ac \Sigma_\gamma}{1 + c^2 \Sigma_\gamma}  } - 8 \frac{c_1 c_3^2}{c_2^3} \\
    &= -\frac{ac\Sigma_\gamma (a^2-c^2) \frac{ Y_\gamma \Sigma_\gamma}{\gamma^2}}{(1+c^2 \Sigma_\gamma)(1+c^2 \Sigma_\gamma + (a^2 - c^2) \frac{Y_\gamma \Sigma_\gamma}{\gamma^2})} + \frac{ac \frac{ Y_\gamma \Sigma_\gamma}{\gamma^2}}{1+c^2 \Sigma_\gamma + (a^2 - c^2) \frac{Y_\gamma  \Sigma_\gamma}{\gamma^2}} \\
    &\qquad - 8 \alpha \frac{\frac{\Sigma_\gamma Y_\gamma}{\gamma^2} (ac)^3 \Sigma_\gamma^2 \paren{1 - \frac{Y_\gamma}{\gamma^2} }^2}{\paren{(a^2 - c^2) \frac{\Sigma_\gamma Y_\gamma}{\gamma^2} + (1+c^2 \Sigma_\gamma)}^3}.
\end{align*}
Again using \eqref{eq: gamma bounds taylor filter gain}, we may bound
\begin{align*}
    \abs{\xi} & \leq 64 \abs{a}\abs{c} \frac{ \Sigma_\gamma Y_\gamma}{\gamma^2}  \paren{\Sigma_\gamma \abs{a^2-c^2} + 1 + (ac \Sigma_\gamma)^2}.
\end{align*}
\hfill $\blacksquare$
\begin{proposition}\label{prop: scalar advrobust covariance DARE upper bound obsility}
Let $\Sigma_\star$ be the solution to 
\[
\dare\paren{a, c, 1, 1},
\]
and $\Sigma_\gamma$ be the solution to 
\[  
    \dare\paren{a(1 + \Lambda_\gamma), c (1+\Lambda_\gamma), 1, 1},
\] 
where 
\[
    \Lambda_\gamma = \frac{Y_\gamma (1+L_\gamma^2)}{\gamma^2 - Y_\gamma (1+L_\gamma^2)}
\] Define $\bar Y$ such that $\bar Y \geq \norm{Y_\gamma}$ for all $\gamma$ such that the conditions of \eqref{eq: adv rob observer gain} are satisifed. Suppose that $\abs{c} > 0$. 
and $\beta = \max\curly{1, \frac{\bar Y(1+((\abs{a} +1)/\abs{c})^2}{\gamma^2 - \bar Y(1+((\abs{a} +1)/\abs{c})^2} \max\curly{\abs{a}, \abs{c}} + 1}$. Then for
\begin{align*}
    \gamma^2 \geq  \bar Y(1+((\abs{a} +1)/\abs{c})^2 + \frac{3}{2}  \abs{c}^{-1} \paren{\abs{c} + 1} \max\curly{\norm{a}, \norm{c}}  \bar Y(1+((\abs{a} +1)/\abs{c})^2,
\end{align*}
the following bound holds:
\begin{align*}
    \norm{\Sigma_\gamma - \Sigma_\star} &\leq 16 \frac{\bar Y(1+((\abs{a} +1)/\abs{c})^2}{\gamma^2 - \bar Y(1+((\abs{a} +1)/\abs{c})^2} \max\curly{\abs{a}, \abs{c}}\paren{1 + \abs{c}^{-1}}  \paren{\abs{c}+1}^2  \norm{\Sigma_\star}.
\end{align*}

\end{proposition}
As with \Cref{prop: advrobust DARE upper bound ctrlbility}, the proof of \Cref{prop: scalar advrobust covariance DARE upper bound obsility} follows the proof of Proposition~3 in \cite{mania2019certainty}. The modifications to reach \Cref{prop: advrobust DARE upper bound ctrlbility} are the same as those to reach \Cref{prop: scalar advrobust covariance DARE upper bound obsility}, followed by substituting in $\ell=1$. 

\predUBscalar*
\textit{Proof:}
Observe that if we denote $\Lambda_\gamma = \frac{Y_\gamma(1+L_\gamma^2)}{\gamma^2 - Y_\gamma (1+L_\gamma^2)}$, then we may write
\begin{align*}
    \Sigma_\gamma = (1 + \Lambda_\gamma)^2(a+ L_\gamma c) \Sigma_\gamma (a+L_\gamma c)^\top (1+ \Lambda_\gamma) + (a+\Lambda_\gamma)^2(1+L_\gamma^2).
\end{align*}
Defining $\tilde \Sigma_\gamma = \frac{ \Sigma_\gamma}{(I+\Lambda_\gamma)^2}$, we have
\begin{align*}
    \tilde \Sigma_\gamma &=  \dare( (a(1+\Lambda_\gamma), c(1+\Lambda_\gamma), 1,  1).
\end{align*}
By the triangle inequality,
\begin{align*}
    \norm{\Sigma_\gamma - \Sigma_\star} \leq \norm{\Sigma_\gamma - \tilde \Sigma_\gamma} + \norm{\tilde \Sigma_\gamma - \Sigma_\star}.
\end{align*}
As in \Cref{sec: upper bound}, we may bound $\norm{\tilde \Sigma_\gamma - \Sigma_\star}$ when both $A\Lambda_\gamma$ and $C\Lambda_\gamma$ are small. To bound $\norm{\Sigma_\gamma-\tilde \Sigma_\gamma}$, note that
\[
    \Sigma_\gamma - \tilde \Sigma_\gamma = (\Lambda_\gamma^2 + 2 \Lambda_\gamma) \tilde \Sigma_\gamma 
\]
Define $\bar Y$ such that $\bar Y \geq Y_\gamma$ for all $\gamma$ such that the conditions of \eqref{eq: adv rob observer gain} are satisfied, and note that for all $L_\gamma$ that render the closed loop system stable, we must have $\abs{L_{\gamma}} \leq \frac{\abs{a}+1}{\abs{c}}$. Then
\[
    \norm{\Sigma_\gamma - \tilde \Sigma_\gamma} \leq 4 \frac{\bar Y(1+((\abs{a} +1)/\abs{c})^2}{\gamma^2 - \bar Y(1+((\abs{a} +1)/\abs{c})^2}\abs{\tilde \Sigma_\gamma},
\]
as long as $\gamma^2 \geq \frac{3}{2} \bar Y(1+((\abs{a} +1)/\abs{c})^2$. 
Combining the above results with \Cref{prop: advrobust covariance DARE upper bound obsility} and the previous results in \cref{s: state prediction}, the theorem follows. 
\hfill $\blacksquare$

\end{document}